\documentclass[epj]{svjour}

\usepackage[usenames,dvipsnames]{xcolor}
\usepackage{graphicx}
\usepackage{dcolumn}
\usepackage{bm}
\usepackage{hyperref}
\usepackage[utf8]{inputenc}
\usepackage{amsmath}
\usepackage{amssymb}

\allowdisplaybreaks

\newcommand{\nn}{\nonumber}
\renewcommand{\vec}[1]{{\bf #1}}

\newcommand{\iunit}{{\rm i}}

\newcommand{\rvec}{\vec{r}}

\newcommand{\etal}{\emph{et al.}}

\newcommand{\vnabla}{\boldsymbol{\mathbf\nabla}}

\newcommand{\NLO}{\text{NLO}}
\newcommand{\NiiLO}{\text{N2LO}}

\usepackage{dsfont}

\usepackage[boxruled,vlined]{algorithm2e}

\graphicspath{{}}

\begin{document}

\title{Iterative approaches to the self-consistent nuclear energy density
       functional problem.}
\subtitle{Heavy ball dynamics and potential preconditioning.}
\titlerunning{Iterative approaches to the self-consistent nuclear EDF problem.}

\author{W. Ryssens\inst{1,2}, M. Bender\inst{2} and P.-H. Heenen\inst{3}}
\institute{ Center for Theoretical Physics, 
            Sloane Physics Laboratory, Yale University, New Haven, CT 06520, USA.
            \and
           IPNL, Universit\'e de Lyon,
           Universit\'e Lyon 1, CNRS/IN2P3,
           F-69622 Villeurbanne, France .
           \and
           PNTPM, CP229, Universit\'e Libre de Bruxelles, B-1050 Bruxelles, Belgium}

\abstract{
Large-scale applications of energy density functional (EDF) methods depend on 
fast and reliable algorithms to solve the associated non-linear self-consistency 
problem. When dealing with large single-particle variational spaces, existing 
solvers can become very slow, and their performance dependent on manual 
fine-tuning of numerical parameters. In addition, convergence can sensitively 
depend on particularities of the EDF's parametrisation under consideration.
Using the widely-used Skyrme EDF as an example, we investigate the impact 
of the parametrisation of the EDF, both in terms of the operator structures 
present and the size of coupling constants, on the convergence of numerical 
solvers. We focus on two aspects of the self-consistency cycle, which are the 
diagonalisation of a fixed single-particle Hamiltonian on one hand and the 
evolution of the mean-field densities and potentials on the other. 
Throughout the article we use a coordinate-space representation, for which 
the behaviour of algorithms can be straightforwardly analysed.
We propose two algorithmic improvements that are easily implementable in 
existing solvers, heavy-ball dynamics and potential preconditioning. We 
demonstrate that these methods can be made virtually parameter-free, requiring 
no manual fine-tuning to achieve near-optimal performance except for isolated
cases. The combination of both methods decreases substantially the CPU 
time required to obtain converged results. The improvements are illustrated for 
the MOCCa code that solves the self-consistent HFB problem in a 3d coordinate 
space representation for parametrisations of the standard Skyrme EDF at 
next-to-leading order in gradients and its extension to next-to-next-to-leading 
order.
}

\maketitle

%%%%%%%%%%%%%%%%%%%%%%%%%%%%%%%%%%%%%%%%%%%%%%%%%%%%%%%%%%%%%%%%%%%%%%%%%%%%%%%%

\section{Introduction}

Many-body techniques based on energy density functionals (EDFs) offer
a microscopic description of both ground-state and excited-state properties
of atomic nuclei~\cite{Bender03}. These methods have in common that the total
binding energy is calculated from a functional of one-body densities that
are constructed from auxiliary product states. On the most basic level,
the self-consistent Hartree-Fock (HF), HF+BCS, or Hartree-Fock-Bogoliubov
(HFB) equations are solved self-consistently to determine a product
state whose densities minimise the total binding energy for a given EDF.
On a more advanced level, correlations beyond the mean field are
described as either a superposition of particle-hole excitations on top
of such a reference state, or by mixing different reference states
constructed with some systematically varied auxiliary conditions.

Over time, many different implementations have been set up, differing in
their choice of numerical representation and in the nature of the EDF. Compared
 with other microscopic methods, the computational cost of
self-consistent mean-field methods scales rather favourably
with system size, rendering the entirety of the nuclear chart accessible.
The self-consistent equations present a nonlinear optimization
problem, requiring iterative techniques to obtain their solution. Considering
that every iteration requires the (approximate) diagonalisation of either a
single-particle Hamiltonian or a quasiparticle Hamiltonian, computational
requirements can nevertheless be substantial when employing large
single-particle bases. As a result, computational resource requirements can
still become a limiting factor, especially when aiming at systematic
calculations across large sets of nuclei.

Large single-particle bases arise naturally in coordinate space approaches. 
Coordinate space approaches have the attractive feature that they offer
easily controllable numerical accuracy, in both infrared and ultraviolet
cutoffs, and this independently of the nuclear configuration~\cite{Ryssens15}.
For an harmonic oscillator (HO) basis, the other widely-used choice, control of
convergence is much more delicate~\cite{Arzhanov16}. However, in full 3d 
geometry already modest choices of spatial discretization lead to matrices whose
 storage exceeds by far 
the typical memory capacity offered by the latest high-performance computing 
facilities. Fortunately, for many 
applications one can avoid dealing with full matrices, as one is only interested
in a limited set of single-particle states with the lowest energy.

In terms of numerical cost, a coordinate-space representation is particularly
competitive for EDFs that yield a local mean-field Hamiltonian. The
non-relativistic Skyrme EDF is arguably the most widely one used among
these. In this case the structure of the EDF is motivated
by the expectation value of contact interactions with gradients and is built
out of products of local densities and currents with internal and external
derivatives. Significant efforts have been made to push the adjustment of 
its parameters such that there are fits that provide an overall
high-quality description of certain classes
of observables~\cite{Goriely09,Kortelainen14}.
In the course of these studies, however, it has become clear that it
is unlikely that any further significant improvement can be made within
the standard form of the Skyrme EDF. This, in turn, motivates the ongoing
study of its systematic extension.

The standard form of the Skyrme EDF includes terms up to second order
in gradients, which in present terminology is called next-to-leading order
(NLO). A possible and already explored way to extend it is to add
higher-order bilinear gradient terms with four (next-to-next-to-leading order,
N2LO) and six (N3LO) gradients \cite{Carlsson08,Raimondi11,Becker17}. 
Already the first exploratory test calculations with our codes involving 
such terms have indicated that widely-used algorithms for solving
the self-consistent problem for standard Skyrme EDFs can fail to 
converge efficiently for these extended functionals, as we will discuss
below.

The aim of this paper is to analyse the origin of the differences of
behaviour between the parametrisations and to single\ out which terms of 
the EDFs govern the achievable convergence rate. We will then show how to 
improve the convergence properties of mean-field calculations, with two main 
requirements: to decrease the computing time and to set up algorithms whose 
convergence does not require a case-by-case fine tuning of numerical parameters 
that renders systematic calculations tedious. We focus on two aspects of the 
solution of the self-consistent mean-field equations in coordinate space. 
The first one
is the diagonalisation of a given fixed mean-field Hamiltonian. We will
call this part of the calculation the \emph{diagonalisation} subproblem. It 
is usually achieved with some variant of the gradient-descent method in which a 
limited set of single-particle wave functions are evolved until they converge to
the eigenstates with the lowest eigenvalues~\cite{Maruhn14,Bonche05}. 
The second aspect is the evolution of the potentials entering the mean-field
Hamiltonian, which depend on one-body densities constructed from the eigenstates
of the mean-field Hamiltonian. We will refer to this part as the
\emph{self-consistent field} (SCF) subproblem.  This second aspect of the 
self-consistent problem is often limited to a linear mixing of densities 
between two iterations. Two further subproblems have to be dealt with when 
solving the mean-field equations: we will address the treatment of 
constraints during the SCF iterations in a second paper, while our present
strategy to treat the pairing subproblem is briefly evoked in appendix A.

We propose two improvements that are easy to implement in existing solvers:
\emph{Heavy ball dynamics} for the diagonalisation of the mean-field Hamiltonian
and \emph{potential preconditioning} for the evolution of the potential.
These proposed improvements were implemented and tested with the MOCCa 
code~\cite{RyssensPhd}. This mean-field solver is based on the same principles 
as the public EV8 code~\cite{Bonche05,Ryssens15a}, but supersedes 
its functionalities by offering a wide range of symmetry options that give 
access to a larger range of applications.

While significant parts of the discussion are tailored to the Skyrme EDF
equations in coordinate-space representation, the algorithms we discuss
and the proposed improvements do neither depend on the choice of EDF nor 
on the choice of numerical representation and can also be easily adapted
to other frameworks. Indeed, some groups also developed Cartesian 
3d coordinate-space HF solvers that can handle the non-local exchange part 
of the finite-range Gogny interaction \cite{Matsuse98,Nakatsukasa13S}. The
additional integrals, however, substantially increase the computational cost,
as would including pairing correlations.
Therefore, systematic HFB calculations with this interaction are habitually 
performed in a HO-basis representation. The matrix elements of Gaussian 
forces then take a separable form which allows for very reasonable
precision at moderate computational cost. Such representation in 
general favours iteration algorithms that evolve directly the Thouless
matrix representing the HFB state \cite{Mang76,Egido95,Robledo11} instead
of the strategy we will discuss below. For effective 
Hamiltonians that cannot be easily mapped on Gaussians, however, the 
advantages of using a HO basis are less evident. For example, the 3d HF 
solver designed for the realistic low-momentum interaction $V_{\text{low} k}$ 
described in Ref.~\cite{vanDalen14} uses the same plane-wave basis that also
underlies Cartesian coordinate-space representations~\cite{Ryssens15}.

This paper is organised as follows. In Section~\ref{sec:selfconsistent}
we introduce the general form of the Skyrme functional used here as well
as the self-consistent problem. We study the diagonalisation of the 
single-particle Hamiltonian in Section~\ref{sec:linearsub} and the iterative
treatment of mean-field densities and mean-field potentials in
Section~\ref{sec:dmixing}. Finally, numerical tests of the proposed
improvements are presented in Section~\ref{sec:numericaltests}.

%%%%%%%%%%%%%%%%%%%%%%%%%%%%%%%%%%%%%%%%%%%%%%%%%%%%%%%%%%%%%%%%%%%%%%%%%%%%%%%

\section{The self-consistent Skyrme-HFB problem on a coordinate space mesh}
\label{sec:selfconsistent}

\subsection{The SCF equations with the two-basis method}

The energy of the nuclear configuration is determined by an energy functional
that consists of five terms~\cite{Bender03,Jodon16}
\begin{equation}
\label{eq:Etot}
E_{\text{tot}}
  =     E_{\text{kin}}
      + E_{\text{Sk}}
      + E_{\text{Coul}}
      + E_{\text{cm}}
      + E_{\text{pair}}
\, .
\end{equation}
These are the kinetic energy $E_{\text{kin}}$, the Skyrme EDF $E_{\text{Sk}}$
modelling the strong interaction between nucleons in the particle-hole
channel, the Coulomb energy $E_{\text{Coul}}$ resulting from the
electromagnetic repulsion between protons, a centre-of-mass correction
$E_{\text{cm}}$ and a pairing EDF modelling the strong interaction between
nucleons of the same  species in the particle-particle channel.
These five terms are provided by functionals of normal and anomalous one-body
densities that are calculated from an auxiliary many-body product state.

Depending on the treatment of pairing correlations, this state is either 
a Slater determinant (HF) or a Bogoliubov quasiparticle vacuum (HF+BCS 
and HFB), which is represented in an underlying basis of single-particle
orbitals $\phi_{j}(\vec{r}) =  \langle \vec{r} | a^\dagger_j | - \rangle$
of dimension $N_b$. Here and in the following, except when needed, we do not 
specify the isospin labels that introduce a block structure into all 
matrices in the single-particle basis.

Within a given single-particle basis, the single-particle Hamiltonian $\hat{h}$
and the pairing tensor $\Delta$ are defined by the functional derivatives of the
EDF with respect to the normal density matrix $\rho$ and the anomalous density
matrix $\kappa$~\cite{RingSchuck}
\begin{align}
\label{eq:defh}
h_{ij}
& =  \frac{\delta E }{\delta \rho_{ji}}
  = h^{*}_{ji}  \, ,
      \\
\label{eq:defdelta}
\Delta_{ij}
&=  \frac{\partial E }{\delta \kappa_{ij}^*}
= - \Delta_{ji} \, .
\end{align}
Together they form the HFB quasiparticle Hamiltonian $\mathcal{H}$
\begin{eqnarray}
\label{eq:HFBhamil}
\mathcal{H}
& = &
\begin{pmatrix}
   h - \lambda  &
  \Delta\\
 -\Delta^* &
 -h^* + \lambda
\end{pmatrix} \, ,
\end{eqnarray}
where the Fermi energy $\lambda$ is a Lagrange multiplier whose value has to be
adjusted to fix the average number of particles to the targeted number of
neutrons ($N$) or protons ($Z$), respectively. The diagonalisation of
$\mathcal{H}$ yields a set of eigenvectors $(U_{\mu} V_{\mu})^T$
\begin{equation}
\label{eq:HFB}
\mathcal{H}
\begin{pmatrix}
       U_{\mu} \\
       V_{\mu}
\end{pmatrix}
=
E_{\mu}^{\rm qp}
\begin{pmatrix}
       U_{\mu} \\
       V_{\mu}
\end{pmatrix}
\end{equation}
that represent the quasi-particle wave functions with quasi-particle
energy $E_{\mu}^{\rm qp}$, and which we call \textit{HFB basis}.

One can solve the problem by diagonalising $\mathcal{H}$ in the full space of 
dimension $2N_b$, which is for instance done in~\cite{Pei14}. When $N_b$ is 
large, however, it becomes challenging to store the relevant matrices, let alone 
diagonalise them. The problem can remain feasible however, if one introduces a 
cutoff of the pairing interaction, reducing the number of quasiparticles to 
those that meaningfully contribute to the auxiliary state.

The two-basis method, first introduced in \cite{Gall94}, offers a
straightforward way to introduce such a cutoff. The method relies on the
construction of the single-particle states that diagonalise $\hat{h}$
\begin{equation}
\label{eq:eigenh}
\hat{h} | \psi_{j} \rangle
= \epsilon_{j} \, | \psi_{j} \rangle
\end{equation} 
that is called the HF basis. Within this basis, a
pairing cutoff can easily be defined as a function of the Fermi energy
$\lambda$ and the diagonal matrix elements of $\hat{h}$. This effectively
limits the relevant space of the HFB problem to a small number 
of states in the HF basis. We denote the effective size of the HF basis 
by $\Omega$. The numerical burden is then completely
shifted to the construction of the HF basis in the full single-particle space.

For this reason, we focus on the self-consistent construction of the HF 
basis in the following. While an appropriate algorithm for the partial 
diagonalisation of $\mathcal{H}$ is an essential ingredient of a successful 
calculation, we assume that a reliable and fast implementation is available. In 
practice, the only output of such algorithm that is required for the 
purpose of our study is the set of matrix elements $\rho_{ij}$ and 
$\kappa_{ij}$ in the HF basis. These can then be used to construct the 
mean-field densities as defined in Section~\ref{sect:Skyrme}.

%-------------------------------------------------------------------------------

\subsection{Representation in 3D coordinate space}
While leaving the discussion as general as possible, we assume
a 3d representation in coordinate space. For the spatial discretization, we 
choose an equidistant Cartesian mesh with $N_m = N_x \times N_y \times N_z$ 
coordinate space coordinates $\vec{r}_{ijk}$, where the total number of points 
is determined by the distance between the discretization points and the volume 
of the box. Any function $f(\vec{r})$ on the mesh is then represented 
by its values at the discretization points
\begin{equation}
f(\vec{r})
\rightarrow f(x_i,y_j,z_k) = f(\vec{r}_{ijk}) \equiv f_{ijk} \,  .
\end{equation}
In particular, the single-particle wave functions $|\psi\rangle$ are 
represented as spinors $\psi(\vec{r})$
\begin{align}
|\psi\rangle & \rightarrow \psi(\vec{r}_{ijk}) =
\begin{pmatrix}
\psi (\vec{r}_{ijk}, \sigma = +) \\
\psi (\vec{r}_{ijk}, \sigma = -)
\end{pmatrix} \, ,
\end{align}
meaning that for each nucleon species one can construct $2 N_m$ different
single-particle wave functions on the mesh. A typical choice is 
$N_X = N_Y = N_Z = 40$ \cite{Ryssens15}, which implies a basis of 
$128000$ single-particle wave functions for each nucleon species.

The integral of an arbitrary function over the volume of the box is given by
the $N_m$-point rectangular rule~\cite{Ryssens15}
\begin{equation}
\label{eq:mesh:int}
\int_V \! d^3 r \; f(r)
\to \sum_{ijk} f_{ijk} .
\end{equation}
The calculation of the derivatives of functions is equivalent to a matrix
multiplication. In the case of the Laplacian of a function $f(\vec{r})$ at
the point $\rvec_{ijk}$, one has to calculate
\begin{align}
\label{eq:mesh:Laplacian}
\lefteqn{\big[ \Delta f \big] (\rvec_{ijk})
} \nn \\
& = \big[  \big( \partial_x^2 + \partial_y^2 + \partial_z^2 \big) \,
    f \big] (\rvec_{ijk})
    \nonumber \\
& = % \sum_{ijk}
    \Big( \sum_{p = 1}^{N_x} D^{(2)}_{ip} f_{pjk}
         +\sum_{q = 1}^{N_y} D^{(2)}_{jq} f_{iqk}
         +\sum_{r = 1}^{N_z} D^{(2)}_{kr} f_{ijr}
    \Big)  ,
\end{align}
where the matrices $D^{(2)}$ represent the second derivative in a given
Cartesian direction. There are several possible choices for their form.
We use Lagrange mesh derivatives~\cite{Ryssens15,Baye86a}, for which
the $D^{(2)}$ are full matrices, as they provide very accurate results
already for very coarse meshes, i.e.\ a modest number of mesh points $N_m$.
The Lagrange derivatives have the clear advantages that partial
integrations are exact up to machine precision, and that higher-order 
derivative matrices can be calculated as products of matrices of first-order 
derivatives. Together with the quadrature rule of Eq.~\eqref{eq:mesh:int}, they 
implicitly define an underlying basis of plane waves in a box~\cite{Ryssens15},
such that this special case of coordinate-space representation can be 
unambiguously treated with the standard techniques of linear algebra.

A widely-used alternative are finite difference formulas. These correspond
to sparse matrices $D^{(2)}$ that can be very efficiently multiplied, but
on a given mesh are much less accurate than Lagrange-mesh derivatives~\cite{Ryssens15}. The
reasons are that partial derivatives are not exact and that higher-order
derivative matrices are not numerically equivalent to repeated applications
of the first-order derivative matrix. This can be compensated for by
increasing the number of mesh points in for a given volume. For NLO
Skyrme EDFs, there is also the time-tested possibility to arrive at very
satisfying results by solving the HFB equations using finite-difference
formulas and to recalculate all observables with Lagrange derivatives
after convergence~\cite{Ryssens15}. For terms with four derivatives in the
Skyrme N2LO EDF, however, this procedure becomes less reliable.

%--------------------------------------------------------------------------
%
\subsection{The Skyrme functional}
\label{sect:Skyrme}

We discuss in detail the two terms from Eq.~\eqref{eq:Etot} that
are relevant for our discussion: the kinetic energy $E_{\rm kin}$ and the 
Skyrme energy $E_{\rm Sk}$.
For a discussion of the other terms, we refer to Ref.~\cite{Ryssens15a}.

The Skyrme part $E_{\rm Sk}$ of the EDF is built out of local one-body
densities. For the purpose of our discussion, it is sufficient to limit
ourselves to the central and spin-orbit parts at NLO, for which we will
analyse time-even and time-odd terms, and the time-even part of the N2LO
functional with additional central terms as proposed in \cite{Becker17}.
Further terms, such as tensor terms at NLO~\cite{Lesinski07,Bender09,Hellemans12},
additional density dependences of NLO terms~\cite{Goriely10}, and three-body
interactions up to NLO \cite{Sadoudi13,Sadoudi13b} have been considered
in the literature, but as far as the numerical solution of the self-consistent
mean-field equations is concerned, these do not behave differently from the
terms discussed here.

\paragraph{Densities.}
Assuming that all single-particle states either represent a neutron or a
proton, the full one-body density matrix for the nucleon species $q=n$, $p$
in coordinate space can be split into a scalar and a vector in spin space
\cite{Perlinska04}
\begin{align}
\rho_q (\vec{r}\sigma, \vec{r}' \sigma')
&= \sum_{jk} \rho_{kj} \, \psi^*_j(\vec{r}'\sigma') \, \psi_k(\vec{r},\sigma)
 \nonumber \\
&= \tfrac{1}{2} \, \rho_q(\vec{r}, \vec{r}') +
  \tfrac{1}{2} \, \langle \sigma' |\hat{\boldsymbol{\sigma}} | \sigma \rangle
  \cdot \vec{s}_q(\vec{r}, \vec{r}') \, ,
\end{align}
The $\rho_{jk}$ are the elements of the density matrix determined by the 
solution of the HFB problem in the restricted set of $\Omega$ states of the HF 
basis. 

The local mean-field densities used to define the Skyrme EDF at NLO in its
form relevant for our discussion are
\begin{subequations}
\label{eq:densities:NLO}
\begin{align}
\rho_q(\vec{r})
&= \rho_q(\vec{r}, \vec{r}') \big|_{\vec{r} = \vec{r}'} \, ,
   \\
\tau_q(\vec{r})
&= \nabla \cdot \nabla' \,
   \rho_q(\vec{r}, \vec{r}')\big|_{\vec{r} = \vec{r}'} \, ,
    \\
J_{q, \mu \nu}(\vec{r})
&= -\tfrac{i}{2} \big(\nabla_{\mu}-\nabla'_{\mu} \big) \,
   s_{q, \nu} (\vec{r}, \vec{r}') \big|_{\vec{r} = \vec{r}'} \, ,
    \\
\vec{s}_q(\vec{r})
&= \vec{s}_q(\vec{r}, \vec{r}') \big|_{\vec{r} = \vec{r}'} \, ,
    \\
\vec{T}_q(\vec{r})
&= \nabla \cdot \nabla' \,
   \vec{s}_q(\vec{r}, \vec{r}') \big|_{\vec{r} = \vec{r}'} \, ,
    \\
\vec{j}_q(\vec{r})
&= -\tfrac{i}{2} \big(\nabla_{\mu}-\nabla'_{\mu} \big)
   \rho_q(\vec{r}, \vec{r}') \big|_{\vec{r} = \vec{r}'} \, .
\end{align}
\end{subequations}
The local density $\rho_q(\vec{r})$,
kinetic density $\tau_q(\vec{r})$ and spin-current tensor density
$J_{q, \mu \nu}(\vec{r})$ are even under time-re\-ver\-sal, whereas the
spin density $\vec{s}_q(\vec{r})$, kinetic spin density $\vec{T}_q(\vec{r})$
and current density $\vec{j}_q(\vec{r})$ are odd. As a consequence, the
latter are zero when time-reversal invariance is imposed.

The N2LO Skyrme functional requires the introduction of
additional local densities that either contain additional derivatives
or have a more complicated tensor structure than the NLO densities.
There are several possible choices~\cite{Ryssens18}; we use here the
convention of Ref.~\cite{Becker17}, where the densities that contribute
in time-reversal-invariant systems are given by
\begin{subequations}
\label{eq:densities:N2LO}
\begin{align}
\tau_{q, \mu\nu} (\vec{r})
& = \nabla_{\mu} \, \nabla_{\nu}' \, \rho_{q}(\vec{r}, \vec{r}')
          \big|_{\vec{r} = \vec{r}'} \, ,
          \label{eq:taudef}
          \\
K_{q, \mu\nu \kappa} (\vec{r})
& = \nabla_{\mu} \, \nabla_{\nu}' \, s_{q,\kappa}(\vec{r}, \vec{r}')
    \big|_{\vec{r} = \vec{r}'} \, , \\
V_{q, \mu \nu} (\vec{r})
&=  - \tfrac{\iunit}{2} \left(\nabla_{\mu}-\nabla'_{\mu} \right) \,
    (\nabla \cdot \nabla') \,
                    s_{q, \nu} (\vec{r}, \vec{r}')
                    \big|_{\vec{r} = \vec{r}'} \, ,
   \\
\label{eq:def:Q}
Q_{q}(\vec{r})
&=  \Delta \, \Delta' \, \rho_{q} (\vec{r}, \vec{r}')
                    \big|_{\vec{r} = \vec{r}'} \, .
\end{align}
\end{subequations}
The densities $Q_q(\vec{r})$ and $V_{q,\mu \nu}(\vec{r})$ are time-even, whereas
$\tau_{q,\mu \nu}(\vec{r})$ and $K_{q, \mu\nu \kappa}(\vec{r})$ are 
neither time-even nor time-odd. Note that the NLO densities $\tau_q(\vec{r})$ 
and $\vec{T}_q(\vec{r})$ are contractions of the full tensor densities 
$\tau_{q, \mu \nu}(\vec{r})$ and $K_{q, \mu\nu \kappa}(\vec{r})$ 
needed for the N2LO functional.

\paragraph{The kinetic energy.}

The kinetic energy $E_{\rm kin}$ can be written as a functional of the local
kinetic density $\tau_q(\vec{r})$ as
\begin{equation}
\label{eq:Ekin}
E_{\rm kin}
= \sum_{q=n,p} \int \! d^3r \, \frac{\hbar^2}{2 m_q} \, \tau_q(\vec{r}) \,,
\end{equation}
where $m_q$ is the mass of the nucleon species $q = n$, $p$.

\paragraph{The Skyrme energy.}
In what follows, we analyse the behaviour
of time-even and time-odd terms at the NLO level on the one hand, and the 
behaviour of the time-even terms in an extended functional with N2LO terms on the
other hand
\begin{align}
\label{eq:skyrme:energy}
E_{\text{Sk}}^{\NLO}
& =  \int \! d^3r \,
      \Big[
        \mathcal{E}^{(0)}_{\text{Sk}, e}(\vec{r})
      + \mathcal{E}^{(2)}_{\text{Sk}, e}(\vec{r})
      + \mathcal{E}^{(0)}_{\text{Sk}, o}(\vec{r})
      + \mathcal{E}^{(2)}_{\text{Sk}, o}(\vec{r})
      \Big] ,
      \\
E_{\text{Sk},e}^{\NiiLO}
& =  \int \! d^3r \,
      \Big[
        \mathcal{E}^{(0)}_{\text{Sk}, e}(\vec{r})
      + \mathcal{E}^{(2)}_{\text{Sk}, e}(\vec{r})
      + \mathcal{E}^{(4)}_{\text{Sk}, e}(\vec{r})
      \Big] ,
\end{align}
where the superscripts $(0)$, $(2)$, and $(4)$ indicate the order in
derivatives and the subscripts $e$ and $o$ if the terms are bilinear in
time-even or time-odd local densities, respectively. The various energy
densities up to NLO read
\begin{eqnarray}
\label{eq:SkTeven:0}
\mathcal{E}^{(0)}_{\rm Sk,e} (\vec{r})
& = & \sum_{t=0,1}
      \Big[
               C^{\rho\rho}_t \, \rho_t^2 (\vec{r})
             + C^{\rho\rho\rho^{\alpha}}_t \rho_0^\alpha (\vec{r}) \, \rho_t^2 (\vec{r})
      \Big] \, ,
      \\
\label{eq:SkTeven:2}
\mathcal{E}^{(2)}_{\rm Sk,e} (\vec{r})
& = & \sum_{t=0,1}
      \Big[
               C^{\rho\tau}_t \, \rho_t (\vec{r}) \, \tau_t (\vec{r})
             + C^{\rho \Delta \rho}_t \, \rho_t (\vec{r}) \, \Delta \rho_t (\vec{r})
      \nonumber \\
&   &
             + C^{\rho \nabla \cdot J}_t  \rho_t (\vec{r}) \, \nabla \cdot \vec{J}_t (\vec{r})
      \nonumber \\
&   &
             - C^{sT}_t \sum_{\mu, \nu} J_{t, \mu \nu} (\vec{r}) \, J_{t, \mu \nu} (\vec{r})
      \Big] \, , \\
\label{eq:SkTodd:0}
\mathcal{E}^{(0)}_{\text{Sk,o}}(\vec{r})
& = & \sum_{t=0,1}
      \Big[ C^{ss}_t \vec{s}_t^2 (\vec{r})
          + C^{ss\rho^\alpha}_t \rho_0^\alpha (\vec{r}) \, \vec{s}_t^2  (\vec{r})
      \Big] \, ,
      \\
\label{eq:SkTodd:2}
\mathcal{E}^{(2)}_{\text{Sk,o}}(\vec{r})
& = & \sum_{t=0,1}
      \Big[
        C^{s T}_t \vec{s}_t (\vec{r}) \cdot \vec{T}_t (\vec{r})
      + C^{s \Delta \vec{s}}_t \vec{s}_t (\vec{r}) \cdot \Delta \vec{s}_t (\vec{r})      \nonumber \\
&   &
      - C_t^{\rho \tau} \vec{j}_t^2 (\vec{r})
      + C_t^{\rho \nabla J}  \vec{s}_t (\vec{r}) \cdot \nabla \times \vec{j}_t (\vec{r})
      \Big]
\, ,
\end{eqnarray}
where proton and neutron densities have been recoupled to isoscalar ($t=0$)
and isovector ($t=1$) ones, for example
$\rho_0(\vec{r}) = \rho_n(\vec{r}) + \rho_p(\vec{r})$ and
$\rho_1(\vec{r}) = \rho_n(\vec{r}) - \rho_p(\vec{r})$. In the last term
of Eq.~\eqref{eq:SkTeven:2} the summation is over Cartesian components of
the tensor density, while the rank-1 contraction $\vec{J}_q$ of the
spin-current density $J_{q,\mu \nu}$ is given by~\cite{Lesinski07}
\begin{equation}
J_{q,\mu}(\vec{r})
= \sum_{\nu \kappa} \epsilon_{\mu \nu \kappa} J_{q,\nu \kappa}(\vec{r}) \, .
\end{equation}
We have written Eqs.~\eqref{eq:SkTeven:2} and~\eqref{eq:SkTodd:2} with the
usual convention that terms that have to be combined in order to ensure
Galilean invariance of the EDF have the same coupling constant
\cite{Hellemans12}.

Similarly, the part of the N2LO extension of the Skyrme functional 
depending on time-even densities can be written as
%
%\begin{strip}
\begin{align}
\label{eq:EDF:4:e}
  \mathcal{E}^{(4)}_{\text{Sk,e}}(\vec{r}) 
= & \sum_{t=0,1}
     \Big[
           C_t^{\Delta \rho \Delta \rho} (\Delta \rho_t(\vec{r}))^2
       +   C^{\tau \tau}_t \tau_t^{2}(\vec{r}) \nonumber \\
&      +   C^{\rho Q}_t \rho_t(\vec{r}) \, Q_t(\vec{r})
       + 2 \, C^{\tau_{\mu\nu} \tau_{\mu\nu}}_t \sum_{\mu,\nu} \tau_{t,\mu \nu}^2(\vec{r})
        \nonumber \\
&      - 2 \, C^{\tau\nabla\nabla\rho}_t \sum_{\mu,\nu} \tau_{t,\mu\nu}(\vec{r})
                       \nabla_{\mu} \nabla_{\nu} \rho_t(\vec{r})  \nonumber  \\
&      + 2 \, C^{KK}_t  \sum_{\mu,\nu,\kappa} \left(K_{t,\mu \nu \kappa}(\vec{r}) \right)^2
       \nonumber \\
&   - 2 \, C^{JV}_t \sum_{\mu,\nu} J_{t,\mu\nu}(\vec{r}) V_{t,\mu\nu}(\vec{r})\Big] \, ,
\end{align}
where all sums are over Cartesian components of the tensors and where
we have opted to label the coupling constants $C_t$ for each term
separately, even though many are identical to guarantee the Galilean invariance
of the EDF~\cite{Becker17}.

\paragraph{The single-particle Hamiltonian.}
To simplify the notation, we introduce the density-vector $\mathsf{R}$ as
a helpful shorthand for the set of mean-field densities. For the two cases
we discuss below, it is given by
\begin{align}
\mathsf{R}^{\NLO}_q
& = ( \rho_q, \tau_{q}, J_{q,\mu\nu}, \vec{s}_q, \vec{T}_q, \vec{j}_q, )
    \, ,
    \\
\mathsf{R}^{\NiiLO}_q
& = ( \rho_q, \tau_{q,\mu\nu}, J_{q,\mu\nu},
       K_{q,\mu \nu\kappa}, V_{q,\mu\nu}, Q_{q} ) \, ,
\end{align}
where we use the indices $a$, $b$, \ldots to refer to the individual components 
of $\mathsf{R}_{q,a}(\vec{r})$, $q=n$, $p$.

With this density vector, the expression for the energy can be rewritten
as $E_{\rm tot}(\mathsf{R})$. The individual terms in the single-particle
Hamiltonian are then obtained by rewriting the variation of the energy
with respect to the full density matrix in Eq.~\eqref{eq:defh} as the sum
of variations with respect to the individual densities 
$\mathsf{R}_{q,a}(\vec{r})$ as
\begin{equation}
\label{eq:variation}
h_q(\vec{r}\sigma, \vec{r}'\sigma')
= \sum_{a} \int \! d^3r'' \,
  \frac{\delta E}{\delta \mathsf{R}_{q,a}(\vec{r}'')}
  \frac{\delta \mathsf{R}_{q,a}(\vec{r}'')}
       {\delta \rho_q(\vec{r}' \sigma', \vec{r}\sigma)}
 \, .
\end{equation}
The derivative of the energy with respect to the density 
$\mathsf{R}_{q,a}(\vec{r})$ can be identified as an associated mean-field 
potential $\mathsf{F}_{q,a}(\vec{r})$
\begin{equation}
\mathsf{F}_{q,a}(\vec{r})
\equiv \frac{\delta E}{\delta \mathsf{R}_{q,a}(\vec{r})}
\end{equation}
that in general depends on the density vectors $\mathsf{R}_{p,a}(\vec{r})$ 
and $\mathsf{R}_{n,a}(\vec{r})$ of both nucleon species.

In analogous fashion, we introduce the potential-vector
$\mathsf{F}$, which for the two types of EDF we consider here reads
\begin{align}
\mathsf{F}^{\NLO}_{q}
& = \left( \mathsf{F}^{\rho}_{q},
           \mathsf{F}^{\tau}_{q},
           \mathsf{F}^{J}_{q,\mu\nu},
           \mathsf{F}^{\vec{s}}_q,
           \mathsf{F}^{T}_q,
           \mathsf{F}^{\vec{j}}_q
    \right) \, ,
    \\
\mathsf{F}^{\NiiLO}_{q}
& = \left( \mathsf{F}^{\rho}_{q},
           \mathsf{F}^{\tau}_{q,\mu\nu},
           \mathsf{F}^{J}_{q,\mu\nu},
           \mathsf{F}^{K}_{q,\mu \nu\kappa},
           \mathsf{F}^{V}_{q,\mu\nu},
           \mathsf{F}^{Q}_{q}
    \right) \, .
\end{align}
The corresponding single-particle Hamiltonians can then be written as
a function of these potential-vectors
\begin{align}
\label{eq:singleh:NLO}
\hat{h}^{\NLO}_{q}[\mathsf{F}]
= & \,
      \mathsf{F}^{\rho}_q (\vec{r})
    + \mathsf{F}^{\vec{s}}_q (\vec{r})\cdot \boldsymbol{\sigma}
    - \iunit \mathsf{F}^{\vec{j}}_{q} \cdot \boldsymbol{\nabla}
    \nonumber \\
&    - \iunit \sum_{\mu, \nu} \mathsf{F}^{J}_{q, \mu \nu}(\vec{r}) \nabla_{\mu} \sigma_{\nu}
    - \vnabla \cdot \mathsf{F}^{\tau}_{q}(\vec{r}) \vnabla
    \nonumber \\
&   - \sum_{\kappa} \vnabla \cdot  \mathsf{F}^T_{q \kappa}
      \vnabla \sigma_{\kappa}  \, ,
    \\
\label{eq:singleh:N2LO}
\hat{h}^{\NiiLO}_{q}[\mathsf{F}]
= & \, 
      \mathsf{F}^{\rho}_q(\vec{r})
    + \mathsf{F}^{\vec{s}}_q (\vec{r})\cdot \boldsymbol{\sigma}
    - \iunit \mathsf{F}^{\vec{j}}_{q} \cdot \boldsymbol{\nabla} \nonumber \\
&   - \frac{\iunit}{2} \sum_{\mu, \nu} \Big[
          \mathsf{F}^{J}_{q, \mu \nu}(\vec{r}) \nabla_{\mu} \sigma_{\nu} +
          \nabla_{\mu} \mathsf{F}^{J}_{q, \mu \nu}(\vec{r}) \sigma_{\nu} \Big]
    \nonumber \\
&  - \sum_{\mu, \nu} \left[  \nabla_{\mu} \mathsf{F}^{\tau}_{q, \mu \nu}(\vec{r})
         \nabla_{\nu}
    + \sum_{\kappa} \nabla_{\mu} \mathsf{F}^T_{\mu \nu \kappa}
        \nabla_{\nu} \sigma_{\kappa} \right]
       \nonumber \\
&  + \iunit \sum_{\mu, \nu, \kappa} \nabla_{\kappa} \mathsf{F}^{V}_{q,\mu \nu}
         \nabla_{\nu} \nabla_{\mu} \sigma_{\kappa} + \Delta \mathsf{F}^{Q}_q \Delta   \, ,
\end{align}
where we have ordered the various terms according to the order of gradients
that act on the single-particle wave functions, from lowest~(0) to
highest~(4). The operator structure of the individual terms results from the 
second functional derivative in the chain rule in Eq.~\eqref{eq:variation}. We 
also simplified the single-particle Hamiltonians in Eqs. \eqref{eq:singleh:NLO} 
and \eqref{eq:singleh:N2LO} by using that the gradients of certain potentials 
vanish for the EDFs used here. For more details on the EDF at the NLO level, we
 refer the reader to~\cite{Bender03,Hellemans12}. A detailed discussion of the
full EDF at N2LO level, including the time-odd terms, will be given 
elsewhere~\cite{Ryssens18}.

These equations show that the derivatives of the single-particle wave functions 
are needed at two different stages of the calculation, when summing up the local
densities $\mathsf{R}_q$, and when calculating the action of the single-particle
Hamiltonian $\hat{h}_q$ on the single-particle states. To minimise the 
computational cost, it is customary to calculate all necessary derivatives only
once each time a new set of single-particle states has been determined during 
the iterative process and to store them for later use during the iteration.

The gradients contained in the kinetic and Skyrme
terms of the EDF determine to a large extent the behaviour of the
strategies employed for the diagonalisation subproblem and for the SCF iterations.
They enter into these terms in two distinct ways: either as an \emph{external
gradient} acting on a local density, for example $\Delta \rho(\vec{r})$,
or as an \emph{internal gradient} that is contained within the definition
of a local density, as for example in the definition of $\tau(\vec{r})$,
Eq.~\eqref{eq:taudef}.
Internal derivatives within the densities determine the operator structure of
$\hat{h}_q$, whereas external derivatives contribute to the expressions for
the potentials $\mathsf{F}_q(\vec{r})$. As we discuss in what follows,
the operator structure of the single-particle Hamiltonian (and hence
internal derivatives) determines the convergence of the diagonalisation
subproblem, whereas external derivatives of the densities $\mathsf{R}_{q,a}(\vec{r})$ 
contribute to the expressions for the potentials $\mathsf{F}_{q,a}(\vec{r})$.

%------------------------------------------------------------------------------
%
\subsection{The self-consistent problem}
\label{sec:SCP}

The single-particle Hamiltonian of the nucleon species $q$ is a function of the 
vector $\mathsf{F}_q$, which depends on the vectors of the densities 
$\mathsf{R}_p$ and $\mathsf{R}_n$ of protons and neutrons, which in turn are 
constructed from the single-particle wave functions. The equations to be solved
 are:
\begin{equation}
\hat{h}
\big( \mathsf{F} \big[ \mathsf{R} ( |\psi_1 \rangle,
|\psi_2 \rangle \ldots,) \big] \big) \, |\psi_j \rangle
= \epsilon_j  |\psi_j \rangle  \, .
\end{equation}
Since the single-particle Hamiltonian depends on its solution, the problem
is highly non-linear and can only be solved by an iterative process.
All quantities that depend on the iteration number $i$ will be marked by a
superscript $(i)$:
\begin{displaymath}
| \psi^{(i)}_j \rangle, \;
\mathsf{R}^{(i)}, \;
\mathsf{F}^{(i)}, \;
h^{(i)} \, .
\end{displaymath}
The usual strategy followed to solve the mean-field equations is composed of
several separate, but linked, steps.

The first step is the (partial) diagonalisation of $\hat{h}^{(i)}$, to
obtain a set of single-particle orbitals $|\psi_{l}^{(i+1)}\rangle$. With a 
fixed mean-field potential-vector $\mathsf{F}$, this is a linear algebra 
problem. We call this the \emph{diagonalisation} subproblem.

In order to obtain the matrix elements of $\rho$ and $\kappa$, the second
step is the restricted diagonalisation of the HFB Hamiltonian
$\mathcal{H}$ of Eq.~\eqref{eq:HFBhamil}. We call this the
pairing subproblem, which is discussed briefly in Appendix A.

The third step is the evolution of the single-particle Hamiltonian from
one iteration to the next. Suppose the single-particle Hamiltonian
$h^{(i)}$ has been constructed in iteration $(i)$. Its eigenstates, which we
label as $|\psi_j^{(i+1)} \rangle$, then provide the starting point for the 
next iteration.

The ensemble of these steps defines a map $G^{\mathsf{F}}$ on the space of 
$\mathsf{F}$:
\begin{equation}
G^{\mathsf{F}}: \mathsf{F}^{(i)} \rightarrow
G^{\mathsf{F}}(\mathsf{F}^{(i)}) = \mathsf{F}^{(i+1)}_{|\psi\rangle}\, ,
\end{equation}
where $\mathsf{F}^{(i+1)}_{|\psi\rangle}$ is calculated from the mean-field 
densities $\mathsf{R}^{(i+1)}$ using the formulas presented in 
Section~\ref{sect:Skyrme}.
Since the potentials $\mathsf{F}$ depend on the densities $\mathsf{R}$, we
can alternatively formulate this process as a map on the space of the densities
\begin{equation}
\label{eq:fixedpoint}
G^{\mathsf{R}}:
\mathsf{R}^{(i)} \rightarrow G^{\mathsf{R}}(\mathsf{R}^{(i)})
= \mathsf{R}^{(i+1)}_{|\psi\rangle} \, ,
\end{equation}
where this time the density $\mathsf{R}^{(i+1)}_{|\psi\rangle}$ is calculated 
from the single-particle states $|\psi^{(i+1)}\rangle$ \emph{after} 
diagonalisation of $h^{(i)}$.

Self-consistency is achieved when the potentials and densities get mapped 
onto themselves, meaning that
\begin{equation}
G^{\mathsf{F}}(\mathsf{F}^{(\infty)}) = \mathsf{F}^{(\infty)} \, , \qquad
G^{\mathsf{R}}(\mathsf{R}^{(\infty)}) = \mathsf{R}^{(\infty)} \, .
\end{equation}
The evolution of the mean-field densities and potentials with the
iterations should thus be aimed at finding the fixed points
$\mathsf{F}^{(\infty)}$ and $\mathsf{R}^{(\infty)}$ of the maps
$G^{\mathsf{R}}$ and $G^{\mathsf{F}}$. We demonstrate in 
Section~\ref{sec:dmixing} that taking directly
\begin{equation}
\mathsf{F}^{(i+1)} = \mathsf{F}^{(i+1)}_{|\psi\rangle} \, , \qquad
\mathsf{R}^{(i+1)} = \mathsf{R}^{(i+1)}_{|\psi\rangle} \, ,
\end{equation}
does in general not lead to a stable iterative process. A more robust
procedure to determine $\mathsf{F}^{(i+1)}$ and $\mathsf{R}^{(i+1)}$ from the
potential and density vectors at iteration $(i)$ is needed to achieve a stable 
iterative scheme. As already proposed in the introduction, we call the evolution
 of the potentials and densities the \textit{self-consistent-field} (SCF) 
iteration, which is the frequently used terminology in atomic 
physics~\cite{Pulay80}.

It is customary to introduce constraints on the expectation values of specific
one-body operators into the mean-field equations, specifying for instance 
the multipole moments of the nuclear density. In practice, such 
constraints add terms to the single-particle Hamiltonian but otherwise 
affect neither the diagonalisation nor the SCF subproblems. We postpone the 
discussion of strategies how to efficiently treat constraints in the context of 
the algorithms discussed here to a future publication. 

Since the mean-field Hamiltonian $\hat{h}^{(i)}$ changes at each iteration, it
is not necessary to perform an accurate diagonalisation at each 
SCF iteration. The densities and potentials can be constructed from approximate
 eigenstates of $\hat{h}^{(i)}$. In electronic structure physics, such procedure
 reduces the CPU time by an order of 
magnitude~\cite{Zhou14}. In the case of a 3d coordinate space representation, 
this is particularly advantageous as the cost of diagonalising $\hat{h}$ dwarfs 
the cost of updating the densities and potentials.

\begin{figure}[t!]
\centerline{\includegraphics[width=0.85\linewidth]{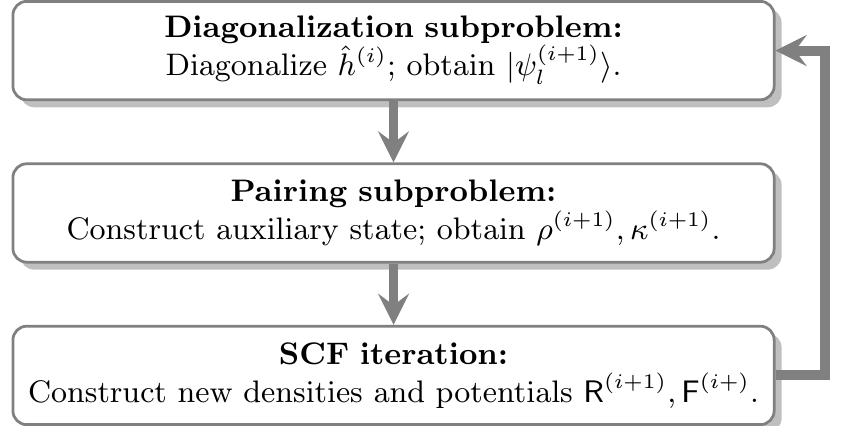}}
\caption{Schematic diagram of the division into subproblems of the general
        solution strategy for the self-consistent problem.}
\label{fig:algogeneral}
\end{figure}

The division of the resolution of the mean-field equations in different steps is
illustrated in Fig.~\ref{fig:algogeneral}. While all steps in this figure 
are coupled in a practical implementation, we will discuss them separately in 
the next sections.

%%%%%%%%%%%%%%%%%%%%%%%%%%%%%%%%%%%%%%%%%%%%%%%%%%%%%%%%%%%%%%%%%%%%%%%%%%%%%%%%
\section{The diagonalisation subproblem}
\label{sec:linearsub}

\subsection{Iterative diagonalisation}
Given a single-particle Hamiltonian $\hat{h}^{(i)}$, generated by the 
potentials $\mathsf{F}^{(i)}$, we are looking for its exact eigenstates 
$|\psi_{\ell}^{\rm ex}\rangle$ and its eigenvalues $\epsilon^{\rm ex}_\ell$
\begin{equation}
\label{eq:eigen}
\hat{h}^{(i)} | \psi_{\ell}^{\rm ex} \rangle
= \epsilon_{\ell}^{\rm ex} | \psi_{\ell}^{\rm ex} \rangle  \, .
\end{equation}
To exclude the trivial null solution and to enforce that the single-particle
states are orthonormalised, the numerical solution of Eq.~\eqref{eq:eigen}
has to be supplemented by constraints on orthonormality
\begin{equation}
\label{eq:ortho}
\langle  \psi_k^{\rm ex} | \psi_\ell^{\rm ex} \rangle
= \delta_{k \ell} \, .
\end{equation}
In practice, Eq.~\eqref{eq:eigen} is a matrix equation whose dimension is equal to
the size of the basis used to solve the problem. If this dimension is small 
enough, the problem can be solved by a direct diagonalisation using standard 
library routines. Several codes using an expansion in a HO basis employ this 
strategy~\cite{HFBTHO,HFODD}. This is also feasible in coordinate space 
representation when 1d spherical symmetry is assumed~\cite{lenteur}. We are 
interested in the following in cases where the dimension of the basis is too 
large to permit calculating and storing $\hat{h}^{(i)}$ explicitly. 

Very often one is not interested in the entire spectrum of 
$\hat{h}^{(i)}$ in the numerical basis.\footnote{A counterexample is the 
construction of the reference state for quasiparticle RPA and its various 
extensions.} 
For the calculation of
the normal and pair densities, it is then sufficient to determine the 
$\Omega$ single-particle orbits with single-particle energies below the 
pairing cutoff, see Appendix~\ref{app:twobasis}. For HF calculations, $\Omega$ 
does not have to be much larger than the number of nucleons. When 
pairing correlations are taken into account with an effective pairing 
interaction as defined in Refs.~\cite{Rigollet99,Bender2000}, the value for 
$\Omega$ in heavy nuclei is at most equal to twice the number of nucleons. 
This is several orders of magnitude smaller than the number of basis
states $N_m$ in the coordinate-space representation. In these conditions,
iterative determination of the $\Omega$ lowest single-particle states 
is significantly less demanding in both CPU time and memory requirements than 
direct diagonalisation of $\hat{h}^{(i)}$ in a basis of dimension $N_m$.

Restricting the eigenvalue problem of Eq.~\eqref{eq:eigen} to the $\Omega$
lowest states allows us to recast the problem. Consider the Rayleigh quotient 
associated with the single-particle Hamiltonian
\begin{equation}
\label{eq:Rayleigh}
R_{h}(|\psi_\ell \rangle)
= \frac{\langle \psi_\ell |\hat{h}^{(i)} | \psi_\ell \rangle}
       {\langle \psi_\ell | \psi_\ell \rangle} \, .
\end{equation}
It can be shown~\cite{SaadBook} that the stationary points of $R_{h}$ 
correspond to the eigenstates of $\hat{h}$, and that at these points, 
$R_{h}$ is equal to the corresponding eigenvalue $\epsilon^{\rm ex}_\ell$. 
Further we define
\begin{equation}
R^{\rm tot}_{h} \big( | \psi_1 \rangle, | \psi_2 \rangle, \ldots,
                     | \psi_{\Omega} \rangle
                \big)
= \sum_{\ell=1}^{\Omega} R_h \big( | \psi_\ell \rangle \big) \, .
\end{equation}
With the constraint on orthonormality of the single-particle orbitals, 
Eq.~\eqref{eq:ortho}, any set of $\Omega$ eigenstates of $\hat{h}^{(i)}$ is
a stationary point of $R^{\rm tot}_{h^{(i)}}$. In particular, we have
for a set of indices $\ell_{1}$, \ldots, $\ell_{\Omega}$
\begin{equation}
R^{\rm tot}_{h}(\left|\psi^{\rm ex}_{\ell_1}\right\rangle,
                \left|\psi^{\rm ex}_{\ell_{2}} \right\rangle,
    \ldots, \left| \psi^{\rm ex}_{\ell_{\Omega}}\right\rangle)
= \sum_{m=1}^{\Omega} \epsilon^{\rm ex}_{\ell_{m}}
\end{equation}
where different $\ell$ label the many local minima of $R^{\rm tot}_{h}$.
Each corresponds to a different combination of $\Omega$ eigenstates of
$\hat{h}$ in the complete $N_m$-dimensional space.
The absolute minimum of  $R^{\rm tot}_{h}$ is obtained for the $\Omega$
single-particle states corresponding to the lowest eigenvalues of $\hat{h}$.
For these we have
\begin{align}
\label{eq:Rayleigh:sum}
\text{min }_{\left|\psi_{\ell_1} \right \rangle, \left|\psi_{\ell_2} \right \rangle,
    \ldots, \left|\psi_{\ell_{\Omega}} \right \rangle} R^{\rm tot}_{h}
    &= \sum_{m=1}^{\Omega} \epsilon^{\rm ex}_m \, .
\end{align}
The eigenvalue problem in Eq.~\eqref{eq:eigen} can be recast as a constrained 
optimization problem
\begin{equation}
\begin{split}
\left\{ \left| \psi_1^{\rm ex}  \right \rangle,
\left|\psi_2^{\rm ex}  \right \rangle, \ldots,
   \left|\psi_{\Omega}^{\rm ex} \right \rangle \right\}
      &= \text{arg min} \left[ R^{\rm tot}_{h}\right] \, , \\
    \text{subject to } & \langle \psi^{\rm ex}_k | \psi^{\rm ex}_\ell \rangle
    = \delta_{k \ell} \, .
\end{split}
\label{eq:consopt}
\end{equation}
From Eq.~\eqref{eq:consopt}, it follows that the eigenvalue problem can be
solved by iterative methods. We introduce a second iteration counter $j$
that labels the evolution of the single-particle wave functions
\begin{equation}
| \psi^{(i,j)}_1 \rangle, | \psi^{(i,j)}_2 \rangle,
   \ldots, | \psi^{(i,j)}_{\Omega} \rangle.
\end{equation}
The natural starting point for the iterative diagonalisation is the set of
single-particle orbitals from the previous SCF iteration,
$|\psi_\ell^{(i,0)} \rangle = | \psi^{(i-1)}_\ell \rangle$.

In the context of self-consistent mean-field models for which the 
diagonalisation is embedded into an SCF iteration, an approximate 
diagonalisation of $\hat{h}$ is in general sufficient: instead of limiting the 
number of iterations by a convergence criterion the number of iterations is 
fixed to a maximum number $j_{\rm max}$.

The simplest iterative optimization technique is the method of gradient 
descent~\cite{NumRecipes}, which is frequently used for calculations involving
EDFs~\cite{Bonche05,Mang76,Davies80}, although sometimes with modifications 
as in~\cite{Maruhn14,Reinhard82}. 

We first discuss this algorithm and analyse its performance
for Skyrme functionals at NLO and N2LO. Then, we will show how to 
improve on its convergence with a minor change in the 
algorithm that leads to a scheme dubbed \textit{heavy ball dynamics} 
in the literature.

%--------------------------------------------------------------------------

\subsection{Gradient descent and heavy-ball dynamics for quadratic forms}
\label{sec:linear}

Focussing on the problem of diagonalisation, we first sketch both gradient 
descent and heavy-ball dynamics for a sche\-ma\-tic problem that often 
serves as a laboratory for the analysis of minimisation 
algorithms~\cite{NumRecipes,Goh2017,Qian99}.

Consider a quadratic form for a positive definite and symmetric matrix
$A$ and a vector $\vec{b}$ on a (unspecified) vector space~\cite{Goh2017}
\begin{equation}
\label{eq:quadratic}
f(\vec{x})
= \frac{1}{2} \vec{x}^T A \vec{x} - \vec{b}^T \vec{x} \, .
\end{equation}
The minimum of $f$ is reached for
\begin{equation}
\label{eq:f:min}
\vec{x}^{(\infty)}
= A^{-1} \vec{b} \, .
\end{equation}
The minima of $f$ coincide with the zeros of its first derivative
with respect to $\vec{x}$
\begin{equation}
\label{eq:derivquad}
\frac{\partial f}{\partial \vec{x}} = A \vec{x} - \vec{b} \, .
\end{equation}
minimising the function $f$ is thus equivalent to solving the system of
coupled linear equations posed by setting Eq.~\eqref{eq:derivquad} to zero.

An iterative solution of the minimisation is most easily analysed using a 
transformed variable $\vec{y}$
\begin{align}
\label{eq:y:def}
\vec{y} = \vec{x} - A^{-1} \vec{b}  \, ,
\end{align}
so that we have
\begin{align}
f(\vec{y}) = \tfrac{1}{2} \, \vec{y}^T A \vec{y}\, ,& \qquad 
\frac{\partial f}{\partial \vec{y}} = A \vec{y} \, .
\end{align}
At the solution of the minimisation problem,
 $\vec{x} = \vec{x}^{(\infty)}$, $\vec{y} = 0$. An efficient iterative scheme 
should thus provide an evolution of the transformed variable, such that 
$\vec{y} \rightarrow 0$ as quickly as possible. We will demonstrate below 
that the spectrum of eigenvalues of $A$, denoted by $\lambda_{n}$, 
strongly influences the possible convergence rate. In particular, the 
condition number of $A$ will play a crucial role. It is defined as the 
ratio between the largest and the lowest eigenvalues
\begin{equation}
\label{eq:condition:number}
\kappa^{A}
= \frac{\lambda_{\rm max}}{\lambda_{\rm 1}} \, .
\end{equation}

\paragraph{Gradient descent.}
Starting from an initial guess $\vec{x}^{(0)}$, the gradient-descent update 
from iteration $j$ to iteration $j+1$ is given by
\begin{align}
\vec{x}^{(j+1)}
&= \vec{x}^{(j)} - \alpha \frac{\partial f}{\partial \vec{x}} 
\bigg|_{\vec{x} = \vec{x}^{j}}  \nonumber \\
&= \vec{x}^{(j)} - \alpha \big( A\vec{x}^{(j)} - \vec{b} \big)  \, ,
\label{eq:lineargradient}
\end{align}
where $\alpha$ is a numerical parameter still to be specified. The evolution
in terms of the variable $\vec{y}$ can be developed  into the eigenvectors 
$\vec{x}_k^{A}$ of $A$
\begin{equation}
\vec{y}^{(j)} = \sum_{k} a_{k}^{(j)} \vec{x}_k^{A} \, .
\label{eq:distance}
\end{equation}
Inserting this into Eq.~\eqref{eq:lineargradient}, we obtain for the  expansion 
coefficients $a_{k}^{(j)}$
\begin{equation}
a_k^{(j+1)} =  (1 - \alpha \lambda_k) \, a_k^{(j)} \, .
\end{equation}
One can obtain the expansion coefficients at
iteration $(j)$ from the initial value at iteration $(0)$ by applying
$j$ times $(1 - \alpha \lambda_k)$
\begin{equation}
\label{eq:sequence}
a_k^{(j)}
= (1 - \alpha \lambda_k)^{j} \, a_k^{(0)} \, .
\end{equation}
Under the condition that
\begin{equation}
\label{eq:gd:alpha}
0 < \alpha  < \frac{2}{\lambda_{\rm max}} \, ,
\end{equation}
the value of $| 1 - \alpha\lambda_k |$ is smaller than one, and 
the components of $\vec{y}$ in Eq.~\eqref{eq:sequence} decay exponentially 
to zero at a rate that increases for larger eigenvalues. For this reason,
the coefficient $a_{1}^{(j)}$ tends to zero the slowest out of all $a_k$.

The overall rate of convergence is thus determined by $-\alpha\lambda_1$. 
Combined with Eq.~\eqref{eq:gd:alpha}, this means that
it is the spread in eigenvalues of $A$ that governs the convergence of the 
gradient-descent method. When the condition number, Eq.
\eqref{eq:condition:number}
of $A$ is large, gradient descent converges only slowly to the 
optimum of the function $f$.

\paragraph{Heavy-ball dynamics.}

In Ref.~\cite{Polyak64}, B.~T.~Polyak proposed the heavy-ball 
algorithm to improve on the convergence behaviour of the gradient-descent 
method for the iterative solution of systems of linear equations. The 
heavy-ball update scheme is given by
\begin{align}
\vec{x}^{(j+1)}
& = \vec{x}^{(j)} - \alpha \big( A\vec{x}^{(j)} - \vec{b} \big)
    + \mu \, \delta \vec{x}^{(j)} \, ,
    \nonumber \\
\delta\vec{x}^{(j)}
& = \big( \vec{x}^{(j)} - \vec{x}^{(j-1)} \big) \, .
\end{align}
Compared to gradient descent, there is an  additional
$\mu \, \delta \vec{x}^{(j)}$ term, which is usually called the \emph{momentum
term}, while $\mu$ is called the \emph{momentum parameter}. This term introduces
 a memory effect that allows for speeding up the iterative process.

The analysis of this method is more involved than that of the gradient-descent
method. As discussed in Refs.~\cite{Goh2017,Qian99}, the algorithm converges to 
the minimum of $f$ on the condition that
\begin{align}
0 \leq  \mu  < 1  & \, ,  \qquad
0 < \alpha <  \frac{2}{\lambda_{\rm max}} (1 + \mu) \, .
\label{eq:momcondition}
\end{align}
The components of the variable $\vec{y}$ 
of Eq.~\eqref{eq:y:def} evolve according to
\begin{align}
a_{k}^{(j+1)}
& =  a_{k}^{(j)}  +  v^{(j)}_{k} \, ,
    \nonumber \\
v^{(j)}_{k}
& = - \alpha \, \lambda_k  \, a_{k}^{(j)}
    + \mu \big( a_{k}^{(j)} - a_{k}^{(j-1)} \big) \, .
\label{eq:momeigen}
\end{align}
The evolution determined by Eq.~\eqref{eq:momeigen} is analogous to the one
of a damped harmonic oscillator, as outlined in Appendix~\ref{app:damposc}. 
Three different regimes can be distinguished depending on the value of $\mu$: 
the motion of $a_{k}^{(j+1)}$ can either be \textit{underdamped}, 
\textit{overdamped} or \textit{critically damped}. For a given value of $\alpha$, 
the critical momentum value $\mu^{\rm crit}_k$ separating the regimes is 
given by
\begin{equation}
\mu^{\rm crit}_k = \left(1 -\sqrt{\alpha \lambda_k}\right)^2 \, .
\label{eq:mucrit}
\end{equation}
If $\mu > \mu^{\rm crit}_k$, the motion of $a_{k}$ is underdamped: $a_{k}$
undergoes slowly decaying oscillations around zero as a function of the
iterations. By contrast, if $\mu < \mu^{\rm crit}_k$, the motion of $a_{k}$
is overdamped. In this case, the coefficient $a_k$ does not oscillate, but
decays exponentially to zero as for gradient descent. When
$\mu = \mu^{\rm crit}_k$, the motion is critically damped. For this particular 
value of the momentum parameter $\mu$, the coefficient $a_{k}$ decays to zero 
significantly faster than in the other two cases.

However, the critical value for $\mu^{\rm crit}_k$ is different 
for every $a_k$. It can be shown that the optimal compromise in
terms of convergence rate of the errors between small eigenvalues and
large eigenvalues is given by~\cite{Qian99}
\begin{equation}
\label{eq:muoptA}
\mu^{A}_{\rm opt}
= \left(\frac{\sqrt{\kappa^A} - 1}{\sqrt{\kappa^A}+ 1}\right)^2 \, .
\end{equation}
The optimal value of $\alpha$ to be used in conjunction with this value of 
the momentum parameter is~\cite{Goh2017}
\begin{align}
\alpha^A_{\rm opt} &= 
\left(\frac{2}{\sqrt{\lambda_{\rm max}}+\sqrt{\lambda_1}} \right)^2  \nonumber\\
&=  \frac{2}{\lambda_{\rm max}}(1+ \mu^A_{\rm opt}) -
 \frac{1}{\kappa^{A}}\left(\frac{2}{\sqrt{\lambda_{\rm max}}+\sqrt{\lambda_1}}\right)^2 
 \, ,
\end{align}
where the second equality can be shown with straightforward algebra. This 
optimal value for $\alpha$ is virtually equal to the upper limit
dictated by Eq.~\eqref{eq:momcondition}, because in most cases of interest 
$\kappa^A$ is large. 

The optimal iterative parameters of the method are thus closely related to 
the condition number $\kappa^A$ of $A$, Eq~\eqref{eq:condition:number}: when 
$\kappa^A$ is high, the momentum parameter $\mu$ should be taken close to 1. 
In such case, the size of $\alpha$ can be nearly doubled compared to gradient 
descent, resulting in much faster convergence.

However, the advantage of heavy-ball dynamics over gradient descent goes 
beyond the factor two on the upper limit of $\alpha$, as will be 
demonstrated with practical examples in Section~\ref{sec:numericaltests}.  
One can prove that heavy-ball dynamics~\cite{Goh2017,NesterovBook} with optimal 
parameters $(\alpha^A_{\rm opt},\mu^A_{\rm opt})$, achieves the fastest 
convergence rate that is theoretically possible across a range of similar 
methods: it is impossible to do better using algorithms that 
only incorporate information on the first-order derivatives of the quadratic
form given by Eq.~\eqref{eq:quadratic}.

%------------------------------------------------------------------------------
%
\subsection{The gradient descent method applied to the diagonalisation 
subproblem}
\label{sec:graddesc}

The schematic problem discussed in Section~\ref{sec:linear} is a 
simplification of the self-consistent mean-field problem that we want 
to solve. Nevertheless, both gradient descent and heavy-ball dynamics 
can be employed with only minor modifications.

The gradient of the Rayleigh quotient, Eq.~\eqref{eq:Rayleigh}, 
$R^{\rm tot}_h$ with respect to the $\ell$-th single-particle wave function 
is given by
\begin{equation}
\label{eq:gradrayleigh}
\frac{\partial R_h^{\rm tot}}
     {\partial | \psi_\ell \rangle} \bigg|_{\left| \psi_\ell \right\rangle
= |\psi^{(i,j)}_\ell \rangle}
= ( \hat{h}^{(i)} - \epsilon_\ell^{(i,j)} ) \, | \psi_\ell^{(i,j)}\rangle \, ,
\end{equation}
taking into account that $|\psi_\ell^{(i,j)}\rangle$ is normalized and where
$\epsilon_\ell^{(i,j)}$ is the $\ell$-th diagonal matrix element of 
$\hat{h}^{(i)}$ 
\begin{equation}
\label{eq:defepsilon}
\epsilon_\ell^{(i,j)}
= \langle \psi^{(i,j)}_\ell | \hat{h}^{(i)} | \psi_\ell^{(i,j)} \rangle \, .
\end{equation}
The gradient-descent update from iteration $j$ to $j+1$ for the
$\ell$-th single-particle wave function is then given by
\begin{equation}
\label{eq:gradscheme}
|\phi^{(i,j+1)}_{\ell}\rangle
= |\psi^{(i,j)}_{\ell}\rangle  - \frac{dt}{\hbar} \,
  \big( \hat{h}^{(i)} - \epsilon_\ell^{(i,j)} \big) | \psi^{(i,j)}_{\ell} \rangle
\, .
\end{equation}
The $|\phi^{(i,j+1)}_{\ell}\rangle $ do not constitute an orthonormal set, even
if the $|\psi^{(i,j)}_{\ell}\rangle$ do. For this reason, the algorithm 
needs to be complemented by an explicit orthonormalisation step that 
transforms the set of $|\phi_{\ell}^{(i,j)}\rangle$ into an orthonormal set 
$|\psi_{\ell}^{(i,j)}\rangle$ by, for instance, a Gram-Schmidt process. 
This step can be seen as a projection on a feasible set~\cite{NumRecipes}; 
a more precise name for the algorithm would in fact be \emph{projected gradient
descent}. An alternative to the orthonormalisation step is to add a set of
Lagrange constraints to the optimization problem in Eq.~\eqref{eq:consopt},
as for example done in Ref.~\cite{Gan01}.

In the context of nuclear mean-field methods, the gra\-dient-descent 
evolution of the single-particle wave functions was originally derived from 
the first-order approximation to the operator of evolution in imaginary 
time~\cite{Davies80}. For this historical reason we have replaced the 
step size $\alpha$ by $dt/\hbar$. The gradient descent method is often 
called \textit{imaginary time evolution}~\cite{Bonche05}. In other fields 
of physics, such appellation is used for more advanced techniques to approximate
the exponentiation of the single-particle Hamiltonian~\cite{Bader13}.

The similarities between the simpler problem of Section~\ref{sec:linear} and
the mean-field problem are revealed by comparing 
Eqs.~\eqref{eq:gradrayleigh} and \eqref{eq:lineargradient}: the gradient 
of the Rayleigh quotient is similar to that of a quadratic form, where
$\hat{h}^{(i)}$ plays the role of the matrix $A$. There are however two 
important differences: the first one is that the second term in 
Eq.~\eqref{eq:gradrayleigh} varies from one iteration to the next, 
whereas the vector $\vec{b}$ was kept fixed in 
Eq.~\eqref{eq:lineargradient}.
The second difference is that the self-consistent
problem requires the evolution of many single-particle wave functions, 
constrained to be orthonormal, whereas we considered only one vector 
$\vec{x}$ in Section~\ref{sec:linear}. Because of these differences, we 
cannot provide an analytical formula of the evolution of the 
single-particle wave functions, whereas for the simplified problem we 
are able to write down Eq.~\eqref{eq:sequence}.

As for the schematic problem of Section~\ref{sec:linear}, the gradi\-ent-descent 
method only converges for a limited range of values of $dt$. For 
fast convergence, it is advisable to use the largest feasible value. Since
for all $\ell$ one has $\epsilon_{\ell}^{(i,j)} > \epsilon^{\rm ex}_1$, the
value of $\epsilon^{\rm ex}_{\rm max} - \epsilon^{\rm ex}_{1}$ is an upper
bound for the largest eigenvalue of $(\hat{h}^{(i)} - \epsilon_{\ell}^{(i,j)})$.
By analogy to the schematic problem, we obtain a condition on $dt$
\begin{equation}
\label{eq:dtgradcondition}
\frac{dt}{\hbar}
<  \frac{2}
        {\epsilon^{\rm ex}_{\rm max} - \epsilon^{\rm ex}_{1}} \, ,
\end{equation}
see also the discussion in Refs.~\cite{Ryssens15,Bonche05,Davies80,Reinhard82}.
Unfortunately, Eq.~\eqref{eq:dtgradcondition} does not offer a practical
way of selecting $dt$ at the first iteration. While  $\epsilon^{(i,0)}_1$ is in
general a sufficient estimate of $\epsilon^{\rm ex}_{1}$, 
no information is available on $\epsilon_{\rm max}^{\rm ex}$.

A robust way to judge the convergence of the diagonalisation subproblem is 
evaluating the weighted dispersion of the single-particle 
energies $(dh^2)^{(i,j)}$, defined as
\begin{equation}
\label{eq:dh2}
(dh^2)^{(i,j)}
= \frac{1}{A}\sum_{\ell=1}^{\Omega} \rho_{\ell\ell}
  \Big[ \langle \psi_\ell^{(i,j)} | (\hat{h}^{(i)})^{\dagger} \hat{h}^{(i)}
        | \psi_\ell^{(i,j)}\rangle - (\epsilon_j^{(i,j)})^2 
  \Big] \, .
\end{equation}
Weighting the individual contributions by the diagonal matrix element
of the density matrix $\rho_{\ell\ell}$ \cite{Ryssens15a} limits the sum
to those single-particle states that contribute to observables. Small (large) 
values of this quantity indicate that the single-particle 
wave functions at iterations $(i,j)$ are good (bad) approximations to the 
eigenstates of $\hat{h}^{(i)}$.
%
%-----------------------------------------------------------------------------
%
\subsection{The choice of $dt$ and the largest eigenvalue of $\hat{h}^{(i)}$}
\label{sec:largesth}

The upper end of the spectrum of $\hat{h}^{(i)}$ depends not only on the 
parametrisation employed and the numerical representation of the 
single-particle wave functions, but in a weaker way also on the particular
configuration of potentials at iteration $(i)$. For a general case, a precise 
analysis is not straightforward.  Some inferences can be made based on the 
operator structure of the single-particle Hamiltonian, which in turn is 
determined by the operator structure of the local densities through 
Eq.~\eqref{eq:variation}.

For NLO Skyrme EDFs, such analyses have been made before, see 
for instance \cite{Ryssens15a,Reinhard82}. In that case, the maximal 
eigenvalue of $\hat{h}^{(i)}$ is dominated by the kinetic energy of the
corresponding eigenstate. The variation of the energy with respect 
to $\tau(\vec{r})$ results in the presence of a Laplacian operator
in the single-particle Hamiltonian $\hat{h}^{(i)}$. 
Whereas other terms from Eq.~\eqref{eq:SkTeven:2} also contribute to the 
potential $\mathsf{F}^{\tau}(\vec{r})$, and a Laplacian operator also appears 
in the time-odd term multiplied by $\mathsf{F}^{T}(\vec{r})$, the contribution 
from the kinetic energy, Eq. \eqref{eq:Ekin}, can be expected to be by far 
the dominant one, as otherwise one would work with a parametrisation whose
effective mass takes an unrealistic value in one of the various spin-isospin 
channels.

The largest kinetic energy is found for the oscillatory mode with the 
shortest wavelength that can be represented in the chosen numerical 
representation. In a co\-or\-di\-nate-space representation using 
an equidistant mesh as done here, the analysis is straightforward:
the state that oscillates the most quickly has one node less than
there are mesh points. On a mesh with spacing $dx$, the momentum
of a state $|\psi^{\vec{k}_{\rm max}}\rangle$ with wavelength $2 dx$
therefore provides an upper bound for the maximum momentum $\vec{k}_{\rm max}$
\begin{equation}
|\vec{k}_{\rm max}|^2 \lesssim 3 \frac{\pi^2}{dx^2} \, ,
\label{eq:maxk}
\end{equation}
where the factor $3$ arises from the contributions of all Cartesian directions.
In that case the maximal eigenvalue 
of $\hat{h}^{(i)}$ scales with the mesh spacing as
\begin{align}
  \epsilon^{\rm NLO}_{\rm max} 
& \simeq
   -\frac{\hbar^2}{2m}
   \langle \psi^{\vec{k}_{\rm max}} |  \Delta|\psi^{\vec{k}_{\rm max}}\rangle  
   = \frac{\hbar^2}{2m} \frac{3\pi^2}{dx^2} \, .
  \label{eq:NLOscaling}
\end{align}
Inserting relation~\eqref{eq:NLOscaling} into 
Eq.~\eqref{eq:dtgradcondition}, one finds that the maximal allowed value of 
$dt$ decreases quadratically with the mesh spacing, as was discussed 
in~\cite{Bonche05,Ryssens15a}.

For N2LO functionals, the analysis is much less clear cut. Compared to the 
NLO case, there are additional contributions with two gradients to the 
single-particle Hamiltonian, Eq.~\eqref{eq:singleh:N2LO}, that have a more 
complicated tensor structure than the NLO terms. 
More importantly for our analysis, there are also terms with three or four 
gradients acting on the wave function in Eq.~\eqref{eq:singleh:N2LO}. 
The contribution of those containing three gradients to the highest 
eigenvalue of $\hat{h}^{(i)}$ scales as $|\vec{k}_{\rm max}|^3$, and 
those with four gradients even scale with $|\vec{k}_{\rm max}|^4$.
Because of the contribution from the kinetic energy, 
$\mathsf{F}^{\tau}(\vec{r})$ can be 
expected to be positive and the dominant one among terms with two gradients. 
By contrast, the potentials specific to N2LO can all have either sign.
For given $\vec{k}_{\rm max}$, the dominant contribution to the highest 
eigenvalue of $\hat{h}^{(i)}$ depends then on the relative sizes and signs 
of the various N2LO potentials when compared to $\mathsf{F}^{\tau}(\vec{r})$.

In the most unfavourable scenario, $\mathsf{F}^{Q}(\vec{r})$ is positive 
and of a size that makes  
$\langle \psi^{\vec{k}_{\rm max}} |\mathsf{F}^{Q} \Delta \Delta
| \psi^{\vec{k}_{\rm max}}\rangle$ a large positive number, such that 
the largest single-particle energy that can be represented on the mesh 
scales with $dx^{-4}$ instead of $dx^{-2}$, leading to larger values
$\epsilon^{\rm N2LO}_{\rm max}$ than what is estimated for NLO functionals, 
Eq.~\eqref{eq:NLOscaling}. The step size $dt$ of the gradient descent for 
the diagonalisation subproblem then has to be reduced accordingly, slowing
down the convergence rate.

\begin{figure}[t!]
\centerline{\includegraphics[width=0.8\linewidth]{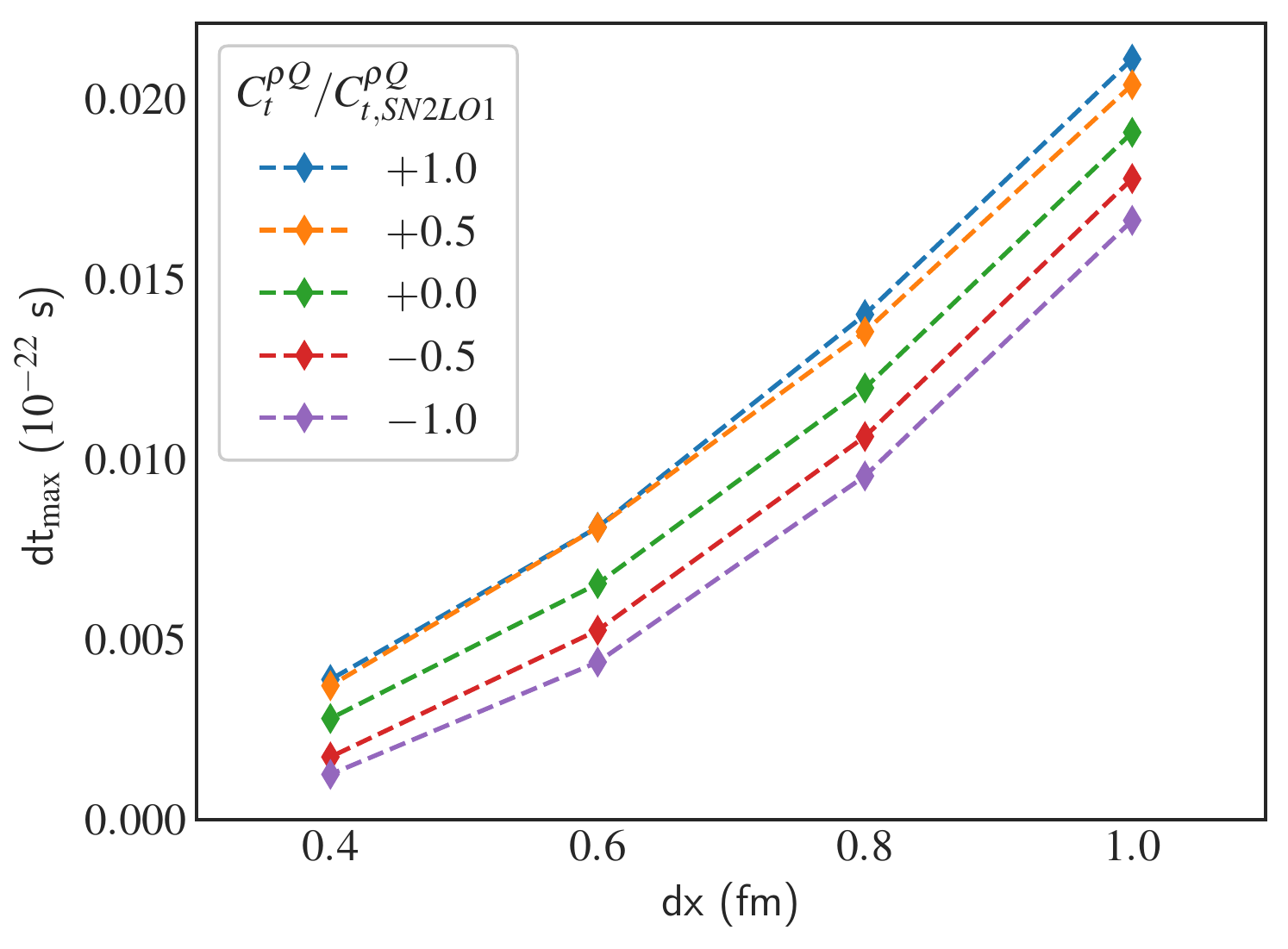}}
\caption{Maximum value of
$dt$ for which gradient-descent calculations of $^{40}$Ca converge for a
modified SN2LO1 interaction (see text) as a function of $dx$, for different
values of the N2LO coupling constants $C^{\rho Q}_0$ and $C^{\rho Q}_1$.}
\label{fig:dt}
\end{figure}

More importantly, because of the parametrisation-dependence of the 
relative sizes and signs of the potentials multiplying three and four 
gradients, for future N2LO parametrisations we have to expect a wide 
variation in the largest eigenvalues of $\hat{h}^{(i)}$ on a given mesh.
This, in turn, will lead to a large spread in the performance of gradient 
descent for the diagonalisation subproblem. To illustrate this point, 
we have constructed a set of toy N2LO parametrisations
based on the SN2LO1 parametrisation~\cite{Becker17}, for which all of 
the N2LO coupling constants have been set to zero, except for 
$C^{\rho Q}_0$ and $C^{\rho Q}_1$.
Figure~\ref{fig:dt} shows the largest value of $dt$ for which
a gradient-descent calculation for $^{40}$Ca converges, as a function of the
mesh discretization $dx$. When $C^{\rho Q}_0 = C^{\rho Q}_1 = 0$, the toy
parametrisation does not contain N2LO terms anymore, such that the curve 
represents the typical behaviour found at NLO.
For non-zero values of the N2LO coupling constants, and depending on their 
sign, the maximal allowable value of $dt$ either increases or decreases with 
respect to the NLO curve. This variation is quite large; at $dx=0.8$~fm, there 
is a factor of about two difference in $dt$ (and hence convergence rate) when 
changing the sign of $C^{\rho Q}_0$ and $C^{\rho Q}_1$. Even for this modest 
range of coupling constants, the sensitivity of the maximal value $dt$ is 
rather high, a situation that is not desirable at all when $dt$ has to be set 
by hand.

%-----------------------------------------------------------------------------

\subsection{Heavy-ball dynamics for the diagonalisation subproblem}
\label{sec:hbdiagonal}
The heavy-ball update from iteration $j$ to iteration $j+1$ for the
$\ell$-th single-particle wave function is given by\footnote{We recall that 
the index $i$ labels the iterations of the SCF 
subproblem where the single particle Hamiltonian $\hat{h}^{(i)}$ is updated 
and $j$ labels the iterations of the diagonalisation of $\hat{h}^{(i)}$.}
\begin{align}
| \delta \psi^{(i,j)}_{\ell} \rangle
&= |\psi^{(i,j)}_{\ell}\rangle  - |\psi^{(i,j-1)}_{\ell}\rangle 
\qquad \qquad \text{(if } j\not= 0) \, , \\
|v_{\ell}^{(i,j)} \rangle  &=
     - \frac{dt}{\hbar}\left( \hat{h}^{(i)} - \epsilon_\ell^{(i,j)} \right) 
| \psi^{(i,j)}_{\ell} \rangle  + \mu | \delta \psi_{\ell}^{(i,j)} \rangle \,,\\
|\phi^{(i,j+1)}_{\ell}\rangle
&= |\psi^{(i,j)}_{\ell}\rangle  + |v^{(i,j)}_{\ell}\rangle \, .
\label{eq:momscheme}
\end{align}
For the first iteration, $j=0$, one initializes 
$|\delta \psi^{(i,0)}_{\ell} \rangle$ with 
$|\delta \psi^{(i-1,j_f)}_{\ell}\rangle$  from the last iteration $j_f$
performed with the previous Hamiltonian $\hat{h}^{(i-1)}$.
An orthonormalization of the single-particle wave functions 
$|\phi_{\ell}\rangle \rightarrow |\psi_{\ell}\rangle$  is also needed at 
each iteration. Compared to the gradient-descent method, this algorithm does 
not require extra CPU time, but only extra storage of the auxiliary 
variables $| \delta \psi^{(i,j)}_{\ell} \rangle$.

\begin{figure}[t!]
\centerline{\includegraphics[width=0.8\linewidth]{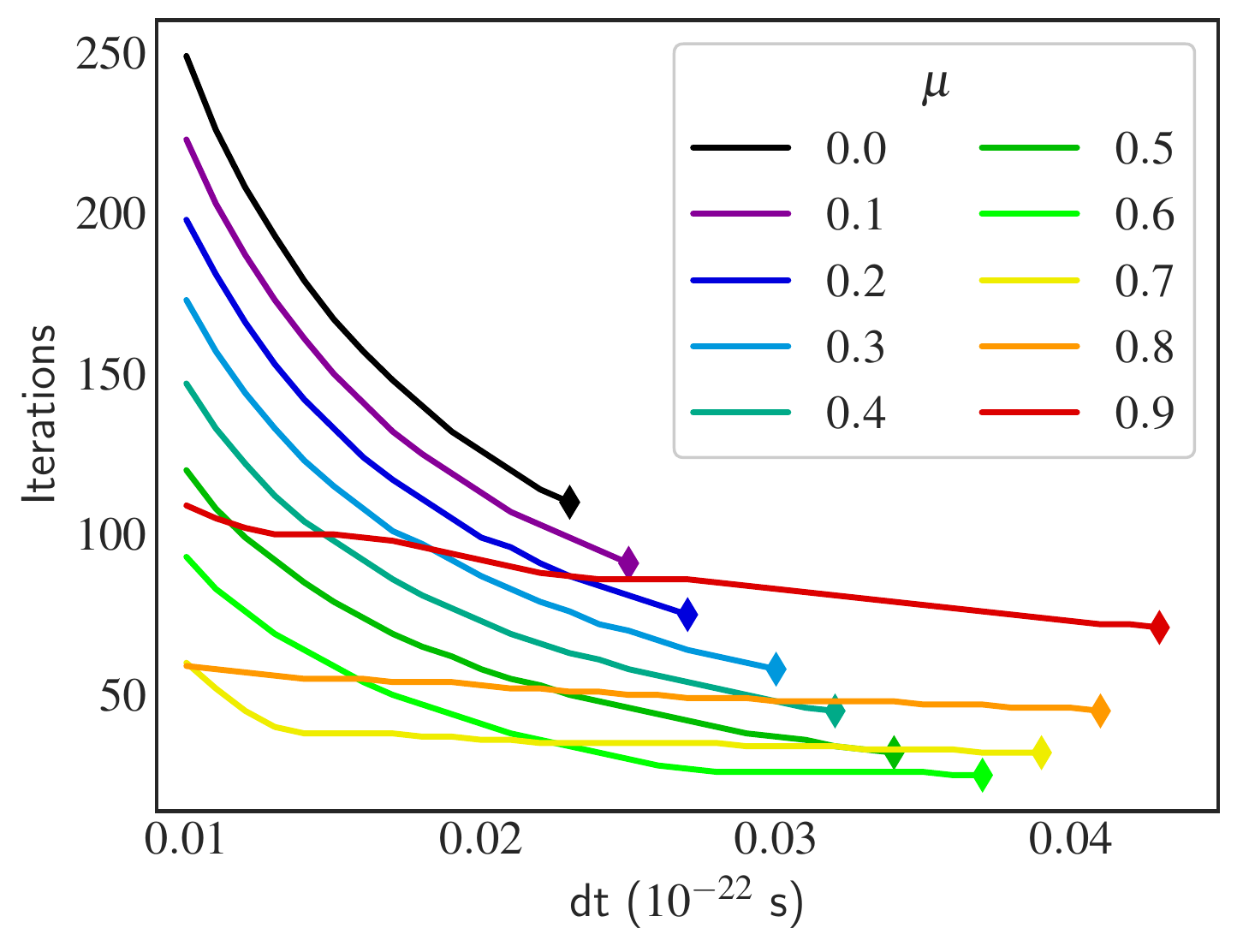}}
\caption{ Number of iterations needed to reach a weighted dispersion 
          $(dh^2)^{(0,j)} = 10^{-5}$ MeV$^2$ as a function of $dt$ and $\mu$, 
          without updating mean-field densities and potentials. The nucleus 
          under consideration is $^{16}$O in a box with $N_x = N_y = N_z = 24$ 
          at $dx=0.8$ fm using the SLy4 parametrisation of the Skyrme NLO EDF. 
          All calculations were initialized using spherical Nilsson orbitals. 
          The last point on every line, marked by a diamond, is the highest 
          value of $dt$ for which we were able to converge the calculation.}
\label{fig:iterations}
\end{figure}

We expect from Eq.~\eqref{eq:momcondition} that the algorithm converges under
the condition that
\begin{equation}
\label{eq:dthbcondition}
0 \leq \mu < 1\, , \qquad
\frac{dt}{\hbar}
< \frac{2 \, (1 + \mu)}
       {\epsilon^{\rm ex}_{\rm max} - \epsilon^{\rm ex}_{1} } \, .
\end{equation}
Figure~\ref{fig:iterations} illustrates the dependence of the convergence speed
on the specific choices made for $dt$ and $\mu$ for the case of the 
diagonalisation subproblem of $^{16}$O. For a given $\mu$, the amount of 
iterations needed to converge is smallest for values of $dt$ just below the 
ceiling value of Eq.~\eqref{eq:dthbcondition}. As $\mu$ increases, the 
ceiling value for $dt$ increases as well, until 
for $\mu = 0.9$ it almost doubles compared to the gradient descent case. The 
overall minimum of iterations needed is, however, achieved for a specific 
value of $\mu$, here close to 0.6.

The heavy-ball method has the same drawback as the gradient-descent method:
the optimal selection of $dt$ depends on the numerical basis and the
form and parametrisation of the EDF. To complicate the situation, 
the additional parameter $\mu$ also impacts the convergence rate. In the next 
section, we propose an algorithm to select appropriate values
of the iterative parameters dynamically during the optimization process.

%--------------------------------------------------------------------------
%
\subsection{Dynamical update of the iteration parameters}
\label{sec:dynparam}

We are not aware of a theoretical study to find the optimal parameters for
either gradient descent or heavy-ball dynamics applied to an optimization 
problem similar to that of Eq.~\eqref{eq:consopt}. We therefore propose, 
without detailed analysis, a few simple formulas that are inspired by the 
results of Section~\ref{sec:linear} and which can be used to dynamically 
estimate appropriate numerical parameters at every SCF iteration $(i)$, 
eliminating all need for human fine-tuning.

For both algorithms, we first need an estimate for the largest eigenvalue 
of the single-particle Hamiltonian, $\epsilon^{\rm ex}_{\rm max}$. 
For its determination, we introduce an extra single-particle wave 
function $|\psi^{(i)}_{\text{aux}}\rangle$ that does not participate in 
the evolution of the single-particle states the nuclear observables are
calculated from. Introducing a third iteration counter (k), we evolve 
this auxiliary wave function with the so-called power-iteration scheme
\cite{SaadBook} 
\begin{equation}
|\psi^{(i,k+1)}_{\text{aux}}\rangle 
= \frac{1}
 {\langle \psi^{(i,k)}_{\text{aux}} |(\hat{h}^{(i)})^{2} |\psi^{(i,k)}_{\text{aux}} \rangle}  
\hat{h}^{(i)} |\psi^{(i,k)}_{\text{aux}} \rangle \, .
\label{eq:poweriteration}
\end{equation}
For $k \rightarrow \infty$, $|\psi^{(i,k)}_{\text{aux}}\rangle$ will 
converge to the eigenstate with largest possible eigenvalue 
$\tilde{\epsilon}^{(i)}_{\rm max}$ of $\hat{h}^{(i)}$. In practice, 
one does not have to reach large accuracy to obtain an estimate of
\begin{equation}
\label{eq:epsilontilde}
\tilde{\epsilon}^{(i)}_{\rm max} 
\equiv \langle \psi^{(i,\infty)}_{\text{aux}} | \hat{h}^{(i)} 
|\psi^{(i,\infty)}_{\text{aux}} \rangle
\end{equation}
that serves our purpose. 
In general, a few iterations of Eq.~\eqref{eq:poweriteration} suffice to 
determine $\tilde{\epsilon}^{(i)}_{\rm max}$ to within a few MeV. 
The converged auxiliary state from the previous SCF iteration 
$(i-1)$ can be used as starting point at iteration $(i)$. As 
$|\psi^{(i)}_{\text{aux}}\rangle$ is a highly oscillatory state, at 
the very first SCF iteration it can be initialized using a random number 
generator. 

For heavy-ball dynamics, we also need to estimate the momentum parameter. 
By analogy to Eq.~\eqref{eq:muoptA}, we propose to set
\begin{equation}
\label{eq:muopth}
\mu^{(i)}_{\rm dyn}
=  \left(\frac{\sqrt{\kappa^{(i)}} - 1}{\sqrt{\kappa^{(i)}}+ 1}\right)^2 \, ,
\end{equation}
where $\kappa^{(i)}$ is the analogue of the condition number of the matrix
$A$ in Section~\ref{sec:linear}. Because of the simultaneous optimization
of many eigenvalues under an orthogonality constraint, we cannot 
use Eq.~\eqref{eq:condition:number}. After some numerical experimentation, 
we have adopted the prescription
\begin{align}
\label{eq:effchoice}
\kappa^{(i)} 
= \frac{\tilde{\epsilon}^{(i)}_{\rm max} - \epsilon^{(i)}_{\rm 1}}
       {E^{\rm qp, (i)}_{\rm min}} \, ,
\end{align}
where $\tilde{\epsilon}_{\rm max}$ is given by Eq.~\eqref{eq:epsilontilde}
and $E^{\rm qp, (i)}_{\rm min}$ is the smallest positive quasiparticle 
energy of the HFB Hamiltonian. In our experience, Eq.~\eqref{eq:effchoice}
provides a robust estimate for 
$\mu^{(i)}_{\rm dyn}$ when fed into Eq.~\eqref{eq:muopth}. This formula
has been tailored for the optimization of the diagonalisation problem 
embedded into SCF iteration under the assumption that only the occupied
single-particle states need to be well converged. Other applications will
require a different recipe.

Using Eq.~\eqref{eq:muopth} to calculate the momentum parameter, we set
\begin{equation}
\label{eq:dtopth}
\frac{dt^{(i)}_{\rm dyn}}{\hbar}
= s \frac{2 \, \big( 1 + \mu_{\text{dyn}}^{(i)} \big)}
{\tilde{\epsilon}^{(i)}_{\rm max} - \epsilon_{1}^{(i)}} \, ,
\end{equation}
where we introduced a factor $0< s < 1$ to guarantee that 
Eq.~\eqref{eq:dthbcondition} holds. To obtain a dynamical estimate for the
step size for use in a gradient-descent approach, it is sufficient to set 
$\mu^{(i)}_{\rm dyn} = 0$ in Eq.~\eqref{eq:dtopth}. In practice, we have 
empirically fixed the factor $s$ to  $0.9$ in 
Eq.~\eqref{eq:dtopth}. This empirical choice guaranteed convergence 
in all of the many calculations we performed since implementing 
this scheme.
%-------------------------------------------------------------------------------
\subsection{Comparison to other approaches}

Gradient-based schemes are used by the nuclear physics community for 
decades, with~\cite{Mang76,Davies80} being among the first detailed references
describing the application of such methods to the self-consistent HF and HFB 
problems.

More advanced iterative methods have been developed since. One example
is the nonlinear conjugate-gradient method~\cite{Egido95}. Like the scheme
described in Ref.~\cite{Mang76}, it is often used to address the 
diagonalisation and SCF subproblems simultaneously. Like many other 
advanced optimisation schemes~\cite{NumRecipes}, this method relies on an 
adaptive step size, which needs to be chosen in some optimal way at every
iteration. Its determination is usually accomplished by some variation of 
a line-search method, which requires multiple evaluations of the total 
energy, each time for a different set of single-particle states. Compared 
to the simple gradient scheme described above, in coordinate-space 
representation, the necessary multiple construction of derivatives of 
single-particle states and summation of densities increases the numerical 
cost per iteration by a large factor, as these are the most costly tasks.
Heavy-ball dynamics does not have this drawback, as it does not require 
any evaluation of the total energy or of single-particle matrix elements 
in addition to those already needed for the unmodified gradient descent.

As a last remark, we note that the heavy-ball update scheme presented 
here is 
quite similar to that of Car-Parrinello dynamics~\cite{Car85}, a widely-used 
method in atomic density functional theory~\cite{Hutter12}. In Car-Parrinello 
dynamics, the analogue of the momentum parameter $\mu$ is usually called the 
fictitious electron mass, and has the same impact on the convergence rate.
To the best of our knowledge, the only application of this method to a
nuclear problem is described in Ref.~\cite{Matsuse98}. 

%------------------------------------------------------------------------------
%
\subsection{Summary and MOCCa implementation}

As mentioned before, for calculations performed on a 3d mesh, the 
diagonalisation subproblem is much more CPU intensive than the evolution of the 
mean-field potentials and densities because it requires to determine the 
derivatives of the single-particle wave functions. In practice, the most 
appropriate choice is to restrict the diagonalisation of $\hat{h}^{(i)}$ to a 
single iteration, $j_{\rm max} = 1$.  For this particular choice, the extra 
iteration index $j$ is superfluous and we will drop it in what follows.
The auxiliary variables in Eq.~\eqref{eq:momscheme} become
\begin{equation}
|\delta \psi^{(i)}_{\ell} \rangle 
= |\psi^{(i)}_{\ell} \rangle - |\psi^{(i-1)}_{\ell} \rangle \, .
\end{equation}
A single iteration $(i)$ of the heavy-ball algorithm can be summarised 
as

\begin{enumerate}
\item  Calculate $\tilde{\epsilon}_{\rm max}^{(i)}$, 
       Eq.~\eqref{eq:epsilontilde}, by evolving $|\psi^{(i,k)}_{\rm aux}\rangle$ 
       according to Eq.~\eqref{eq:poweriteration}, starting from 
       $|\psi^{(i-1)}_{\rm aux}\rangle$.
\item Calculate $dt_{\rm dyn}^{(i)}$ and $\mu_{\rm dyn}^{(i)}$ using Eqs.
        \eqref{eq:dtopth}, \eqref{eq:muopth}.
\item Obtain all $|\phi^{i+1}_{\ell}\rangle$ from Eq.~\eqref{eq:momscheme}.
\item Orthonormalize to obtain the set of $|\psi_{\ell}^{(i+1)} \rangle$.
\item Calculate $|\delta \psi^{(i+1)}_{\ell} \rangle$ for use in the next 
      iteration.
\end{enumerate}
The gradient-descent algorithm consists of the same steps with 
$\mu^{(i)}_{\rm dyn} = 0$.

\begin{figure}[t!]
\centerline{\includegraphics[width=0.8\linewidth]{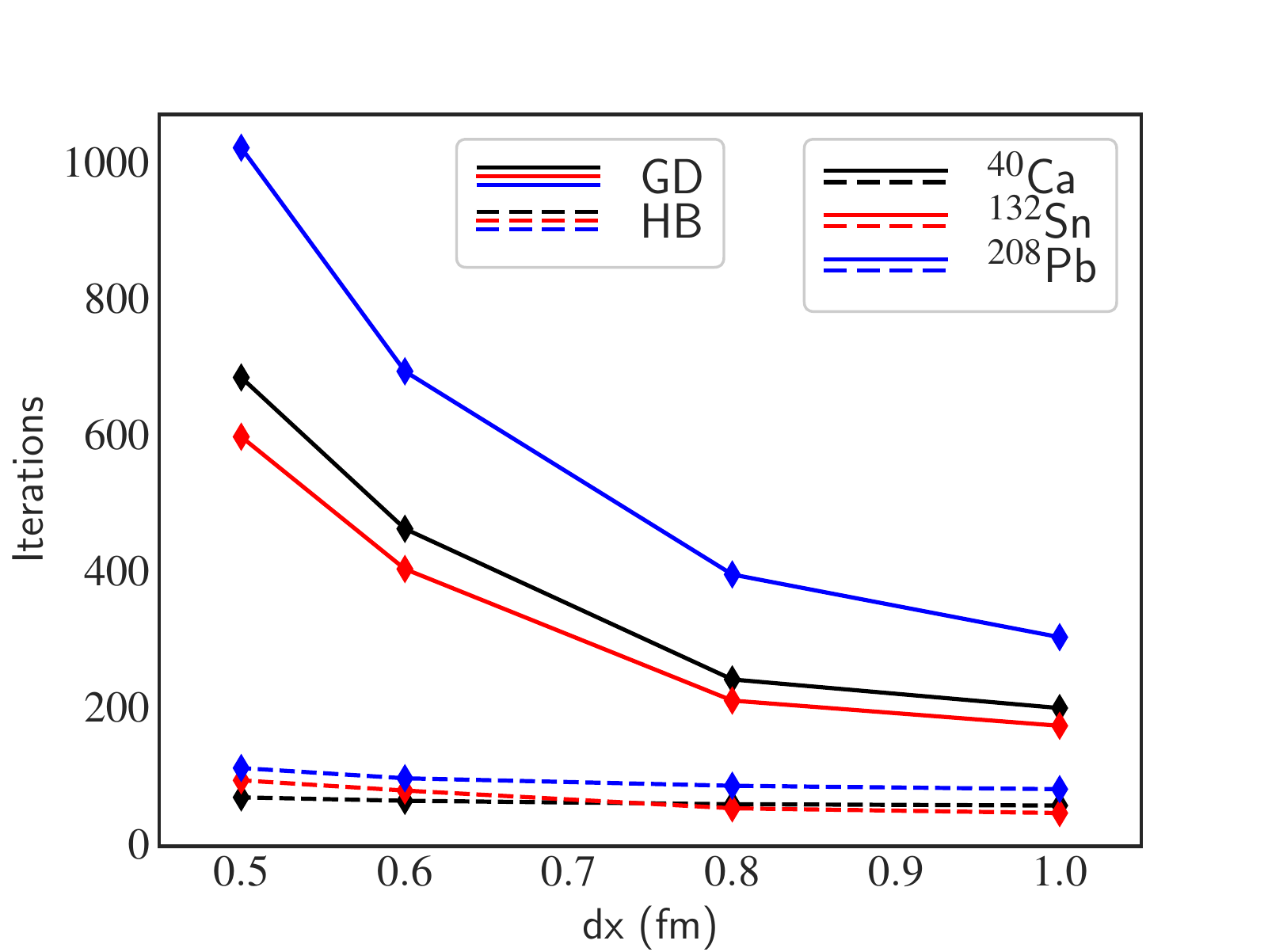}}
\caption{
Number of SCF iterations needed to achieve $(dh^2)^{(i,j)} \leq 10^{-5}$ 
MeV$^2$ for $^{40}$Ca, $^{132}$Sn and $^{208}$Pb using the SLy5s1 
parametrisation of the Skyrme NLO EDF as a function of the mesh 
discretization $dx$.}
\label{fig:spheriter}
\end{figure}

The efficiency of the heavy-ball approach and
the dynamical estimates for the numerical parameters  proposed 
in Section~\ref{sec:dynparam}, is illustrated by Fig.~\ref{fig:spheriter}. 
It shows the number of iterations needed to converge the diagonalisation 
subproblem
for three different spherical nuclei as a function of the mesh discretization 
$dx$. Results obtained with the gradient-descent algorithm and heavy-ball 
dynamics with dynamically estimated values for $dt$ and $\mu$ are compared. 
Recalling that the computational time per iteration is virtually the same in 
both schemes,
the heavy-ball algorithm is several times faster than the gradient-descent 
approach, and this for all mesh discretizations $dx$.
More strikingly, the number of iterations needed to converge the heavy-ball 
algorithm is (almost) independent of the mesh discretization. As discussed 
in Section~\ref{sec:largesth}, for the EDF used here
the largest eigenvalue of the single-particle Hamiltonian increases with the 
inverse of the square of the mesh spacing, Eq.~\eqref{eq:maxk}, increasing the 
spread in eigenvalues of the single-particle Hamiltonian. The maximal value of 
$dt$ decreases for smaller $dx$, which significantly slows down the 
gradient-descent algorithm. Heavy-ball dynamics with dynamically adjusted 
parameters is able to compensate for this spread, validating also the 
recipe provided by Eqs.~\eqref{eq:muopth} and~\eqref{eq:dtopth}.

%%%%%%%%%%%%%%%%%%%%%%%%%%%%%%%%%%%%%%%%%%%%%%%%%%%%%%%%%%%%%%%%%%%%%%%%%%%%%%%%

\section{The SCF Iterations}
\label{sec:dmixing}

\subsection{Position of the problem}

We now investigate the second aspect of the solution of the self-consistent 
mean-field equations, which is the SCF subproblem of evolving the 
single-particle Hamiltonian from one self-consistent iteration to the next. 
For the sake of simple notation, we drop again the isospin index of 
densities and potentials.

As discussed in Section~\ref{sec:selfconsistent}, the aim is to find the
fixed point of the maps $G^{\mathsf{R}}$ and $G^{\mathsf{F}}$. The most
straightforward way to iterate the fixed-point problem of 
Eq.~\eqref{eq:fixedpoint} is to calculate the mean-field densities and 
potentials of iteration $(i+1)$ from the single-particle wave functions, 
obtained by the approximate diagonalisation of $\hat{h}^{(i)}$
in the previous iteration
\begin{align}
\label{eq:directF}
\mathsf{F}^{(i+1)}
& = G^{\mathsf{F}} \left( \mathsf{F}^{(i)}\right)
  = \mathsf{F}_{|\psi\rangle}^{(i+1)} \, ,
    \\
\label{eq:directR}
\mathsf{R}^{(i+1)}
& = G^{\mathsf{R}} \left( \mathsf{R}^{(i)} \right)
  = \mathsf{R}_{|\psi\rangle}^{(i+1)} \, ,
\end{align}
where we use the notation 
$\mathsf{F}^{{(i+1)}}_{|\psi\rangle}, \mathsf{R}_{|\psi\rangle}^{(i+1)}$ to 
emphasize that the right-hand-sides of these relations are the 
potentials and fields 
obtained by direct calculation from the single-particle wave functions at 
iteration $(i+1)$, using the formulas from Section~\ref{sect:Skyrme}.
The scheme defined through 
Eqs.~\eqref{eq:directF} and~\eqref{eq:directR} does not lead to a stable 
iterative process, except in isolated cases. 
To understand why, 
consider the behaviour of the map given by Eq.~\eqref{eq:fixedpoint} in 
the vicinity of a fixed point $\mathsf{F}^{(\infty)}$.  If the error on 
the mean-field potentials is $\delta \mathsf{F}^{(i)}$, one can write
\begin{equation}
\label{eq:fixedlinear}
G^{\mathsf{F}} \left(\mathsf{F}^{(\infty)} + \delta \mathsf{F}^{(i)}\right)
\approx \mathsf{F}^{(\infty)} + 
\frac{\partial G^{\mathsf{F}}}{\partial \mathsf{F}}
\bigg|_{\mathsf{F}_
= \mathsf{F}^{(\infty)}} \delta \mathsf{F}^{(i)} \, .
\end{equation}
When Eq.~\eqref{eq:directF} is used to generate a new set of potentials
$\mathsf{F}^{(i+1)}$, the deviation from the fixed point at the next
iteration can be approximated by
\begin{equation}
\delta \mathsf{F}^{(i+1)}
= \mathsf{F}^{(i+1)} - \mathsf{F}^{(\infty)}
\approx J^{\mathsf{F}} \delta \mathsf{F}^{(i)} \, ,
\end{equation}
where $J^{\mathsf{F}}$ is the Jacobian of the problem for the potential 
$\mathsf{F}$ at the fixed point,
\begin{equation}
J_{ab}^{\mathsf{F}}
= \frac{\partial G_{a}^{\mathsf{F}}}{\partial \mathsf{F}_{b}}
  \bigg|_{\mathsf{F}= \mathsf{F}^{(\infty)}} \, .
\end{equation}
In a similar fashion, we obtain for the mean-field densities $\mathsf{R}^{(i)}$
in the vicinity of the fixed point $\mathsf{R}^{(\infty)}$
\begin{equation}
\delta \mathsf{R}^{(i+1)}
\approx J^{\mathsf{R}} \delta \mathsf{R}^{(i)} \, , \qquad
J_{ab}^{\mathsf{R}}
= \frac{\partial \mathsf{G}_{a}^{\mathsf{R}}}{\partial \mathsf{R}_{b}}
  \bigg|_{\mathsf{R}= \mathsf{R}^{(\infty)}} \, .
\end{equation}
Supposing that the linear approximation holds for all subsequent iterations, 
we obtain after $N$ further iterations
\begin{align}
\delta \mathsf{F}^{(i+N)} \approx
  \left(J^{\mathsf{F}}\right)^N \delta \mathsf{F}^{(i)}
\, , \label{eq:evopot}\\
\delta \mathsf{R}_{q,a}^{(i+N)} \approx
  \left(J^{\mathsf{R}}\right)^N \delta \mathsf{R}^{(i)}
\, . \label{eq:evoden}
\end{align}
This evolution only converges to a fixed point when the matrices
$J^{\mathsf{F}}_q$ and $J^{\mathsf{R}}_q$ are contractive, that is if all of 
their eigenvalues are smaller than one in absolute size. When any eigenvalue 
is larger than one, then the iterative process in general diverges, as 
errors get amplified from one iteration to the next.

The problem of such divergences has been extensively discussed in the context
of self-consistent calculations of atomic systems. In that case, the electron
density can engage in long-wavelength oscillations from one iteration to the
next, which is often called \emph{charge sloshing}. Especially in metallic 
systems, where small changes in input electron density produce large changes
in the output electron density, special measures need to be taken
to safeguard convergence~\cite{LinLin12}. It is the spectrum of the Jacobians
$J^{\mathsf{F}}$ and $J^{\mathsf{R}}$, and through them the form of the 
interaction, that determines convergence: in the atomic case, the divergent 
modes are because of the long-range character of the Coulomb 
potential. For nuclear Skyrme EDFs, we show in the next section that the 
presence of external derivatives of local densities in the EDF, 
Eq.~\eqref{eq:skyrme:energy}, gives rise to divergent short-wavelength 
modes.

\begin{figure}[t!]
\centerline{\includegraphics[width=0.8\linewidth]{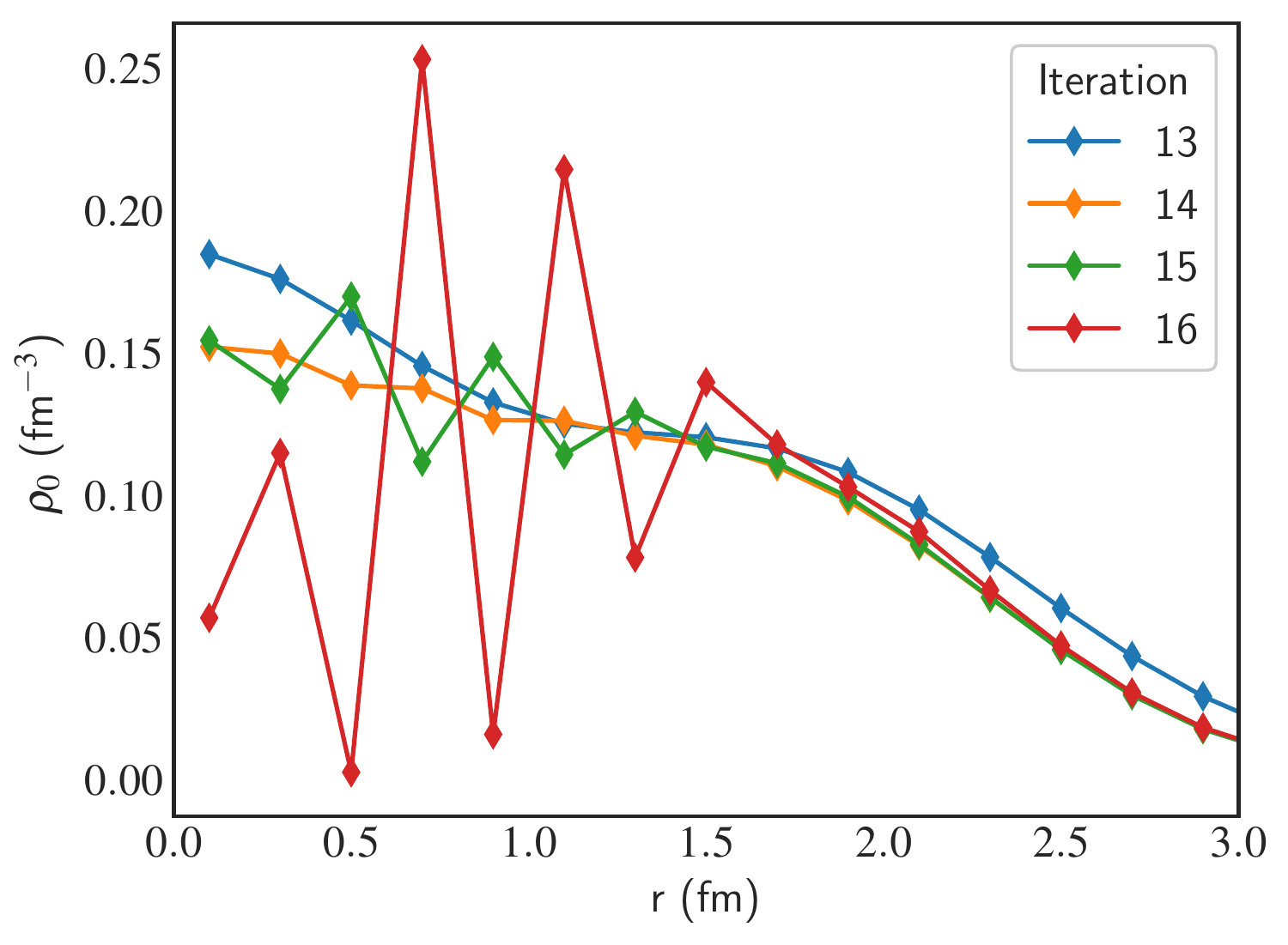}}
\caption{Radial density profile of $^{40}$Ca at iterations 13, 14, 15, and 16
for the SLy4 parametrisation of the Skyrme NLO EDF, with an SCF 
evolution directed by Eqs.~\eqref{eq:directF} and \eqref{eq:directR}.
The diagonalisation subproblem was iterated with heavy-ball dynamics.
At iteration 16, the energy is positive and the iterative process was
stopped. The calculation was initialized using spherical Nilsson orbitals
in a $(N_x, N_y, N_z) = (60,60,60)$ box with a mesh spacing of
$dx=0.2 \; \text{fm}$. We have chosen this mesh spacing for illustrative
purposes only; it is significantly smaller than values that are
typically used in practice ($dx \sim 0.8$ fm). The calculated values of the 
density on the mesh points are connected by straight lines to guide the eye.}
\label{fig:explosion}
\end{figure}

The instability of the iterative scheme of Eqs.~\eqref{eq:directF}
and ~\eqref{eq:directR} because of short-wavelength changes in the density
is illustrated by Fig.~\ref{fig:explosion} for the NLO parametrization
SLy4. The total density of $^{40}$Ca obtained by SCF iterations directly 
using  $\mathsf{F}^{(i+1)}_{|\psi\rangle}$ and 
$\mathsf{R}^{(i+1)}_{|\psi\rangle}$ is shown for iterations 13 to~16. 
At iteration 13 the central density still
looks smooth. During the subsequent iterations, short wavelength
deviations from the fixed point are quickly amplified by
the iterative process: at iteration 14 small unphysical oscillations 
are visible and from  iteration 16 onwards the corresponding total energy 
becomes positive.

Although the evolution of the densities with iteration number $(i)$ is very 
similar, this instability is different in nature from the finite-size 
instabilities that have been studied for the Skyrme EDF in 
Refs.~\cite{Hellemans13,Lesinski06,Pastore2015}. 
The divergence illustrated by Fig.~\ref{fig:explosion} is entirely 
numerical, and its appearance a consequence
of the iterative scheme unintentionally introducing oscillatory 
behaviour in the density that drives the system to a less bound state. 
By contrast, finite-size instabilities occur when the properties of an EDF 
parametrization are such that infinite inhomogeneous nuclear matter is 
more bound than the homogeneous phase and are thus a characteristic of 
a parametrization. All nuclear EDFs have to have such an
instability in the $T=0$, $S=0$ channel, which is responsible for the formation
of finite nuclei. However, it has recently been pointed out that many
parametrisations of the nuclear EDF also exhibit non-physical finite-size 
instabilities in other $S$, $T$ 
channels~\cite{Hellemans13,Lesinski06,Pastore2015,Martini18x,Gonzalez18x}. 
When these 
degrees of freedom can be resolved by the numerical representation used, 
the calculations are driven towards a highly oscillatory state that 
is more bound than a state with conventional density distribution, 
irrespective of the technique used to solve the mean-field equations.
The numerical instability illustrated by Fig.~\ref{fig:explosion}, on the
other hand, 
can be completely eliminated by adapting the iterative scheme, as we will 
show in what follows.

%---------------------------------------------------------------------------

\subsection{Highest eigenvalues of the Jacobians}

Using the chain rule for derivatives, one can write for $J^{\mathsf{R}}$
\begin{equation}
\label{eq:jacobfields}
J_{ab}^{\mathsf{R}}
= \sum_{c} \frac{\partial G(\mathsf{R})_{a}}{\partial \mathsf{F}_{c}}
  \bigg|_{\mathsf{F} = \mathsf{F}^{(\infty)}}
  \frac{\partial \mathsf{F}_{c}}{\partial \mathsf{R}_{b}}
  \bigg|_{\mathsf{R} = \mathsf{R}^{(\infty)}} \, .
\end{equation}
and similarly for $J^{\mathsf{F}}$. To obtain an idea of the behaviour of the 
second partial derivative in Eq.~\eqref{eq:jacobfields}, consider a variation of
the density $\rho(\vec{r})$ of the form
\begin{equation}
\label{eq:rhomodes}
\delta \rho(\vec{r}) = a \, e^{\iunit \vec{k} \cdot \vec{r}} \, ,
\end{equation}
where $a$ is a small constant. For large values of $|\vec{k}|$, the 
variation in the potential $\mathsf{F}^{\rho}(\vec{r})$ can be 
estimated to be
\begin{align}
\label{eq:response}
\delta \mathsf{F}^{\rho}(\vec{r})
\approx & \big[ 2 \, C^{\rho \rho} 
            + (2 + \alpha) \, (1 + \alpha) \,
              C^{\rho \rho \rho^\alpha} \, \rho^\alpha(\vec{r})
        \nonumber \\
        &   - 2 \, C^{\rho \Delta \rho} \, |\vec{k}|^2
            + 2 \, C^{\Delta \rho \Delta \rho} \, |\vec{k}|^4 \big]
      \, \delta \rho(\vec{r}) \, . 
\end{align}
where for sake of simple argument we omit the isospin structure of the 
single-particle Hamiltonian that couples perturbations in the density 
of one nucleon species to changes in the potentials of both species.
Even if $a$ is very small, such density components can make the evolution 
of the potentials divergent. The potentials $\mathsf{F}^{\rho}$ and 
$\mathsf{F}^{\vec{s}}$ are the most volatile, since at a given order of
 functional (NLO, N2LO or N3LO), they contain the highest number of external 
derivatives of a local density, which is either $\rho(\vec{r})$ or
$\vec{s}(\vec{r})$.

In electronic structure calculations, the partial derivative of the output  
densities $G^{\mathsf{R}}(\mathsf{R})$ with respect to the potentials 
$\mathsf{F}$, that constitutes the other part of Eq.~\eqref{eq:jacobfields},
is called the susceptibility~\cite{LinLin12}. It encodes the response of the 
system to a change of the potentials. In the context of EDF approaches, it 
is not obvious how to approximate this response efficiently: it depends 
sensitively on the changes in the single-particle wave functions that 
contribute to the densities. The eigenvalues of the matrices 
$\mathsf{J}^{\mathsf{R}}$ depend in this way on 
the technique used to solve the diagonalisation subproblem. In our case, thanks 
to the iterative diagonalisation by either gradient-descent or heavy-ball
dynamics, the change in the wave functions from one iteration to the next 
can be expected to be limited in size.

As discussed in Section~\ref{sec:linearsub} for the diagonalisation 
subproblem, it is the numerical representation that determines the scaling of 
the eigenvalues of $J^{\mathsf{R}}$ and $J^{\mathsf{F}}$. In coordinate 
space, the largest representable $\vec{k}$ can be estimated by 
Eq.~\eqref{eq:maxk}.
For an NLO functional,
 the largest eigenvalue of the Jacobian scales with $dx^{-2}$. 
For a general N2LO functional, the analysis depends on the balance between the 
contributions in Eq.~\eqref{eq:response}. The first, and so far only, 
available N2LO parametrisation fitted to the properties finite nuclei, 
SN2LO1~\cite{Becker17}, has coupling constants 
$C^{\Delta \rho \Delta \rho}_t$ that are both below 1.0 MeV. With this, 
they are much smaller than the almost one hundred~MeV for this parametrisation's
value of $C^{\rho \Delta \rho}_0$.
As we show in the next section, for SN2LO1, the NLO terms dominate the 
eigenvalues of the SCF Jacobians for typical choices of $dx$.

%------------------------------------------------------------------------------

\subsection{Linear mixing}
\label{subsect:LM}

A simple method to achieve convergence, even in the presence of a Jacobian
with large eigenvalues, is to perform a linear mixing between the updated 
potentials and densities and those from the previous iteration. We replace 
Eq.~\eqref{eq:directF} and Eq.~\eqref{eq:directR} with
\begin{align}
\label{eq:mixf}
\mathsf{F}_a^{(i+1)}
& = \mathsf{F}_a^{(i)}
    + \alpha^{\mathsf{F}} \left[ \mathsf{F}_{a,|\psi\rangle}^{(i+1)}
    - \mathsf{F}_{a}^{(i)}\right]\, , \\
\label{eq:mixr}
\mathsf{R}_a^{(i+1)}
& = \mathsf{R}_a^{(i)} 
    + \alpha^{\mathsf{R}} \left[ \mathsf{R}_{a,|\psi\rangle}^{(i+1)}
    - \mathsf{R}^{(i)}_a \right]\, ,
\end{align}
where $\alpha^{\mathsf{F}}$ and $\alpha^{\mathsf{R}}$ are parameters between 
zero and one, which can be chosen differently for each
density and potential. Using these equations to generate the
evolution of the potentials and densities, the expression equivalent to
Eqs.~\eqref{eq:evopot} and~\eqref{eq:evoden} become after $N$ iterations
\begin{align}
\delta \mathsf{F}^{(i+N)} \approx
\left[ 1 + \alpha_{\mathsf{F}} \left( -1   +
  J^{\mathsf{F}}\right)\right]^N \delta \mathsf{F}^{(i)}
\, , \label{eq:evopot2}\\
\delta \mathsf{R}^{(i+N)} \approx
\left[ 1 + \alpha_{\mathsf{R}} \left( - 1  +
  J^{\mathsf{R}} \right) \right]^N \delta \mathsf{R}^{(i)}
\, . \label{eq:evoden2}
\end{align}
The appropriate size of these mixing constants is related to the maximum
eigenvalues of the Jacobians, as discussed in the previous section. 

To illustrate this, we have constructed two sets of modified versions of 
the SLy4 NLO parametrisation with systematically varied coupling constants
of the time-even gradient terms. For the first set, the isovector terms 
are set to zero, $C^{\rho \Delta \rho}_1 = 0$, while the original coupling 
constant of the isoscalar term $C^{\rho \Delta \rho}_0$ is multiplied with 
a scaling factor between zero and one. The second set fixes 
$C^{\rho \Delta \rho}_0$ at the value of the original parametrisation and 
scales the isovector coupling constant $C^{\rho \Delta \rho}_1$ instead.
For reference, the original SLy4 values 
of the coupling constants are $C^{\rho \Delta \rho}_0 = -76.996$~MeV~fm$^{-5}$
and $C^{\rho \Delta \rho}_1 = 15.657$~MeV~fm$^{-5}$, respectively.

\begin{figure}[t!]
\centerline{\includegraphics[width=0.8\linewidth]{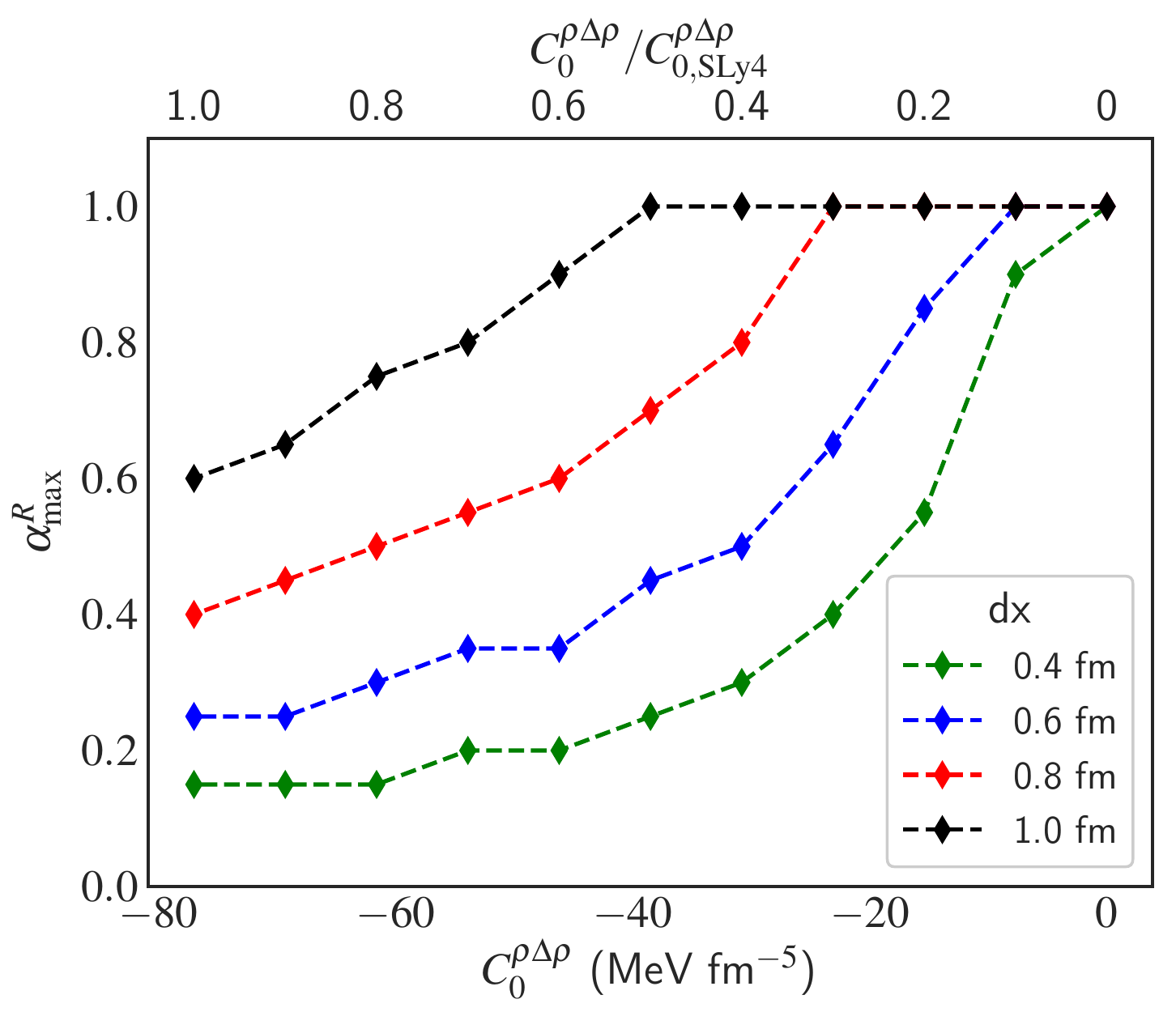}}
\caption{\label{fig:mixingaC0}
Maximum value of $\alpha^{\mathsf{R}}$ that results in a converged solution
for a HF calculation of $^{40}$Ca as a function of the coupling constant
$C^{\rho \Delta \rho}_0$ for the first set of modified SLy4 
parametrisations (see text). The convergence was tested by increasing 
$\alpha^{\mathsf{R}}$ in steps of 0.05, such that values shown are lower 
limits and discontinuous.
}
\end{figure}

Using the first set of modified parametrisations, 
Fig.~\ref{fig:mixingaC0} shows the maximum value of $\alpha^{\mathsf{R}}$ for 
which a calculation for $^{40}$Ca converges for different mesh 
spacings $dx$, keeping $\alpha^{\mathsf{F}} = 1$. For a Skyrme functional with 
$C^{\rho \Delta \rho}_0 = C^{\rho \Delta \rho}_1 = 0$, no mixing is necessary 
to converge the calculation for any choice of mesh spacing $dx$ in our
numerical representation, such that the
SCF update parameter can be set to $\alpha^{\mathsf{R}} = 1$. For finite 
$C^{\rho \Delta \rho}_0$, the value of $\alpha^{\mathsf{R}}$ has to be reduced 
in order to obtain a stable iterative process. For given $dx$,
the largest feasible $\alpha^{\mathsf{R}}$ decreases when increasing
the absolute value of $C^{\rho \Delta \rho}_0$, and it also decreases
when diminishing $dx$ for constant non-zero $C^{\rho \Delta \rho}_0$. For a
typical choice of $dx = 0.8$~fm, $\alpha^{\mathsf{R}}$ has to be smaller than
0.4. For other configurations or other nuclei the ceiling value might be 
slightly different, such that a safe value that works for all cases has to
be chosen even smaller.

\begin{figure}[t!]
\centerline{\includegraphics[width=0.8\linewidth]{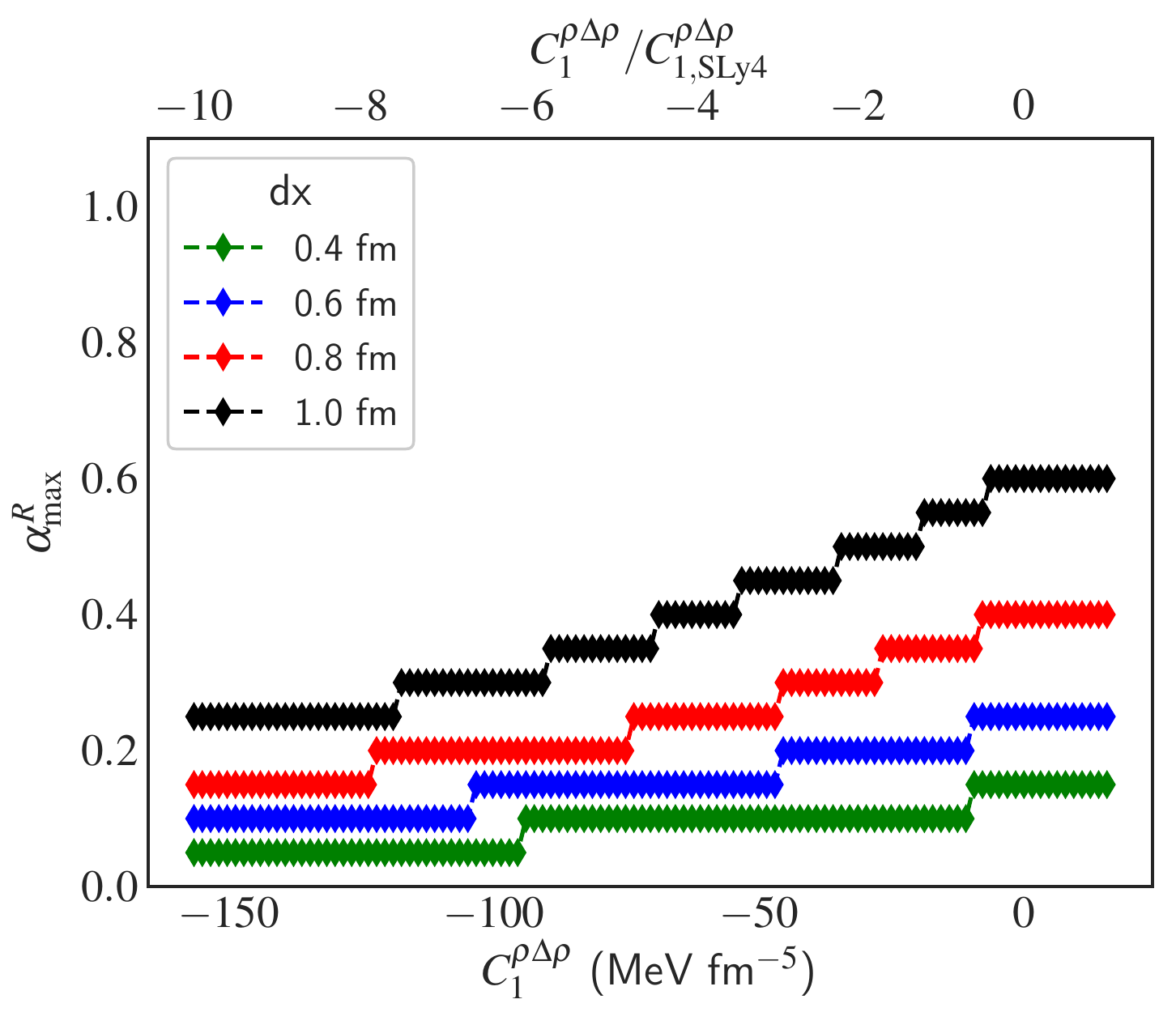}}
\caption{\label{fig:mixingaC1}
Maximum value of $\alpha^{\mathsf{R}}$ that results in a converged solution
for a HF calculation of $^{40}$Ca as a function of the coupling constant
$C^{\rho \Delta \rho}_1$ for the second set of modified SLy4 
parametrisations (see text). The convergence was tested by increasing 
$\alpha^{\mathsf{R}}$ in steps of 0.05, such that values shown are lower 
limits and discontinuous. Note that  
$C^{\rho \Delta \rho}_1/C^{\rho \Delta \rho}_{1,\text{SLy4}} = 0$ corresponds 
to $C^{\rho \Delta \rho}_0/C^{\rho \Delta \rho}_{0,\text{SLy4}} = 1$ in 
Fig.~\ref{fig:mixingaC0}.
}
\end{figure}

Figure~\ref{fig:mixingaC1} shows the largest feasible value of 
$\alpha^{\mathsf{R}}$ for the second set of parametrisations when varying 
$C^{\rho \Delta \rho}_1$ and $dx$. While the values of $C^{\rho \Delta \rho}_0$ 
are very similar for most parametrisations of the Skyrme NLO functional, those 
for $C^{\rho \Delta \rho}_1$ vary on a much wider scale.
The ratio $C^{\rho \Delta \rho}_1/C^{\rho \Delta \rho}_{1,\text{SLy4}}$
cannot be increased by a large factor, though; when reaching values of 
about~2 the non-physical finite-size instabilities discussed in 
Refs.~\cite{Hellemans13,Lesinski06,Pastore2015} set in. There are, 
however, some parametrisations for which $C^{\rho \Delta \rho}_1$ has the 
opposite sign and several times the absolute size compared to SLy4. 
Examples are 
UNEDF0 ($C^{\rho \Delta \rho}_1 = -55.62$~MeV~fm$^{-5}$) \cite{Kortelainen10}, 
UNEDF1 ($C^{\rho \Delta \rho}_1 = -145.38$~MeV~fm$^{-5}$)~\cite{Kortelainen12},
and also UNEDF2 
($C^{\rho \Delta \rho}_1 = -113.16$~MeV~fm$^{-5}$)~\cite{Kortelainen14}. 
The range of $C^{\rho \Delta \rho}_1$ covered in Fig.~\ref{fig:mixingaC1} 
extends to these values. When $C^{\rho \Delta \rho}_1$ becomes comparable to 
$C^{\rho \Delta \rho}_0$ in size and sign, the maximal feasible value of 
$\alpha^{\mathsf{R}}$ has to be reduced below the value dictated by the
isoscalar term alone.

Figures~\ref{fig:mixingaC0} and \ref{fig:mixingaC1} address the possible 
interplay between the size of the coupling constants of the gradient terms 
and the possible choices for numerical parameters that control the convergence 
rate of SCF subproblem for the case of a time-reversal conserving calculation.
When breaking time-reversal invariance, the analysis has also to consider 
the sign and size of the coupling constants $C_t^{s \Delta s}$, $t=0,1$, of 
the gradient terms of the spin density in Eq.~\eqref{eq:SkTodd:2}. The 
corresponding terms behave very similar to the isovector term multiplied by
$C^{\rho \Delta \rho}_1$: too large positive values of the $C_t^{s \Delta s}$
lead to non-physical finite-size instabilities, whereas for very large negative
values of $C_t^{s \Delta s}$ the size of the mixing parameter 
$\alpha^{\mathsf{R}}$ has to be 
adapted. In fact, the non-convergence of calculations for large, negative 
values of $C_t^{s \Delta s}$ reported in Ref.~\cite{Hellemans12} can be 
remedied by lowering $\alpha^{\mathsf{R}}$ below the value used in that
study. When also including terms from a genuine contact tensor force in the 
Skyrme EDF~\cite{Lesinski07,Hellemans12}, there are additional second-order 
gradient terms containing the divergence of the the spin density that 
are to be considered in the analysis. 
For the N2LO Skyrme functional, Eq.~\eqref{eq:EDF:4:e}, with its gradient 
terms that also contribute to other potentials the situation will be further 
complicated, as is the case when considering three-body terms with 
gradients \cite{Sadoudi13}.

\begin{figure}[t!]
\centerline{\includegraphics[width=0.9\linewidth]{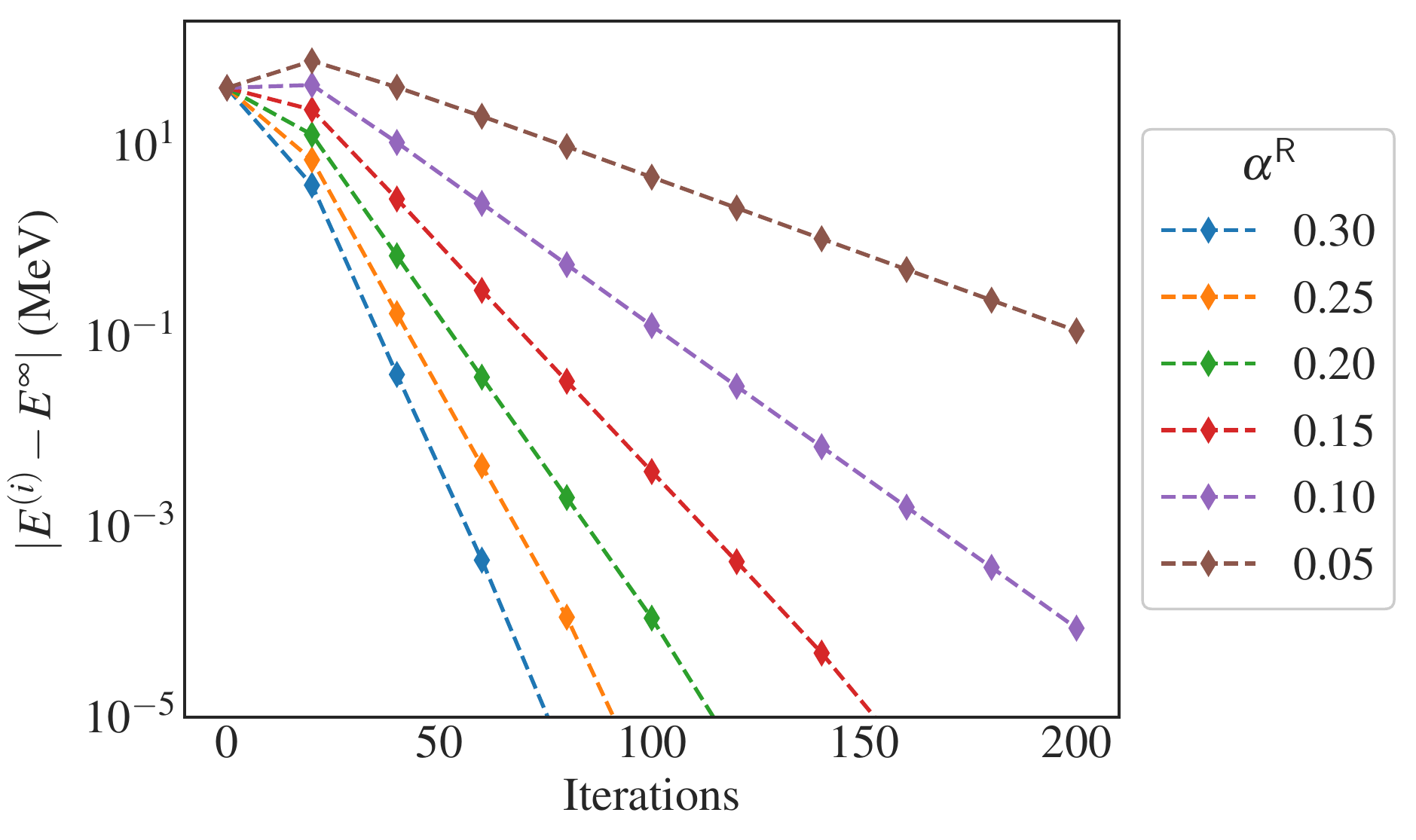}}
\caption{ (Color online)
Difference between the energy at a given iteration $E^{(i)}$ and the energy 
after convergence $E^{(\infty)}$ for different values of 
$\alpha^{\mathsf{R}}$ for a HF calculation of $^{40}$Ca at $dx=0.8 \; \text{fm}$
 with the SLy4 NLO parametrisation.}
 \label{fig:exampmix}
\end{figure}

The parametrisation dependence of the largest possible mixing parameter 
$\alpha^{\mathsf{R}}$ introduces a parametrisation dependence 
of the highest achievable convergence rate of the SCF iteration when using
linear mixing of densities, Eq.~\eqref{eq:mixr}, as smaller values of 
$\alpha^{\mathsf{R}}$ inevitably slow it down. The difference can be quite 
substantial. This is illustrated by Fig.~\ref{fig:exampmix} that displays
the error on the total binding energy of $^{40}$Ca in a calculation with 
the original SLy4 parametrisation for different values of the mixing 
parameter $\alpha^{\mathsf{R}}$ as a function of iteration number. For this 
choice of mesh with $dx = 0.8$~fm, $\alpha^{\mathsf{R}} = 0.3$ converges the 
calculation to the 0.01~keV level in about 70 iterations. In a calculation 
with $\alpha^{\mathsf{R}} = 0.05$, after the same number of iterations the 
energy still changes on the 1~MeV level, and almost 1000 iterations will be 
needed to reach the level of 0.01~keV error.

The main advantage of linear mixing is its simplicity. It has, however,
the serious limitation that the mixing parameter has to be fine-tuned to 
the parametrisation and mesh,  resulting in vastly different convergence rates
 in different circumstances. To the best of our
knowledge, no heuristic exists for selecting parameters prior to a
calculation and manual experimentation is required to guarantee stability
for a given combination of nucleus, mesh and EDF parametrisation
while not needlessly sacrificing CPU time.

%------------------------------------------------------------------------

\subsection{Preconditioning}

All modes of the change of density are treated on an equal footing by linear 
mixing, irrespective of the size of their $|\vec{k}|$. In order to obtain
 an algorithm that allows for propagating long and short wavelength changes
of the densities or potentials differently, we replace Eqs.~\eqref{eq:mixf} 
and~\eqref{eq:mixr} by
\begin{align}
\label{eq:preconpot}
\mathsf{F}_{a}^{(i+1)}
& = \mathsf{F}_{a}^{(i)}
    + P^{\mathsf{F}} \left[ \mathsf{F}_{a, | \psi \rangle}^{(i+1)}
    - \mathsf{F}_{a}^{(i)} \right]\, ,
    \\
\label{eq:preconden}
\mathsf{R}_{a}^{(i+1)}
& = \mathsf{R}_{a}^{(i)}
    + P^{\mathsf{R}} \left[ \mathsf{R}_{a, | \psi \rangle}^{(i+1)}
    - \mathsf{R}_{a}^{(i)} \right] \, .
\end{align}
The matrices $P^{\mathsf{F}}$ and $P^{\mathsf{R}}$ act as 
preconditioners of the fixed-point iteration. Their most useful form
in general depends on the nature of the nuclear EDF, the chosen geometry 
and the numerical representation, and in general has to be a 
compromise between numerical efficiency, robustness, and numerical cost. 
We propose to use a preconditioner of the form
\begin{equation}
\label{eq:preconop}
P^{\mathsf{F}/\mathsf{R}}
= \left(1 - \beta^{\mathsf{F}/\mathsf{R}} \Delta \right)^{-1} \, ,
\end{equation}
where $\beta^{\mathsf{F}}$ and $\beta^{\mathsf{R}}$ are numerical parameters 
and $\Delta$ is the Laplacian. Both $\beta^{\mathsf{F}}$ and 
$\beta^{\mathsf{R}}$ must be positive, in order for the matrices 
$P^{\mathsf{R}}$ and $P^{\mathsf{F}}$ to be positive definite on the mesh.

To illustrate the effect of a preconditioner of the potential, 
$P^{\mathsf{F}}$, take again as an example the response of the potential 
$\mathsf{F}^{\rho}$ of the Skyrme N2LO 
functional, Eq.~\eqref{eq:response}, to a change of density of the 
form~\eqref{eq:rhomodes}
\begin{align}
\delta \mathsf{F}^{\rho, (i)}
%\approx & \, P^{\mathsf{F}} \, \big[ 2 \, C^{\rho \rho} 
%            + (2 + \alpha) \, (1 + \alpha) \,
%              C^{\rho \rho \rho^\alpha} \, \rho^\alpha(\vec{r})
%        \nonumber \\
%        &   + 2 \, C^{\rho \Delta \rho} \, |\vec{k}|^2
%            + 2 \, C^{\Delta \rho \Delta \rho} \, |\vec{k}|^4 \big] \, 
%        \delta \rho(\vec{r}) \, , \nonumber \\
\approx & \,  
        \big[ 2 \, C^{\rho \rho} 
            + (2 + \alpha) \, (1 + \alpha) \,
              C^{\rho \rho \rho^\alpha} \, \rho^\alpha(\vec{r})
               - 2 \, C^{\rho \Delta \rho} \, |\vec{k}|^2
        \nonumber \\
        &  
            + 2 \, C^{\Delta \rho \Delta \rho} \, |\vec{k}|^4 \big] \, 
        \left[ 1 + \beta^{\mathsf{F}} | \vec{k} |^2\right]^{-1} \delta \rho(\vec{r})
\end{align}
For modes with small $|\vec{k}|$, $P^{\mathsf{F}}$ does not impede progress, as
for those modes $P^{\mathsf{F}} \approx 1$. When $|\vec{k}|$ is large, the
preconditioning matrix slows down the oscillatory components,
since $P^{\mathsf{F}} \sim |\vec{k}|^{-2}$ in that case. 

While in principle the preconditioning of the densities in 
Eq.~\eqref{eq:preconden} is a possible alternative to the 
preconditioning of the potentials, it is not as practical and well-behaved: 
there are stringent constraints on
the mean-field densities that are not present for the potentials. The density
$\rho(\vec{r})$, for example, needs to be positive everywhere while its
integral has to be equal to the number of particles. Unlike the linear
mixing of densities, Eq.~\eqref{eq:preconden} does not conserve such
properties from one iteration to the next, requiring additional measures
to rectify this issue. For this reason we set $\beta^{\mathsf{R}} = 0$,
placing the burden of slowing down the oscillatory modes completely on the
preconditioning of the potentials.

We note in passing that the simple form of Eq.~\eqref{eq:preconpot} is but 
one possible way to differentiate between changes in potentials
of different wavelengths. For N2LO parametrisations with large values of 
$C^{\Delta \rho \Delta \rho}_t$ for instance, it might be of interest to 
incorporate an extra Laplacian in the preconditioning matrix. In fact, one
can easily tailor more advanced prescriptions for different types of 
parametrisations or even employ different preconditioners for each individual 
contribution to the various potentials. Using
Eq.~\eqref{eq:preconpot} with an appropriate choice of $\beta^{\mathsf{F}}$,
however has sufficed to stabilize the SCF subproblem in all cases 
we have encountered so far.

\subsection{Practical implementation and choice of $\beta_{\mathsf{F}}$}

Let us first discuss how to implement the preconditioning scheme. The
matrix $P^{\mathsf{F}}$ is a full $(N_x \times N_y \times N_z) \times
(N_x \times N_y \times N_z)$ matrix on the coordinate mesh, and storing 
it as well as performing the matrix multiplication in Eq.~\eqref{eq:preconpot}
would be prohibitively expensive. Its inverse,
$(P^{\mathsf{F}})^{-1} = (1-\beta^{\mathsf{F}} \Delta)$, however, 
has a simple structure on the mesh, since the Laplacian is the sum of 
three separable matrices, see Eq.~\eqref{eq:mesh:Laplacian}. 

We obtain the update $\mathsf{F}_{a}^{\rm update}
\equiv \mathsf{F}_{a}^{(i+1)} - \mathsf{F}_{a}^{(i)}$
of every preconditioned potential in Eq.~\eqref{eq:preconpot} by solving 
the following linear system
\begin{equation}
\label{eq:precon}
\big( 1 - \beta^{\mathsf{F}}\Delta \big) \, \mathsf{F}_{a}^{\rm update} 
= \big[ \mathsf{F}_{a, | \psi \rangle}^{(i+1)} - \mathsf{F}_{a}^{(i)} \big] \, ,
\end{equation}
with a conjugate gradient method~\cite{NumRecipes}. We obtain in this way
the potentials at the next iteration,
$\mathsf{F}_{a}^{(i+1)} = \mathsf{F}_{a}^{(i)} + \mathsf{F}_{a}^{\rm update}$, 
at the cost of only a few
matrix multiplications with $(1 - \beta^{\mathsf{F}} \Delta)$ 
for every mean-field potential.

In practice it is not even necessary to damp all of the mean-field 
potentials with a preconditioned update. It has sufficed in all calculations we
 performed so far to precondition only $\mathsf{F}_q^{\rho}$ and 
also $\mathsf{F}_q^{\vec{s}}$ when time-reversal is broken. In this way, 
we only solve Eq.~\eqref{eq:precon} for at most four real 
functions on the mesh and the numerical cost of the preconditioning is further 
reduced. In the end, the amount of applications of the Laplacian required to 
perform the preconditioning is negligible compared to the applications of the 
Laplacian to the full set of single-particle wave functions at 
every iteration.

\begin{figure}[t]
\centerline{\includegraphics[width=0.8\linewidth]{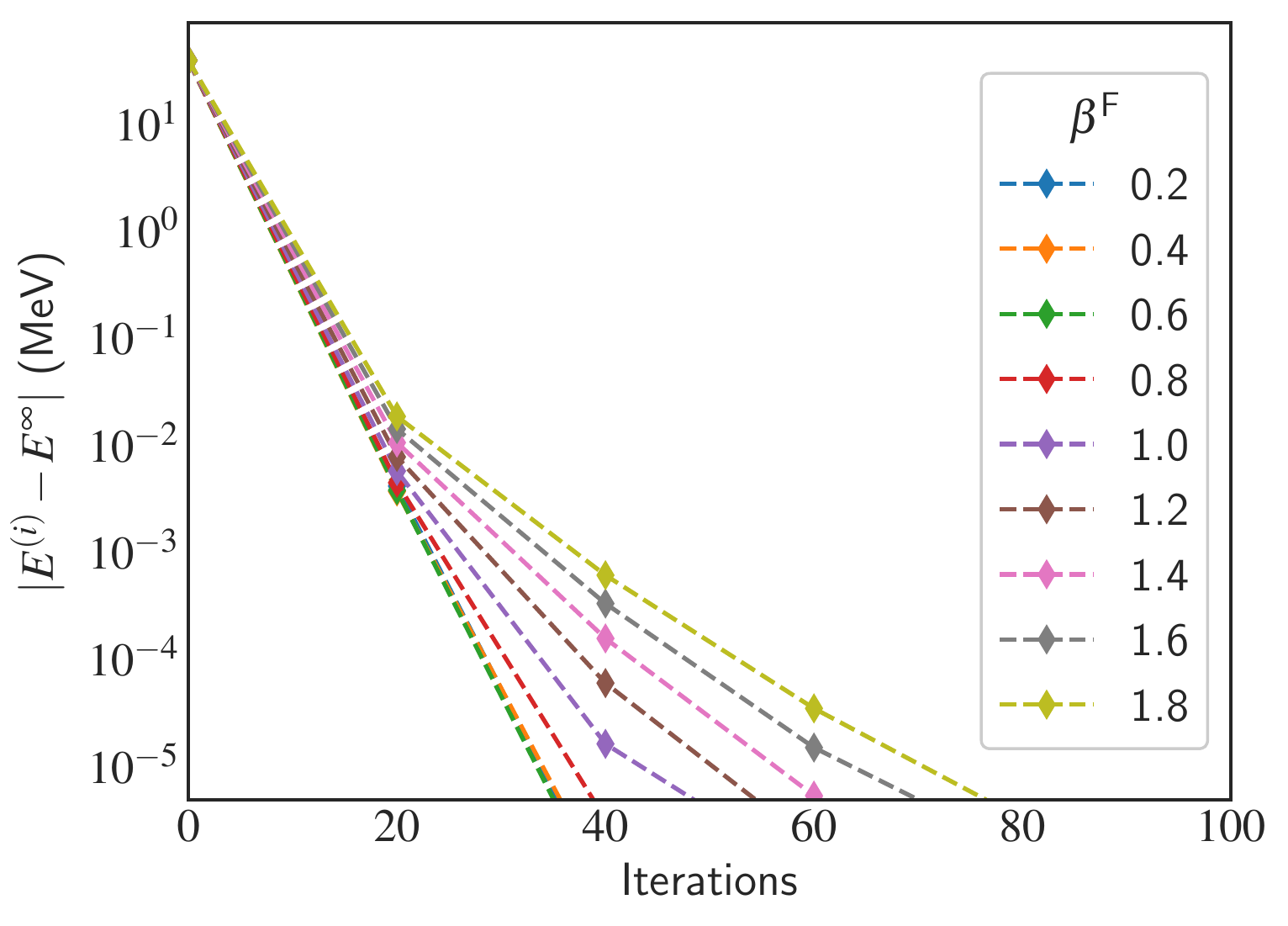}}
\caption{\label{fig:exampprecon}
Same as Fig.~\ref{fig:exampmix}, but from a calculation using the 
potential preconditioner instead of linear mixing,
for different values of the numerical parameter $\beta^{\mathsf{F}}$
entering Eq.~\eqref{eq:precon}.
}
\end{figure}

Finally, let us discuss the choice of $\beta^{\mathsf{F}}$. As with linear
mixing, too small values render the iterative process unstable, whereas
too large values unnecessarily slow down convergence. We have not
succeeded in finding a way to optimize systematically the parameter
$\beta^{\mathsf{F}}$. The convergence speed of the fixed-point iteration is
however significantly less sensitive to a fine-tuning of this parameter than in
the case of linear mixing. Figure~\ref{fig:exampprecon} shows the difference
in the error on energy as a function of the iterations for SLy4 for different 
values of $\beta^{\mathsf{F}}$. For $\beta^{\mathsf{F}} < 0.1$, the iterative 
process is unstable. For larger values of $\beta^{\mathsf{F}}$ the convergence 
rate becomes progressively slower, but at a moderate pace: the overall 
convergence is not very sensitive to the precise value of 
$\beta^{\mathsf{F}}$.

The value $\beta_{\mathsf{F}} = 1.0$ has turned out to be an acceptably fast 
and stable for virtually all nuclei, parametrisations, and choices of 
mesh discretization considered in this paper and all other 
calculations we performed with this scheme so far. The only exception we 
encountered so far is the UNEDF1 parametrisation~\cite{Kortelainen12},
which requires a larger value of $\beta^{\mathsf{F}} \approx 1.8$ to 
reliably converge. This unusual behaviour is again a consequence of
this parametrisation's large negative value of the coupling constant 
$C^{\rho \Delta \rho}_1$ (see Section~\ref{subsect:LM}), such that it 
sets the scale for numerical behaviour instead of $C^{\rho \Delta \rho}_0$, 
which is dominant for most other parametrisations.

%--------------------------------------------------------------------

\subsection{Comments on some alternative approaches}

Self-consistent density functional calculations in atomic
physics have to safeguard the iterative process against
long-wavelength oscillations of the electron density. The
preconditioner proposed by Kerker~\cite{Kerker81} is widely used to
suppress charge sloshing. It can be interpreted as a high-pass 
filter, whereas Eq.~\eqref{eq:preconpot} represents a low-pass filter.

A preconditioner could in principle also be used for the diagonalisation
subproblem. The iterative technique of \textit{damped gradient iteration} 
proposed in Ref.~\cite{Reinhard82,Blum92} can be viewed in this way: it
combines gradient-descent and a pre-conditioning step, similar to 
Eq.~\eqref{eq:preconop}, for the update of the single-particle 
wave functions. This technique is for example used in the Sky3D 
code~\cite{Maruhn14}. The iterative scheme described in 
Ref.~\cite{Robledo11} utilises a preconditioning of quasiparticle 
wave functions in a gradient-based scheme to solve the HFB equations 
in that basis. In a representation like ours, the cost per iteration 
of preconditioning the states is much larger than that of 
preconditioning the potentials as it requires a much larger number
of applications of derivative operators.

The two approaches (linear mixing and preconditioning) presented here use only 
information on the mean-field potentials and densities at iteration $(i)$ to 
construct the potentials and densities at iteration $(i+1)$. Several 
techniques exist that aim to speed-up the convergence of the SCF subproblem 
by accounting for the mean-field potentials and densities from a larger 
number of past iterations. The Broyden method mixes potentials with a memory
 across multiple SCF iterations and has been successfully used to speed-up the
convergence of EDF calculations in the existing solvers HFBTHO and 
HFODD~\cite{Baran08}. Another method, mixing the densities across multiple SCF 
iterations is known either as \emph{Pulay mixing} or the \emph{direct 
inversion in the iterative subspace (DIIS)} method~\cite{Pulay80}. This 
method and several variants of it~\cite{Pulay82,Kudin02} are widely used to 
accelerate self-consistent electronic stucture
calculations. Both methods, Broyden and DIIS, serve the purpose of 
accelerating the SCF convergence and do not by themselves guarantee a stable 
iterative process. These methods can be viewed as providing more advanced 
updates for the right hand sides of Eqs.~\eqref{eq:mixf} and \eqref{eq:mixr}, 
but both still incorporate a mixing parameter, similar to $\alpha^{\mathsf{F}}$ 
and $\alpha^{\mathsf{R}}$, that has to be taken small enough to stabilise the 
iterative process. Although these methods have their qualities, 
we have  limited ourselves to a detailed analysis of the approach that we have 
presented here, as it fulfils all our requirements: it markedly 
improves the convergence speed without significant increase of the 
computational time per iteration or of the memory requirements, while
being rather insensitive to the choice of numerical parameters.

%%%%%%%%%%%%%%%%%%%%%%%%%%%%%%%%%%%%%%%%%%%%%%%%%%%%%%%%%%%%%%%%%%%%%%%%%%%%%%%%

\section{Numerical tests}
\label{sec:numericaltests}

\subsection{Conditions of the tests}

The goal of this section is to compare the convergence 
properties of the schemes that we have developed in the two 
previous sections with what we have used in the past for typical
situations encountered when calculating complex nuclei. The most detailed
tests are performed with the SLy5s1 parametrisation of the Skyrme 
NLO functional~\cite{Jodon16}, which provides a good description of the 
deformation properties of heavy nuclei~\cite{Ryssens18b}. For this 
parametrisation, we will study the convergence for calculations of the ground
state of even-even spherical, axially deformed, triaxially deformed and 
octupole-deformed nuclei, and give also an example from a calculation 
with broken time-reversal symmetry. To examine the differences between the
convergence of NLO and N2LO Skyrme EDFs, some of these calculations have also 
been repeated using the SN2LO1 parametrisation of the latter~\cite{Becker17}
instead.

Four sets of calculations have been performed, using the four possible
combinations of gradient descent (GD) and heavy-ball dynamics (HB) for the
diagonalisation subproblem and of linear mixing (LM) of the densities and a
preconditioned mixing of the potentials (PP) for the SCF subproblem.
These are labelled by combinations of letters, from GD+LM to HB+PP.

The parameters $dt$ and $\alpha^{\mathsf{R}}$ entering GD and LM, 
respectively, have not been optimized, but kept at time-tested
values \cite{Bonche05,Ryssens15a} in order to present an accurate 
image of the methods used in the past. Of course $dt$ is set to a 
different value for each choice of mesh parameters. 
For linear mixing, we have fixed $\alpha^{\mathsf{R}} = 0.25$ in 
Eq.~\eqref{eq:mixr} for all densities and no mixing for the potentials. 
For the HB scheme, we have used the dynamically
estimated values from Eqs.~\eqref{eq:muopth} and \eqref{eq:dtopth}. For 
PP, we have set $\beta^{\mathsf{F}} = 1.0$ in 
Eq.~\eqref{eq:preconop} for $\mathsf{F}^{\rho}_q$ and $\mathsf{F}^{\vec{s}}_q$
as a default value, while not preconditioning any of the other potentials.

All calculations have been initialized with spherical Nilsson orbitals, unless
explicitly stated otherwise. For deformed systems, in order to break the
spherical self-consistent symmetry,  a constraint on the quadrupole moment 
has been introduced during the first ten iterations in order to
break the self-consistent spherical symmetry. In the case of HF+BCS and HFB
calculations, we have added a pairing interaction as described in Appendix
\ref{app:twobasis}.

%-------------------------------------------------------------------------------
\subsection{Convergence indicators}

Several quantities can be used to determine the convergence and its rate 
for a self-consistent calculation. We will mainly use the ones 
discussed earlier in Ref.~\cite{Ryssens15a}. For convenience, we recall 
here their definition.

The weighted dispersion of the single-particle energies as a convergence 
indicator for the linear subproblem has already been introduced in 
Eq.~\eqref{eq:dh2}. As we iterate the diagonalisation subproblem only once 
for every SCF iteration, we will drop the second iteration index in what 
follows, denoting it $(dh^2)^{(i)}$.

The difference between the total energy calculated from the 
EDF~\eqref{eq:Etot} and from the sum of the occupation-weighted 
diagonal matrix elements of the single-particle Hamiltonian $\hat{h}^{(i)}$ 
corrected for rearrangement and pairing energies (see Ref.~\cite{Ryssens15a} 
for the precise definition) also provides a test of convergence of 
self-consistent calculations. At convergence, when the single-particle 
Hamiltonian is diagonal, these quantities are equal up to numerical 
inaccuracies. The difference between the two ways of determining the total 
energy of the nucleus
\begin{equation}
\delta E^{(i)}_{\rm spwf} 
= \big| E_{\rm func}^{(i)} - E^{(i)}_{\rm spwf} \big| \, ,
\end{equation}
can then be used as an indicator of convergence.
Both $(dh^2)^{(i)}$ and $\delta E_{\rm spwf}^{(i)}$ cannot become arbitrarily 
small during the iterative process. The mesh discretization introduces a 
lower bound for both quantities, that in general decreases when using 
smaller values of $dx$~\cite{Ryssens15a}.

Another indicator that we use in our analysis of convergence is the evolution 
of the relative change of the total binding energy $\delta E^{(i)}$ 
from one iteration to the next
\begin{equation}
\label{eq:diffE}
\delta E^{(i)}
= \left| \frac{E^{(i)} - E^{(i-1)}}{E^{(i)}} \right| \, .
\end{equation}
As the iterations progress, the change in energy should decrease, and keep
decreasing until the limits set by machine precision are encountered. While 
a large value of $\delta E^{(i)}$ signals non-convergence, we emphasize that 
a small value does not always signal near-convergence. 

\begin{figure*}[t!]
\centerline{\includegraphics[width=16cm]{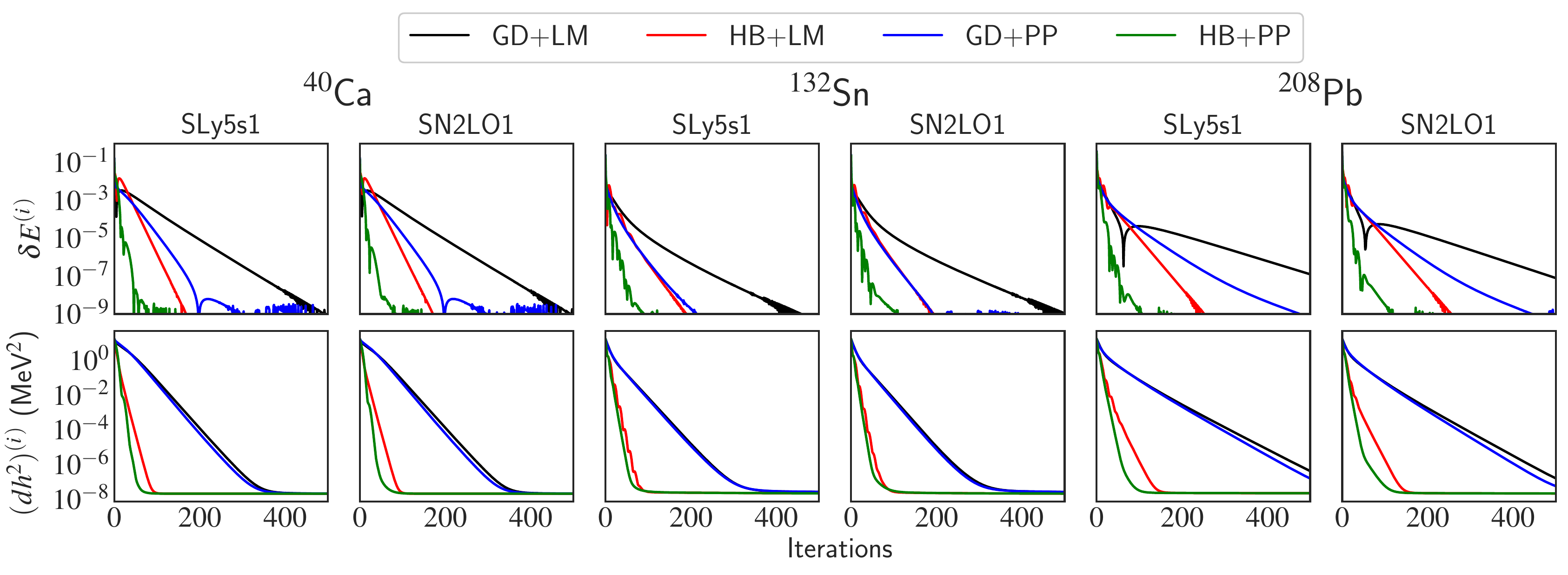}}
%\centerline{\includegraphics[width=5.6cm]{{Spherical.Ca}.pdf}
%            \includegraphics[width=5.6cm]{{Spherical.Sn}.pdf}
%            \includegraphics[width=5.6cm]{{Spherical.Pb}.pdf}}
\caption{\label{fig:spherAll}
Comparison between four combinations of algorithms for the spherical
nuclei $^{40}$Ca, $^{132}$Sn and $^{208}$Pb for the SLy5s1 (left) and 
the SN2LO1 functionals (right), calculated without pairing correlations 
on mesh~(b).
}
\end{figure*}

%-------------------------------------------------------------------------------

\subsection{Multipole deformations and multipole constraints}
\label{sec:constraints}

As argued in Section~\ref{sec:SCP}, we limit the use of constraints in this 
study to the minimum possible. All algorithms discussed here can easily 
accommodate their presence. As constraints introduce additional 
auxiliary conditions to the self-consistent problem that have to be 
satisfied through an iterative process that has its own convergence rate,
they do however unnecessarily complicate the performance comparison 
of the algorithms discussed here. The discussion of the efficient 
algorithmic treatment of constraints in the context of the methods 
used here is deferred to a forthcoming 
paper~\cite{Ryssens18c}. Some of the calculations described in this Section, 
however, necessitate constraints, such that a few comments on their present 
implementation are in order.

Generally speaking, constraints on the expectation va\-lue of an
operator $\hat{O}$ can be imposed by adding a penalty function to the energy. 
In the case of a single constraint, the single-particle Hamiltonian 
at the iteration $(i)$ becomes
\begin{equation}
\label{eq:constrh}
\hat{h}^{(i)}_R = \hat{h}^{(i)} - \lambda^{(i)} \, \hat{O} \, ,
\end{equation}
where $\lambda^{(i)}$ is a Lagrange multiplier. There are two primary 
possibilities to implement constraints: either keeping the Lagrange parameter
$\lambda$ fixed, in which case the code converges to 
a configuration with $dE/d\langle \hat{O} \rangle = \lambda$, or adjusting 
the $\lambda^{(i)}$ during the iterations in such a way that the $N$-body 
expectation value $\langle \hat{O} \rangle$ takes a preset value
at convergence.

We will use both. For constraints on multipole moments of the mass 
density, the Lagrange parameters are adjusted with a method similar to the 
one described in Ref.~\cite{Ryssens15a} until a targeted value for the 
expectation value of $\hat{Q}_{\ell m} \equiv r^{\ell} \, Y_{\ell m}(\vec{r})$
is reached. To plot results, however, we will use the dimensionless 
deformations~\cite{Ryssens15a}
\begin{equation}
\label{eq:betadef}
\beta_{\ell m}
= \frac{4 \pi}{3 R^{\ell} A} \, \langle \hat{Q}_{\ell m}\rangle 
\end{equation}
instead, where $R = 1.2 A^{1/3}$ fm. For systems with triaxial quadrupole 
deformation, we use the quantities~\cite{Ryssens15a}
\begin{equation}
\label{eq:BM:gamma}
\beta
\equiv \sqrt{ \beta_{20}^2 + 2 \beta_{22}^2 } \, ,  \qquad
\gamma 
\equiv \text{atan2} \left( \sqrt{2} \,  \beta_{22} , \beta_{20} \right) \, ,
\end{equation}
to draw maps in the $\beta$-$\gamma$ plane.
The multipole constraints are mainly used to provide background energy 
surfaces that illustrate the path of the evolution of unconstrained 
calculations towards the minimum. An exception are some calculations with 
broken reflection symmetry, where a constraint on the position of the 
centre-of-mass of the nucleus has to be added in order to keep the system
fixed at the centre of the numerical box.

\begin{table}[t!]
\center
\begin{tabular}{cccc}
\hline \noalign{\smallskip}
 & $N_x = N_y = N_z$ & $dx$ (fm) & $dt$ ($10^{-22} \text{s})$ \\
\noalign{\smallskip} \hline \noalign{\smallskip}
(a) & 32 & 1.0 & 0.015 \\
(b) & 40 & 0.8 & 0.012 \\
(c) & 50 & 0.6 & 0.006 \\
(d) & 64 & 0.5 & 0.004 \\
\noalign{\smallskip} \hline
\end{tabular}
\caption{Number of mesh points $N_\mu$ and mesh spacing $dx$ for 
the cubic meshes used to calculate the spherical nuclei. The size of the 
time-step $dt$ for the calculations employing gradient descent is also given.}
\label{tab:mesh}
\end{table}

%-----------------------------------------------------------------------------

\subsection{Spherical nuclei}
\label{sec:spherical}

First tests have been performed for the spherical nuclei $^{40}$Ca, $^{132}$Sn 
and $^{208}$Pb using the NLO and N2LO functionals. The mesh discretization 
$dx$ has been varied according to Table~\ref{tab:mesh}. As shown 
in~\cite{Ryssens15}, mesh~(b) offers an satisfactory compromise between 
numerical accuracy and CPU time for nuclei up to $^{208}$Pb.

Figure~\ref{fig:spherAll} shows the evolution of the relative
change in total energy $\delta E^{(i)}$ and the weighted 
dispersion $(dh^2)^{(i)}$ as a function of the number of iterations for the 
three nuclei and the two parametrisations. Results obtained 
with a given combination of algorithms are very similar for all 
parametrisations and nuclei. The combination GD+LM is always the 
least favourable while HB+PP is the fastest. 

It is important to realize that different indicators are sensitive to
different aspects of the convergence. This can be most clearly seen when 
looking at results from the HB+LM and GD+PP schemes that give intermediate 
results. For HB+LM, the weighted dispersion $(dh^2)^{(i)}$ always falls off 
much quicker than $\delta E^{(i)}$. This indicates that in this case the
late stages of the convergence are dominated by small changes of the 
densities and potentials in the SCF iteration, while the single-particle 
Hamiltonian $\hat{h}^{(i)}$ is already near-diagonal and rediagonalised with 
a single step of the HB scheme after each SCF update. When combining HB with 
PP, then both aspects converge at a similar rapid pace. For GD+PP one finds 
the opposite of HB+LM: $\delta E^{(i)}$ falls of quicker than $(dh^2)^{(i)}$.
In this case, the late stages of the convergence are dominated by the 
diagonalisation subproblem, while thanks to the preconditioner the potentials
and densities are already near the self-consistent solution for the given
set of single-particle states $\{ | \psi^{(i)}_{\ell} \rangle\}$ and can 
be made to follow its slow evolution with a single preconditioned update. 
In that respect, HB+PP and GD+LM are similar as in both schemes the two 
subproblems converge at a similar rate.

\begin{figure}[t!]
\centerline{\includegraphics[width=0.8\linewidth]{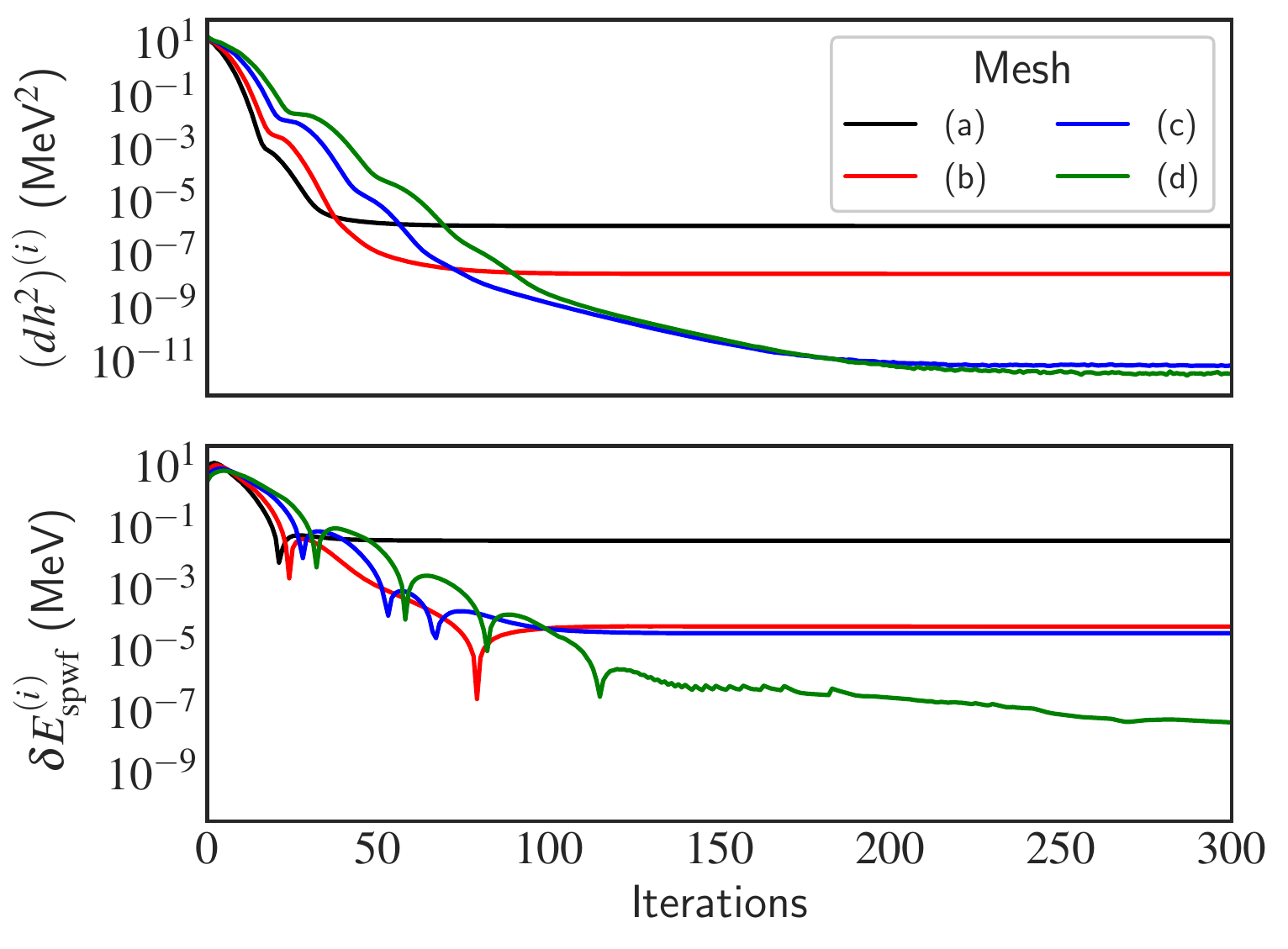}}
\caption{\label{fig:Spvsdisp}
Weighted dispersion (top) and $\delta E_{\rm spwf}^{(i)}$ (bottom) 
as a function of the iterations for the various mesh choices of 
Table~\ref{tab:mesh} and a HF calculation of $^{40}$Ca 
using the HB+PP scheme and the SLy5s1 parametrisation.
}
\end{figure}

\begin{figure}[b!]
\centerline{\includegraphics[width=8.0cm]{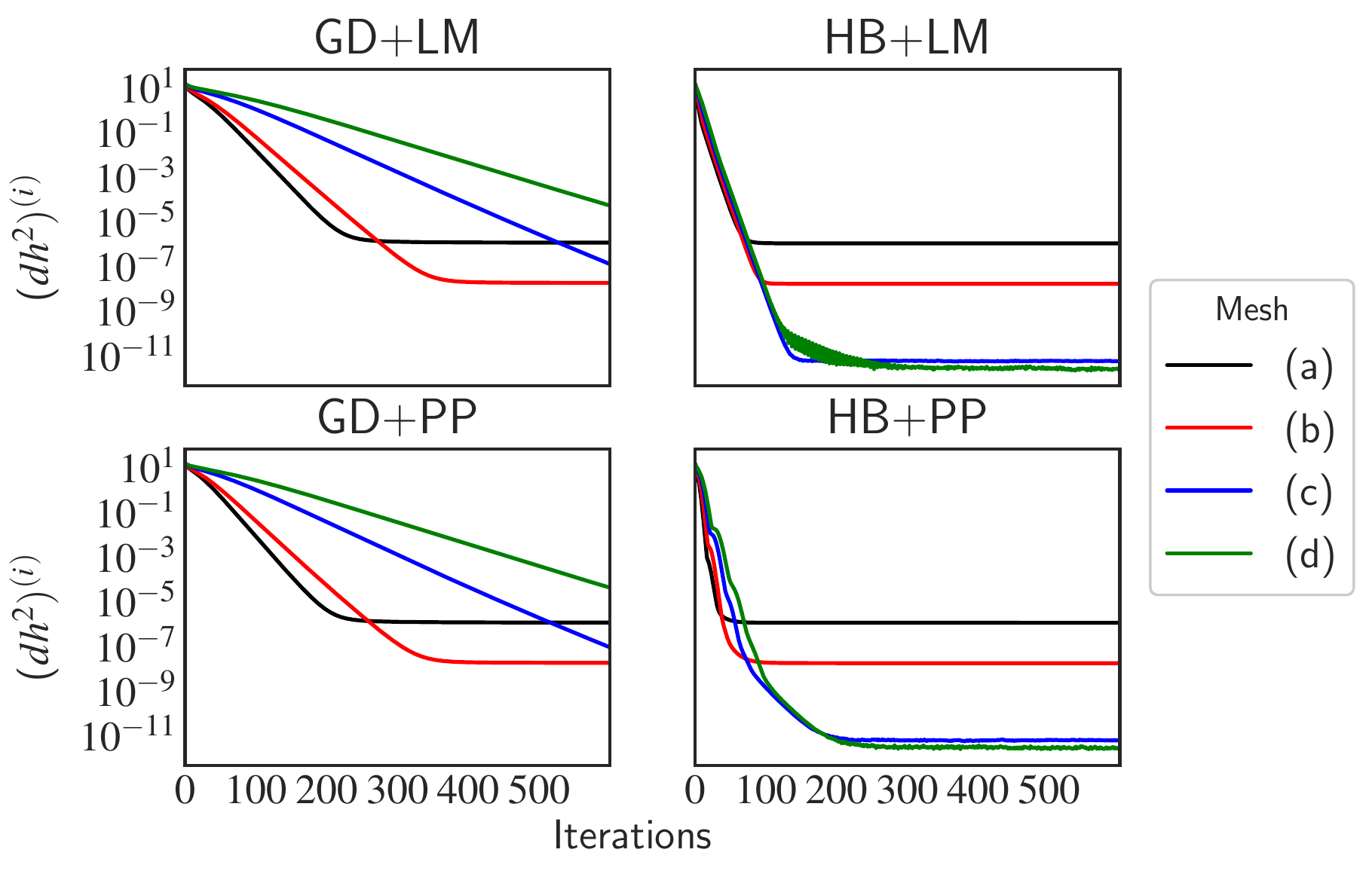}}
\caption{\label{fig:Spvsdisp:all}
Weighted dispersion (top) as a function of the iterations for the various 
mesh choices of Table~\ref{tab:mesh} and a HF calculation of $^{40}$Ca
with the schemes as indicated and the SLy5s1 parametrisation.
}
\end{figure}

Figure~\ref{fig:spherAll} illustrates that one cannot rely on a single quantity
like $\delta E^{(i)}$ as an indicator of convergence, as different
indicators probe different aspects of the convergence of the self-consistent
problem. As already discussed in Section~\ref{sec:linearsub}, the convergence 
rate depends on the representation, as do the smallest values reachable for 
the indicators. This is illustrated by Fig.~\ref{fig:Spvsdisp}, which 
shows the evolution of the weighted dispersion $(dh^2)^{(i)}$ and of 
$\delta E_{\rm spwf}^{(i)}$ for a HB+PP calculation of the spherical
ground state of $^{40}$Ca and the four different mesh discretizations 
$dx$ as indicated in Table~\ref{tab:mesh}. In general, both indicators
reach a plateau that originates from calculating the energy from 
the single-particle states through different intermediate objects that are
not equally resolved on a given mesh. The smaller the value of $dx$, the lower
this plateau, although there is also a lower limit to these plateaus that is
set by the level of truncation and round-off errors of floating-point 
arithmetic. For $(dh^2)^{(i)}$, this limit is reached for mesh~(c).
Both quantities probe the quality of the diagonalisation of $\hat{h}^{(i)}$,
and, hence, behave in roughly the same way, with the plateau reached at 
about the same iteration number. Since these quantities are 
interchangeable for the purpose of our analysis, we focus on the weighted 
dispersion.

\begin{figure}[t!]
\centerline{\includegraphics[width=0.8\linewidth]{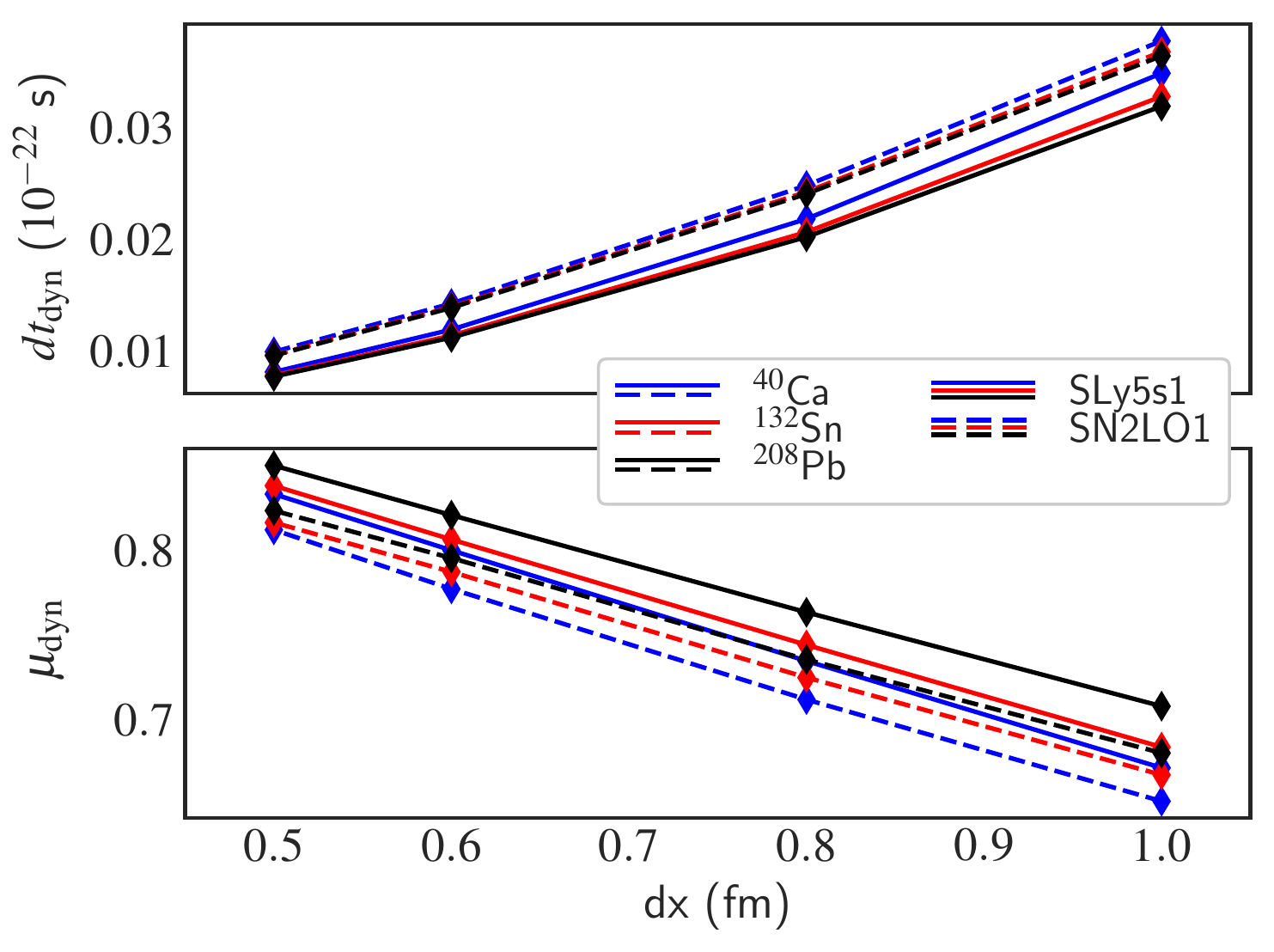}}
\caption{Iterative parameters $dt_{\rm dyn}$ and $\mu_{\rm dyn}$ at the end of
the iterative process, as estimated by Eqs.~\eqref{eq:muopth} and
\eqref{eq:dtopth}, for $^{40}$Ca (blue), $^{132}$Sn (red) and $^{208}$Pb 
(black) using the SLy5s1 (full lines) and SN2LO1 (dashed lines)
parametrisations.}
\label{fig:params}
\end{figure}

As already discussed for Fig.~\ref{fig:spheriter}, the number of iterations
needed to reach the same level of convergence in the HB scheme is almost 
independent of $dx$, while in the GD method it increases dramatically with
decreasing $dx$. Figure~\ref{fig:Spvsdisp:all} illustrates this for the
convergence of the weighted dispersion $(dh^2)^{(i)}$ during a calculation
of $^{40}$Ca with the four different schemes. The schemes employing HB are
clearly superior for meshes with small $dx$, in particular when recalling 
that each individual iteration becomes much more costly when reducing $dx$. 
Interestingly, the convergence of the HB+LM scheme shows almost no $dx$
dependence apart from reaching different plateau levels. By contrast, 
the speed-up from the pre-conditioner when going from HB+LM to HB+PP
re-introduces a very small $dx$ dependence, albeit on a much weaker 
level than what is observed for LM schemes.

A large part of the levelling out of the $dx$ dependence of convergence
when using the HB scheme is made possible by the dynamical adaptation of
its numerical parameters, as already discussed in Section~\ref{sec:dynparam}.
Figure~\ref{fig:params} shows the values of the parameters 
$\mu_{\rm dyn}$ and $dt_{\rm dyn}$ of the HB scheme calculated from 
Eqs.~\eqref{eq:muopth} and \eqref{eq:dtopth} at the last iteration 
of the HB calculations displayed in Fig.~\ref{fig:Spvsdisp:all}.
For a given mesh discretization, the value of $dt_{\rm dyn}$ is 
almost twice as large as the largest value that can safely be used 
in gradient descent. While the time-step $dt_{\text{dyn}}$ shrinks with 
decreasing $dx$, its negative effect on the convergence rate is
counterbalanced by an increase of the parameter $\mu_{\rm dyn}$, reflecting 
an increasing mismatch between the smallest and largest relevant energy 
scale.

%-------------------------------------------------------------------------------

\subsection{Deformed nuclei}
\label{sec:deformed}

Let us now compare the convergence properties of the four methods that we
have defined for three deformed nuclei with qualitatively different energy
surfaces when initialized at a deformation far from the one of the ground 
state. In all cases, the mesh-choice~(b) from Table~\ref{tab:mesh} has 
been used together with the SLy5s1 parametrization. All calculations 
consider pairing correlations in order to allow for a smooth evolution of 
the deformation.

\begin{figure}[t!]
\centerline{\includegraphics[width=\linewidth]{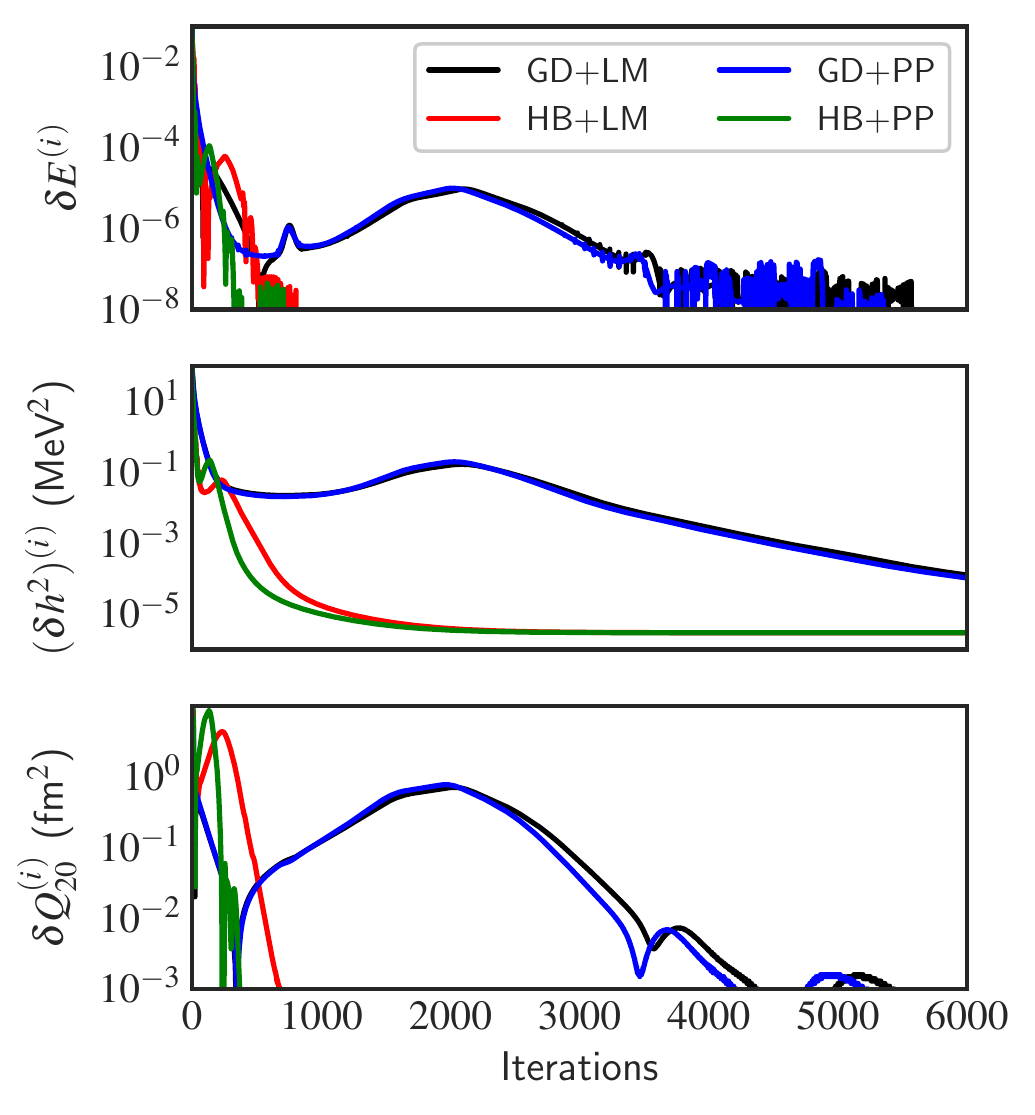}}
\caption{Evolution of the change in energy (top), weighted dispersion (middle)
         and change in axial quadrupole moment (bottom) as a function of
         iterations for $^{240}$Pu using the SLy5s1 parametrisation. }
\label{fig:Pu240}
\end{figure}

As a first example we address the axially-deformed ground state of 
$^{240}$Pu. For this parametrisation, it takes a quadrupole deformation 
of $\beta_{20} \simeq 0.28$ that is accompanied by an energy gain of about 
15~MeV compared to the spherical configuration~\cite{Ryssens18b}. 
Figure~\ref{fig:Pu240} shows the evolution of the relative change in energy 
$\delta E^{(i)}$, the weighted dispersion $(dh^2)^{(i)}$, and the change of 
quadrupole moment 
$\delta Q_{20}^{(i)} \equiv | Q_{20}^{(i)} - Q_{20}^{(i-1)}|$.
during the iteration.

The algorithms including heavy-ball dynamics are again by far superior to 
those based on gradient descent. They provide a reduction in the number of 
iterations of about one order of magnitude. The main source 
for this gain is the much quicker convergence of the deformation, in 
particular when combining heavy-ball dynamics with the potential 
preconditioner, which reduces the number of iterations needed to bring the 
deformation close to the converged value by again a factor of two. This then 
also accelerates the final convergence of the potentials in the SCF problem.

\begin{figure}[b!]
\centerline{\includegraphics[width=\linewidth]{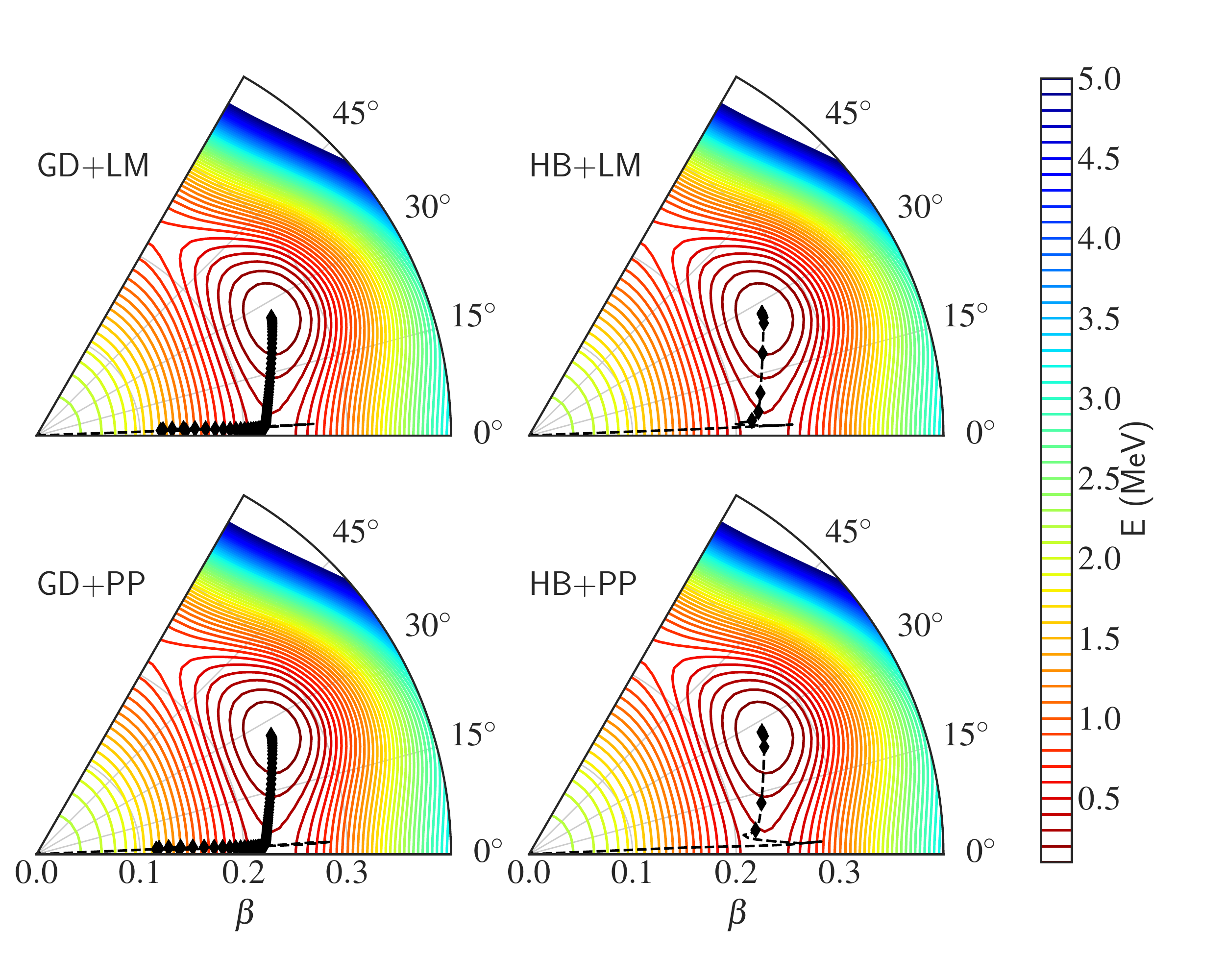}}
\caption{\label{fig:Gepath}
Quadrupole deformation $\beta$ and triaxiality angle $\gamma$
as a function of the iterations of a non-constrained calculation of 
$^{64}$Ge with the four algorithms as indicated, drawn against 
a backdrop of the converged deformation energy surface. 
Contour lines are at every 100~keV and diamonds at every 100th iteration.
Calculations were initialized with the (converged) spherical configuration 
pushed with a constraint to slightly non-axial quadrupole 
deformation for 10 iterations (see text).
}
\end{figure}

The reduction in number of iterations brought by the HB scheme 
compared to the GD method is even more remarkable when searching
for the minimum of a deformation energy surface that is soft in one direction.
As an example we choose $^{64}$Ge. To illustrate the advancement of the 
calculation of the ground state, we have first determined the
(converged) deformation energy surface of this nucleus as a function of $\beta$
 and $\gamma$. Its ground state is triaxial with $\beta \simeq 0.25$ and 
$\gamma = 27^{\circ}$. We have then performed unconstrained calculations 
of the ground state with the four methods that we study. Each calculation 
is initialized with a configuration that has been pushed away from the 
spherical point by applying a constraint on a slightly triaxial 
quadrupole moment for 10 iterations. The paths followed during the 
iterations are indicated by dashed lines in Fig.~\ref{fig:Gepath}, with 
additional markers at every 100 iterations. During the initial 
constrained phase of the calculation, the state is pushed to 
$\beta \simeq 0.28$ at $\gamma \simeq 2.4^{\circ}$,
i.e.\ the end of the straight part of the dashed line. When 
the constraint is released, all calculations go back to smaller values
of $\beta$ at almost constant $\gamma$. What happens after depends on the 
scheme. The methods containing GD go back to much smaller $\beta \simeq 0.12$
within about 200 iterations, before starting to go downhill to larger $\beta$ 
again for a second time. By contrast, the methods containing the HB scheme 
only go back to the bottom of the valley and then turn directly to larger 
$\gamma$ angles.

The difference between HB and GD is that during the SCF iterations
the former arrives so quickly at an approximate diagonalisation of
$\hat{h}^{(i)}$ that the SCF iteration actually moves on an energy 
surface that is quite close to the converged energy surface drawn as the 
background in Fig.~\ref{fig:Gepath} and therefore can follow the valleys 
seen in the plot. By contrast, the GD scheme remains for the first few
hundred iterations so far from the diagonalisation of $\hat{h}^{(i)}$ 
that during this phase the SCF evolution is on a very different energy 
landscape, in which the valley visible in the converged energy surface of 
Fig.~\ref{fig:Gepath} emerges only slowly during the iterative process. 
This is corroborated by the number of iterations needed in each
scheme to reach the turn-off point where the evolution turns into the
direction of $\gamma$. The GD+LM scheme needs 3990~iterations to reach that
point, while for GD+PP it are 3702. This has to be contrasted with 
22~iterations needed for HB+LM and just 17 for HB+PP. But also for the 
latter two schemes finding the minimum in the soft $\gamma$ direction 
requires many more iterations than finding the valley in the steep $\beta$ 
direction. The HB+PP scheme arrives at the minimum after about 700 
iterations, while it takes about 900 for HB+LM, and around 10000 for GD+LM 
and GD+PP. For this kind of calculation, HB dynamics is clearly 
superior to GD, with PP performing slightly better than LM.

\begin{figure}[t!]
\centerline{\includegraphics[width=\linewidth]{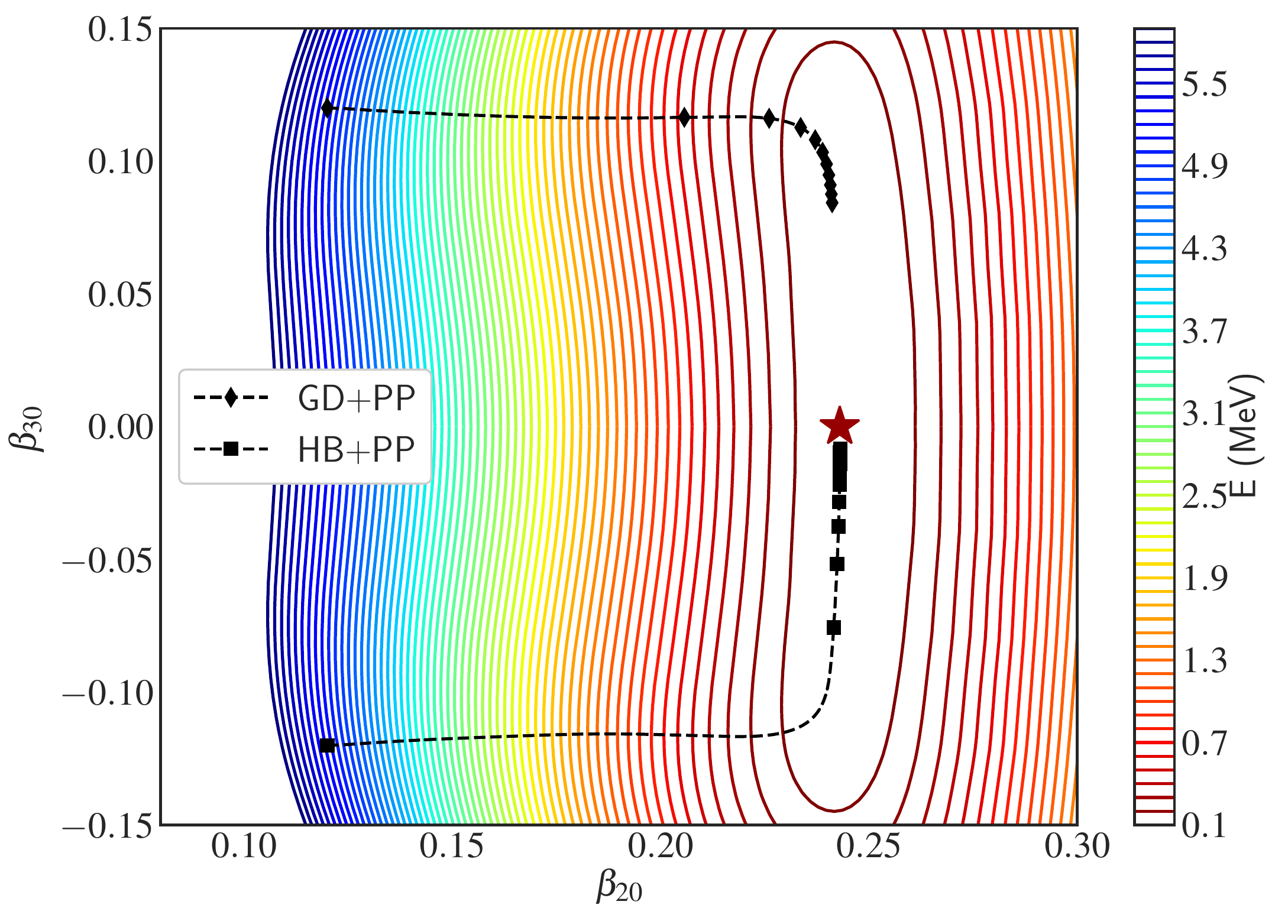}}
\caption{\label{fig:Thpath}
Path in the $(\beta_{20}, \beta_{30})$ plane of a calculation
for $^{232}$Th initialized at $\beta_{20} = |\beta_{30}| = 0.12$ using
the SLy5s1 parametrisation, and employing gradient descent
(diamonds) or heavy-ball dynamics (squares). The HB path is drawn at
negative $\beta_{30}$ for ease of presentation, but is physically
equivalent. Markers are placed at initialization and at every
500th iteration, with the final point placed after 5000 iterations.
Both approaches employed potential preconditioning.
The contour lines are 100 keV apart, while the minimum of the energy surface
at $\beta_{30} = 0.0$ is indicated by a red star.}
\end{figure}

As a final example of the calculation of deformed ground states, we have 
selected $^{232}$Th, a nucleus with an axial reflection-symmetric 
quadrupole deformed ground state that is very soft with respect to octupole 
degrees of freedom~\cite{Ryssens18b}.
At the quadrupole deformation of the minimum, the energy varies by less 
than 100~keV over a wide range of $\beta_{30}$ values up to 0.14. As in 
the previous case, we have first calculated the $(\beta_{20},\beta_{30})$ 
energy surface in order to have a background for the analysis of the
evolution of the non-constrained calculation.
As in the case of $^{64}$Ge, the difference between using the LM or PP 
scheme is minimal compared to the difference between using GD and HB, such 
that we will limit the presentation to the HB+PP and GD+PP 
combinations. Profiting from the symmetry of the deformation energy when
reversing the sign of $\beta_{30}$, Fig.~\ref{fig:Thpath} displays the 
evolution of the GD+PP calculation at positive $\beta_{30}$, whereas results
for HB+PP are drawn at negative $\beta_{30}$. The calculations were both 
run for 5000 iterations, initialized from the same converged state
constrained to $\beta_{20} = \beta_{30} = 0.12$. They follow the
paths indicated by the dashed line, with additional markers set at 
every 500 iterations. As the calculations are this time started from a 
converged state on the slope of the energy surface, they explore almost 
the same energy surface during the calculation, optimizing again first
the energy in steep directions of the surface, which is $\beta_{20}$,
and then slowly follow the tiny gradient in the soft direction.
This admittedly pathological example is a challenge not only because of
the energy surface being flat, but also because convergence requires the
state to adopt a higher symmetry than the one the calculation is initialised
with. After 5000 iterations, the heavy-ball scheme almost arrives at the
minimum, while gradient descent is still very far from it. The
latter, however, cannot be guessed from the convergence indicators, which
at the 5000th iteration take the values 
$\delta E^{(i)} = 3 \times 10^{-9}$~MeV,
$(dh^2)^{(i)} = 4 \times 10^{-7}$~MeV$^2$, and 
$\delta Q_{30}^{(i)} \equiv | Q_{30}^{(i)} - Q_{30}^{(i-1)}|$
= $6 \times 10^{-6}$ fm$^{3}$. 

When organising self-consistent calcula\-tions, some of the issues
of slow convergence of deformation degrees of freedom discussed here can to 
some extent be compensated for by using prior information about the structure 
of the energy surface by initialising a calculation from a reasonably 
converged calculation that has been constrained to a shape similar to the
one of the targeted configuration. Such procedure, however, requires human
time to prepare calculations according to the targeted nucleus or even
parametrisation. An algorithm like HB+PP that homes in quickly on the 
ground state after just having been pushed into roughly the right direction,
as in the case of the example of $^{64}$Ge, brings not just a large reduction
of computational time, but also in human time needed to prepare, run, and 
survey the calculations.

%-----------------------------------------------------------------------------
%
\subsection{Cranked calculations}

At the mean-field level, rotational bands can be described through the 
introduction of a cranking constraint~\cite{RingSchuck,Bonche87}. In many 
cases, it is sufficient to constrain one component of angular momentum 
$\hat{J}_{z}$, such that the mean field entering the HFB 
Hamiltonian~\eqref{eq:HFBhamil} is replaced by
\begin{equation}
\label{eq:Routhian}
\hat{h}^{(i)}_{C}
= \hat{h}^{(i)} - \omega \hat{J}_{z}  \, .
\end{equation}
The rotational frequency $\omega$ plays the role of a Lagrange 
multiplier. The calculations reported here are performed at fixed values of 
$\omega$, such that the constraint acts as an external potential that does not
change during the iterations.

For the calculations discussed so far, time-reversal invariance can be
imposed on the product state, which can be used to simplify the calculations.
The cranking constraint in Eq.~\eqref{eq:Routhian} breaks time-reversal
invariance, such that one has to consider also time-odd densities, the
time-odd terms in the EDF, Eqs.~\eqref{eq:SkTodd:0} and~\eqref{eq:SkTodd:2},
and the corresponding terms in the single-particle 
Hamiltonian, Eq.~\eqref{eq:singleh:NLO}. We note in passing that the SLy5s1
parametrisation that we use here has been constructed such that it does 
not have an unphysical finite-size instability in the time-odd terms, a 
problem encountered for many other parametrisations of the Skyrme 
EDF~\cite{Hellemans12}.

\begin{figure}[t!]
\centerline{\includegraphics[width=\linewidth]{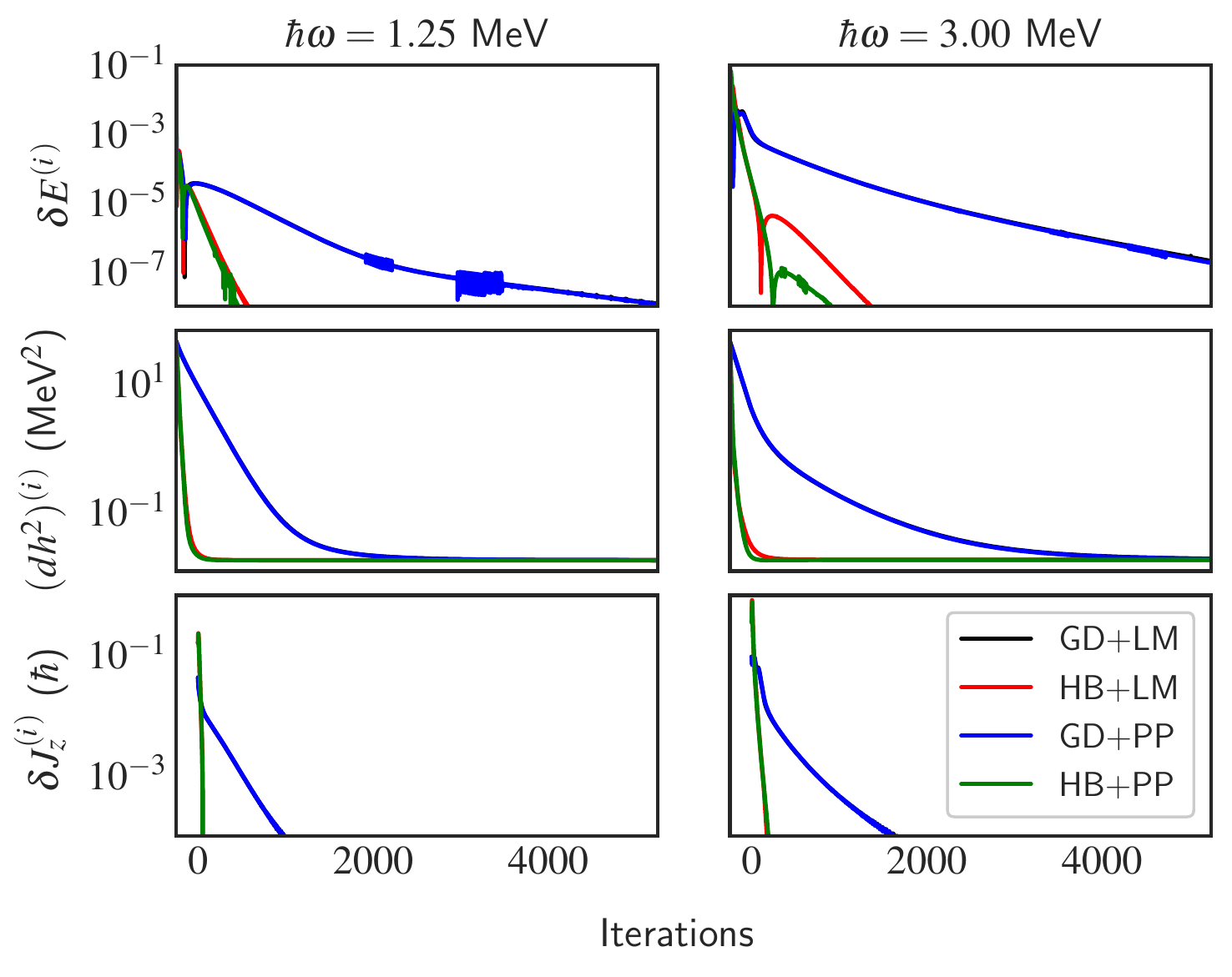}}
\caption{\label{fig:mg24}
Evolution of the relative change in total energy (top), weighted dispersion 
of the single-particle Hamiltonian (middle), and change in expectation of 
angular momentum $\langle J_z \rangle$ bottom) for cranked calculations of 
$^{24}$Mg at $\hbar\omega = 1.25$ MeV (left) and $\hbar \omega = 3.00$ MeV 
(right) using the SLy5s1 parametrisation. 
}
\end{figure}

The light deformed nucleus $^{24}$Mg is a popular testing ground for
the modelling of rotational bands. Figure~\ref{fig:mg24} shows the evolution 
of $\delta E^{(i)}$, $(dh^2)^{(i)}$, and the change in angular momentum 
\begin{equation}
\delta J_z^{(i)} = \big| \hat{J}_z^{(i)} - \hat{J}_z^{(i-1)} \big| \, ,
\end{equation}
during the iterations for two different values of the cranking frequency 
$\omega$. The non-cranked HFB ground-state is prolate and 
time-reversal invariant. It has been used to initialize all calculations. 
To avoid the breakdown of pairing correlations, the HFB method is 
supplemented by the Lipkin-Nogami procedure~\cite{Rigollet99}. 

As for non-cranked calculations, combinations of algorithms using heavy-ball 
dynamics significantly outperform combinations with gradient descent, 
with a reduction of the necessary number of iterations necessary to 
reach similar quality by almost one order of magnitude. The difference
between the HB and GD schemes is especially visible for the angular 
momentum: it converges in a few iterations with heavy-ball dynamics, 
but requires three orders of magnitude more iterations with gradient 
descent. Similarly, $(dh^2)^{(i)}$ falls to the plateau value much quicker 
when using HB dynamics, indicating that $\hat{h}^{(i)}_{C}$ is near-diagonal 
throughout most of the SCF iterations.
Although cranked HFB states are necessarily non-axial,
in this example the deformation changes only minimally when varying
$\omega$ in the range covered. For this reason, potential 
preconditioning does not have a large effect, even though for 
$\hbar\omega=3$~MeV a small speed-up is observed in the change in energy.

\subsection{Odd-mass nuclei}

Within the framework of the two-basis method to solve the HFB equations,
all algorithms described above work in exactly the same way for 
one-quasiparticle states representing odd-mass nuclei as they do for even 
nuclei. The only difference specific to odd nuclei, or any 
multi-quasi\-par\-ticle
state for that matter, concerns the construction of the $U$ and $V$ matrices
when solving the HFB equation~\eqref{eq:HFBhamil} in the subspace provided
by the $| \psi^{(i)}_{k} \rangle$. The tagging of a specific 
configuration, however, introduces auxiliary conditions that have to 
be satisfied by the solution of the self-consistent problem. As the 
analysis of algorithms that follow the targeted configuration through 
large changes of the single-particle spectrum is out of the scope of this 
paper, we address only the simple case of a well-deformed odd-mass nucleus 
that is initialised from a converged false vacuum with same proton and 
neutron number, and therefore similar deformation. The blocking of a 
quasiparticle breaks time-reversal invariance with the same consequences 
as in the case of cranked HFB, and is carried out as described in 
Ref.~\cite{Gall94}.

\begin{figure}[t!]
\centerline{\includegraphics[width=.95\linewidth]{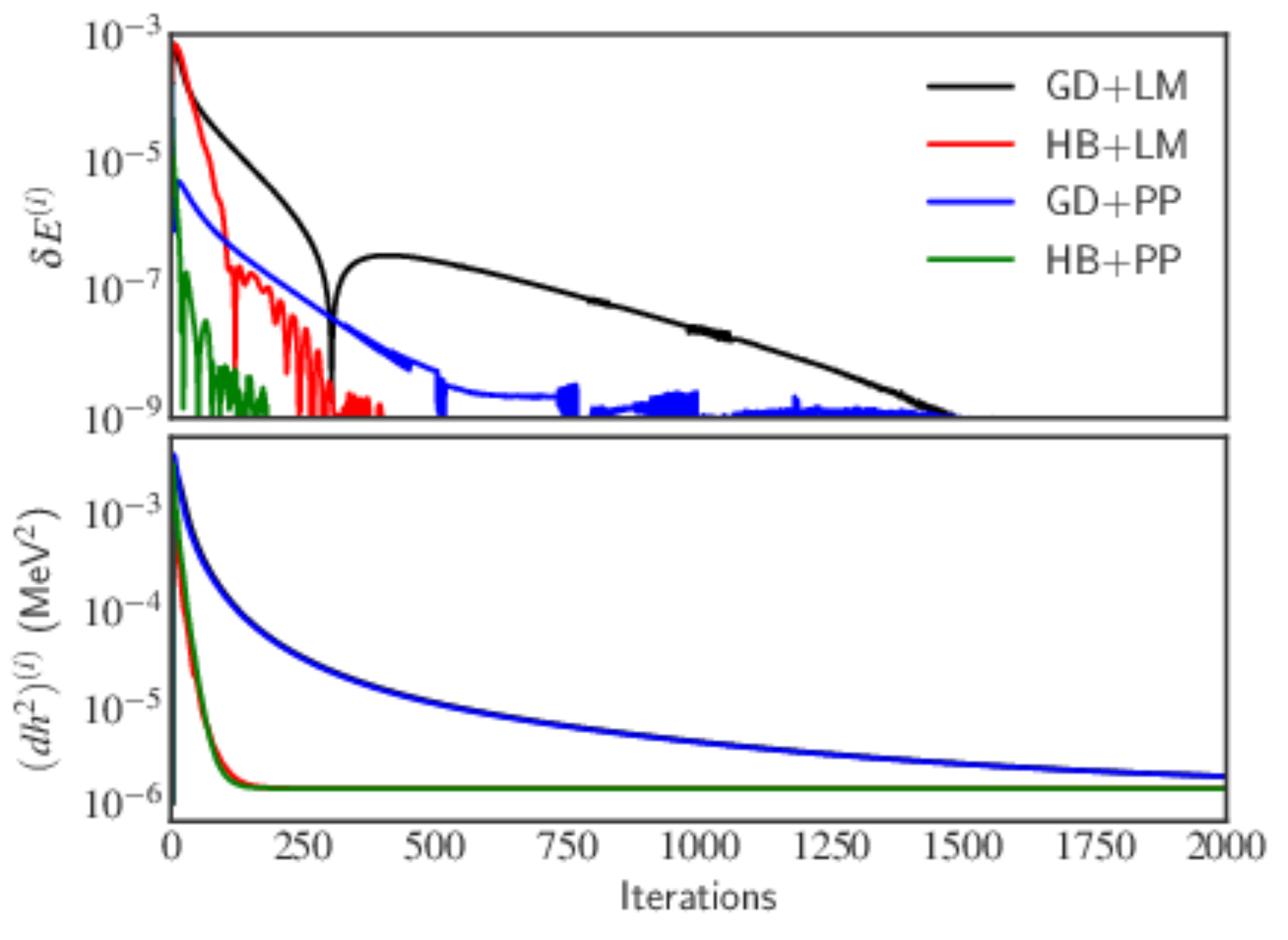}}
\caption{\label{fig:fm251}
Evolution of the relative change in total energy (top) and weighted 
dispersion of the single-particle Hamiltonian (bottom) for a blocked HFB 
calculation of the $K \approx 9/2-$ ground state of $^{251}$Fm. In the lower
panel, the curves for GD+LM and GD+PP are on top of each other.
}
\end{figure}

As example, Fig.~\ref{fig:fm251} shows the evolution of $\delta E^{(i)}$
and $(dh^2)^{(i)}$ during the iterations of the near-axial $K \approx 9/2^-$ 
ground state of $^{251}$Fm, calculated in HFB with the NLO SLy5s1 
parametrisation. Once again,
combinations of algorithms using heavy-ball dynamics outperform 
combinations with gradient descent, leading to a large reduction of the
number of iterations. Very similar to the other cases discussed already,
the heavy-ball scheme quickly near-dia\-go\-na\-lises $\hat{h}^{(i)}$ for the
lowest occupied single-particle states, although $\hat{h}^{(i)}$ 
continues to be updated for much longer by the SCF iterations. When using 
the heavy-ball scheme, the preconditioner also visibly reduces the number 
of necessary SCF iterations compared to linear mixing.

%\vspace{2cm}
%
%%%%%%%%%%%%%%%%%%%%%%%%%%%%%%%%%%%%%%%%%%%%%%%%%%%%%%%%%%%%%%%%%%%%%%%%%%%%%%%

\section{Discussion, summary, and conclusions}

The self-consistent mean-field equations present a non-linear
numerical problem that needs to be solved with iterative methods. Carrying out
a calculation such that it converges reliably and efficiently to the targeted
nuclear configuration requires the control over many interlocking aspects of
the numerical problem. We have discussed here the role of the two 
most essential components of such a calculation.

The first component is the diagonalisation subproblem: the diagonalisation 
of the single-particle Hamiltonian for a given set of mean-field potentials. 
In many cases, iterative diagonalisation algorithms are the most efficient 
ones when dealing with a large single-particle basis.

The second component is constituted by the SCF iterations, whose goal is to 
find a fixed point of the mean-field equations for the densities and 
potentials that determine the single-particle Hamiltonian. This component 
always requires an iterative approach.

For both components, the design of the iterative process algorithms 
has to take into consideration that there are ``soft'' and ``stiff'' directions
in the solution space. The most problematic for the stability of the
numerical solution are highly oscillatory modes: already a tiny accidental 
admixture of such modes during the updates of either the
single-particle states or the potentials can excite the system
and drive it away from convergence. For the Skyrme EDF, the 
degrees of freedom that dominate such behaviour in the 
diagonalisation subproblem on the one hand and the SCF iteration on the
other hand can be easily identified. The former is connected to terms
with external gradients of the local densities in Eqs.~\eqref{eq:SkTeven:2},
\eqref{eq:SkTodd:2}, and~\eqref{eq:EDF:4:e}, whereas the latter is connected
to the number of gradients contained in the local densities defined through
Eqs.~\eqref{eq:densities:NLO} and~\eqref{eq:densities:N2LO}. 
For finite-range interactions, the inadequate treatment of large 
eigenvalues of the single-particle Hamiltonian or of a large Jacobian 
of the SCF iteration will lead to the same convergence problems. For such 
types of EDF, however, the dominant terms that contribute to each are 
not straightforward to identify, in particular when keeping the exact
exchange terms.

In practice, the presence of these problems results in the
need to select one or or even several numerical parameters such
that numerical instabilities are avoided, with the side-effect of slowing
down overall convergence. The feasible (and optimal) values of these
parameters depend in general on the numerical representation, the range
of resolved momenta, the properties of the parametrisation of the 
EDF used in a calculation, and sometimes even on the nucleus. Therefore,
the numerical parameters have either to be adapted case by case or to be 
fixed on the safe side which translates into human cost in the first case 
and in increase of CPU time in the second case.

We have discussed three distinct aspects of the convergence of
self-consistent calculations with Skyrme EDFs in this study, which can be
summarised as follows.

\paragraph{Parametrisation dependence of the diagonalisation subproblem.}

The problematic degrees of freedom for the diagonalisation subproblem are 
related to the largest eigenvalues of the single-particle Hamiltonian, which 
correspond to highly oscillatory single-particle wave functions. Proceeding too
quickly in the iterative diagonalisation can have the side-effect of 
unintentionally amplifying (instead of damping) these oscillations.

For a Skyrme EDF, the coupling to such states is determined by the 
internal gradients present in the local densities out of which the functional 
is built. For NLO parametrisations with realistic effective masses in the 
four spin-isospin channels the  kinetic energy is always the leading term 
in that respect.

The Skyrme EDF at N2LO yields additional contributions, some of which
scale differently because of their higher number of internal gradients such 
as the density $Q(\vec{r})$ of Eq.~\eqref{eq:densities:N2LO}. 
There is no a priori about the relative sign of coupling constants of the 
additional terms, which leads to a complicated interplay between terms that, 
depending on the parametrisation, might either reduce or amplify the
coupling to oscillatory modes compared to the NLO case.

\paragraph{Parametrisation dependence of the SCF subproblem.}

The algorithm for SCF iteration might inadvertently amplify 
oscillatory modes of the densities and mean-field potentials. By 
linearising the evolution, this behaviour can be directly related to the 
largest eigenvalues of the Jacobians of the SCF evolution.

For a Skyrme EDF, these problematic eigenvalues are largely determined by 
the presence of external derivatives of local densities in the EDF.
For most parametrisations at NLO, the isoscalar
$\rho_0(\vec{r}) \Delta \rho_0(\vec{r})$ term
dominates the evolution of the SCF iteration, as among the contributions
to the potentials with two external gradients it is the one with largest 
coupling constant. As this term provides a dominant contribution to the 
surface tension~\cite{Ryssens18b}, for realistic parametrisations its
coupling constant is always of the same sign and takes very similar values.
This is different for the homologues of this term in the other
spin-isospin channels that often vary over a large range, which can 
result in the need to fine-tune numerical parameters.

At N2LO there are again additional terms containing gradients of densities, 
some of which concern other potentials, and some of which have a different 
scaling because of their higher number of external gradients, an 
example being the $\rho(\vec{r}) \Delta \Delta \rho(\vec{r})$ term in 
Eq.~\eqref{eq:EDF:4:e}. Again, there is no a priori about the relative sign 
of the coupling constants of the new terms, which again might require a 
case-by-case fine-tuning of numerical parameters of the SCF iteration. 

These convergence problems have to be distinguished from finite-size 
instabilities of the parametrisations of the EDF as discussed in 
Refs.~\cite{Hellemans12,Hellemans13,Lesinski06,Pastore2015}.
Although their effect on the progression of an iterative calculation is the
same, i.e.\ they trigger oscillations that grow larger at each iteration,
their origin is different: the numerical instability discussed here results 
from the unintended coupling to oscillatory modes that are less bound.
By contrast, a finite-size instability of a given parametrisation results
from states with a highly oscillatory density distribution being more bound 
than states with a physical density distribution. The former can be avoided 
with a suitable choice of numerical parameters, whereas the latter 
is a physical attribute of a parametrisation that might, however, 
not always be well resolved by a given numerical 
representation~\cite{Hellemans13}.

\paragraph{Algorithmic improvements.}

We have discussed two often-used methods, gradient descent and linear 
density mixing and their limitations with respect to the treatment of 
the problematic oscillatory modes in both the diagonalisation 
and SCF subproblems. To alleviate said limitations, we have proposed two 
algorithmic improvements, heavy-ball dynamics and potential preconditioning,
which have several major advantages.

Compared to the gradient-descent approach, the heavy-ball algorithm,
requires significantly less iterations in all cases we have 
investigated so far, and this at a negligible increase of numerical cost 
of each iteration. The improvement is especially striking
when the energy surface is soft around the targeted 
configuration, or when very large eigenvalues of the single-particle 
Hamiltonian are present. In a coordinate-space representation, 
the heavy-ball method removes the inverse scaling of the number of total 
iterations with the mesh parameter that otherwise leads to a 
disproportionate increase of computational cost when reducing $dx$.

In spite of their large difference in performance, the implementation 
of the heavy-ball method differs only little from the gradient-descent 
approach and can be easily implemented into existing solvers using that
method. In the context of self-consistent mean-field calculations,
the number of additional operations and the size of additional storage space 
are negligible compared to what is already needed for the more basic gradient
descent.

The second algorithmic improvement concerns the SCF iteration. Instead of
indiscriminately damping changes at all wavelengths, we propose a
preconditioning scheme for the mean-field potentials that 
permits for more rapid updates of long-wavelength changes compared 
to short-wave\-length ones. This selective damping especially improves 
the convergence of calculations for deformed nuclei, as deformation 
can be allowed to changed more rapidly while maintaining the stability 
of the calculation. As for heavy-ball dynamics, the method is 
straightforward to implement into existing solvers employing linear mixing.
Although the preconditioning step demands significantly more CPU 
operations than linear mixing, its impact on overall CPU time is negligible,
as the cost of the diagonalisation subproblem remains the dominant one.

Finally, we propose a dynamical estimate of the optimal numerical 
parameters for both gradient-descent and heavy-ball dynamics. Also, it
turns out that for existing parametrisations of the Skyrme EDF the 
performance of potential preconditioning for the SCF subproblem is not 
sensitive to the choice numerical parameters. In this way, human fine-tuning 
of parameters is no longer necessary in virtually all cases, 
without sacrificing CPU time.

\paragraph{Outlook.}

The better understanding of the limiting factors of the diagonalisation 
and SCF subproblems allowed us to propose algorithmic improvements that 
result in a reduction of needed CPU time that often reach a factor of ten, as 
well as an almost complete elimination of the need for fine-tuning the numerical
parameters. These advances can now be exploited to perform systematic 
calculations across the nuclear chart at a significantly reduced cost, both in 
terms of CPU and human time. This improvement is particularly welcome for 
the exploration of the possibilities of future N2LO parametrisations, for 
which the performance of gradient descent and linear mixing can 
change dramatically as a result of small modifications of the coupling
 constants.

More specifically, with these improvements CPU-inten\-sive calculations with 
our MOCCa code that simultaneously break multiple point-group symmetries can 
now be performed on a large scale, especially for systems whose energy 
surface is very soft in one shape degree of freedom. This opens the way for 
future systematic investigations of rotational bands based on exotic nuclear 
configurations.

%%%%%%%%%%%%%%%%%%%%%%%%%%%%%%%%%%%%%%%%%%%%%%%%%%%%%%%%%%%%%%%%%%%%%%%%%%%%%%%

\section*{Acknowledgements}

The majority of calculations have been performed at the Computing Centre of
the IN2P3. Another significant part of computational resources have been
provided by the Consortium des {\'E}quipements de Calcul Intensif (C{\'E}CI),
funded by the Fonds de la Recherche Scientifique de Belgique (F.R.S.--FNRS)
under Grant No.~2.5020.11.

\appendix

%%%%%%%%%%%%%%%%%%%%%%%%%%%%%%%%%%%%%%%%%%%%%%%%%%%%%%%%%%%%%%%%%%%%%%%%%%%%%%%

\section{The two-basis method for HFB calculations in coordinate space}
\label{app:twobasis}

Most implementations of the HFB equations represent Eq.~\eqref{eq:HFB}
as a $2 N_b \times 2 N_b$ diagonalisation problem in the
numerical single-particle basis, where the quasi-particle states
$(U_{\mu} V_{\mu})^T$ are given by the columns of the sub-matri\-ces of
the Bogoliubov transformation~\cite{RingSchuck}
\begin{align}
\beta_{\mu}
& =  \sum_{j = 1}^{N_b}
     \big( U_{j \mu}^* \, a_{j} + V_{j \mu}^* a^\dagger_j \big)
\qquad \text{for $\mu = 1, \ldots, N_b$.}
\end{align}
Instead of diagonalising the HFB Hamiltonian in the numerical basis,
it can also be diagonalised in the single-particle basis
$|\psi_{j}\rangle$ that diagonalises the single-particle Ha\-mil\-tonian
\begin{equation}
\label{eq:eigenhapp}
\hat{h} | \psi_{j} \rangle
= \epsilon_{j} \, | \psi_{j} \rangle \, ,
\end{equation}
with eigenvalues $\epsilon_j$ that are usually called single-particle energies.
The HF basis is obtained from the numerical basis by the unitary transformation
$C C^\dagger = C^\dagger C = 1$ that diagonalises
the single-particle Hamiltonian
\begin{align}
 C^\dagger h C   = \epsilon \, ,
\end{align}
where $\epsilon$ is the diagonal matrix with the single-particle energies as
entries.
%\begin{align}
%|\psi_j\rangle
%& = \langle \vec{r} | c^\dagger_j | - \rangle
%  = \langle \vec{r} | \psi_{j} \rangle
%  = \sum_{k=1}^{N_b} C_{jk} \, \phi_k(\vec{r}) .
%%\psi_j(\vec{r})
%%& = \langle \vec{r} | c^\dagger_j | - \rangle
%%  = \langle \vec{r} | \psi_{j} \rangle
%%  = \sum_{k=1}^{N_b} C_{jk} \, \phi_k(\vec{r}) .
%\end{align}
%
In the basis of the eigenstates of the single-particle Hamiltonian, the HFB
matrix takes the form
\begin{eqnarray}
\label{eq:HFBtwobasis}
\tilde{\mathcal{H}}
\begin{pmatrix}
       \tilde{U}_{\mu} \\
       \tilde{V}_{\mu}
\end{pmatrix}
& = &
\begin{pmatrix}
  \epsilon - \lambda  &   \tilde{\Delta}      \\
 -\tilde{\Delta}^*    & - \epsilon + \lambda
\end{pmatrix}
\begin{pmatrix}
       \tilde{U}_{\mu} \\
       \tilde{V}_{\mu}
\end{pmatrix}
= E_{\mu}
\begin{pmatrix}
       \tilde{U}_{\mu} \\
       \tilde{V}_{\mu}
\end{pmatrix}
\, ,
\end{eqnarray}
where the tilde in $\tilde{\mathcal{H}}$ and $\tilde{\Delta}$ indicates that 
these matrices are now calculated in the HF basis. The diagonalisation of 
$\hat{h}$ and $\tilde{\mathcal{H}}$ are of course interlaced, as the density 
matrix in the HF basis $\rho_{jk} = (\tilde{V}^* \tilde{V}^T )_{jk}$ enters the
calculation of the densities that determine $\hat{h}$.

Originally proposed in Ref.~\cite{Gall94}, the two-basis method consists of
determining the quasi-particle basis that diagonalises $\mathcal{H}$ in terms
of the HF basis. It has been used in 3d coordinate-space calculations since
\cite{Gall94,Terasaki96,Ryssens18b}. It has also been implemented into the 1d 
spherical coordinate-space HFB solver \textsc{Lenteur} \cite{lenteur} and the
3d HO code \textsc{HFODD} \cite{Schunck12}, in both cases with an
exact diagonalisation of $\hat{h}$ at each self-consistent field iteration.

Up to this point we just performed a basis transformation. When 
introducing
an effective pairing interaction with a suitably chosen cutoff $f_\ell$
that in the HF basis is a function of the distance $(\lambda - \epsilon_\ell)$
of the single-particle energy $\epsilon_\ell$ from the Fermi energy
$\lambda$, the pairing energy $E_{\text{pair}}$ and pairing gaps
$\tilde{\Delta}_{ij}$ become
\begin{align}
E_{\text{pair}}
& = \tfrac{1}{2} \sum_{ijkl}
     f_i f_j f_k f_l \, \tilde{v}_{ijkl} \,
    \tilde{\kappa}_{ij}^* \, \tilde{\kappa}_{kl} \, ,
     \\
\tilde{\Delta}_{ij}
& = f_i f_j \tfrac{1}{2} \sum_{kl} f_k f_l \, \tilde{v}_{ijkl} \,
    \tilde{\kappa}_{kl} \, ,
\end{align}
where $\tilde{v}_{ijkl}$, $\tilde{\kappa}_{ij}^*$ and $\tilde{\kappa}_{kl}$
are now also calculated in the HF basis.
With such a cutoff, single-particle states above the pairing window have an
occupation zero,such that the elements\footnote{Note that
in our notation $\tilde{\rho}$ is the normal density matrix in the HF basis.
Many papers use $\tilde{\rho}$ for a pair density matrix instead.}
$\tilde{\rho}_{jk} = (\tilde{V}^* \tilde{V}^T )_{jk}$ and
$\tilde{\kappa}_{jk} = (\tilde{V}^* \tilde{U}^T )_{jk}$ of the normal and 
anomalous density matrices that multiply their contribution to
many-body observables are zero. For these states, $\tilde{\mathcal{H}}$
is already diagonal, as the corresponding matrix elements in
$\tilde{\Delta}$ are also zero. It is therefore sufficient to limit
the quasiparticle basis to $\Omega$ states
\begin{align}
\beta_{\mu}
& =  \sum_{j = 1}^{\Omega}
     \big( \tilde{U}_{j \mu}^* \, c_{j}
          + \tilde{V}_{j \mu}^* c^\dagger_j \big)
\qquad \text{for $\mu = 1, \ldots, \Omega$,}
\end{align}
where $\tilde{U}$ and $\tilde{V}$ are the submatrices of the Bogoliubov 
transformation between the HF basis and the HFB basis and $\Omega$ is the size
of a set of single-particle states in the HF basis chosen large enough that it 
encompasses those for which matrix elements $\tilde{V}_{j\mu}$ are non-zero. 
With this, one has mapped the HFB problem from the diagonalisation of a 
$2 N_b \times 2 N_b$ problem to two coupled problems of much smaller 
dimension.

In the calculations presented here, we have systematically employed
the pairing interaction as proposed in Ref.~\cite{Rigollet99} with a cutoff of
5~MeV above and below the Fermi energy.

%%%%%%%%%%%%%%%%%%%%%%%%%%%%%%%%%%%%%%%%%%%%%%%%%%%%%%%%%%%%%%%%%%%%%%%%%%%%%%%%

\section{The link with the damped harmonic oscillator}
\label{app:damposc}

We briefly show the link between the heavy-ball dynamics equations and the
equations of the damped harmonics oscillator. For a more in-depth analysis, 
we refer the reader to Ref.~\cite{Qian99}.

The classical equation of motion of a one-dimensional damped harmonic
oscillator with coordinate $x(t)$ is given by
\begin{equation}
m \frac{d^2 x(t)}{dt^2} + M \frac{dx(t)}{dt}  + \omega^2 x(t) = 0\, .
\label{eq:damposc}
\end{equation}
where $m$ is the mass and $\omega$ the oscillator frequency, while 
$M$ is a constant that governs the friction applied to the oscillator. 
In order to numerically integrate the equation of 
motion~\eqref{eq:damposc}, it has to be discretized. In terms of 
a (small) step size $\delta t$ and an auxiliary variable $V(t)$,  we have
\begin{equation}
\left\{
\begin{split}
V(t+\delta t) &= \frac{ m \delta t}{\left( m + M \delta t \right) } V(t) -
  \frac{ \delta t^2 \omega^2}{\left( m + M \delta t \right) } x(t) \, , \\
 V(t) &= x(t + \delta t) - x(t) \, ,
\end{split}
\right.
\end{equation}
Making a further change of variables
\begin{equation}
\begin{split}
\mu       = \frac{ m \delta t }{\left( m + M \delta t \right) } \, , & \qquad
\Omega^2  = \frac{   \delta t \omega^2 }{\left( m + M \delta t \right) }  \, ,
\end{split}
\end{equation}
and rearranging both equations, we obtain
\begin{equation}
\left\{
\begin{split}
x(t+\delta t) &=  x(t) + V(t+\delta t)  \, , \\
V(t+\delta t) &= - \delta t \Omega^2 x(t) + \mu V(t) \, . 
\end{split}
\right.
\end{equation}
Given appropriate initial conditions, this coupled set of equations can be
integrated. It is identical to Eq.~\eqref{eq:momeigen}, with the 
iteration count taking the role of discrete time, the stepsize $\alpha$ 
taking the role of $\delta t$, and the eigenvalue $\lambda_k$
figuring as the square of the (modified) oscillator frequency $\Omega^2$.

Let us classify the various solutions of Eq.~\eqref{eq:damposc} for an
oscillator starting from $x(0) = x_0$, with $V(0) = 0$. One can verify that
the following prescription
\begin{equation}
x(t) = \frac{x_0}{2} e^{-\gamma t} \left(e^{i \omega_0 t} 
     + e^{-i \omega_0 t} \right)  \, ,
\end{equation}
is a solution to Eq.~\eqref{eq:damposc}, provided that
\begin{align}
\gamma     = \frac{M}{2m}\, , & \qquad %\, ,\\
\omega_0^2 = \frac{\omega^2}{m} - \frac{M^2}{4m^2} \, .
\end{align}
We differentiate three different regimes, depending on the value of
$\omega_0^{2}$
\begin{equation}
\left\{
\begin{array}{ll}
\omega_0^{2} >  0 & \text{underdamped} \, , \\
\omega_0^{2} =  0 & \text{critically damped} \, , \\
\omega_0^{2} <  0 & \text{overdamped} \, . \\
\end{array}
\right.
\end{equation}
When underdamped, $x(t)$ oscillates rapidly around zero with an 
exponentially decaying amplitude. When overdamped, the system decays
exponentially to zero without oscillations. The rate at which is does so
is determined by both $\gamma$ and $\omega^{2}$: when $\omega_0^2$ becomes
more negative, the system decays increasingly slow. For a 
critically damped system, the system also does not oscillate 
around zero, but rather decays directly at a rate of $-\gamma$.

Achieving critical damping should be the goal if we desire a system 
that decays as rapidly as possible. The condition on the damping $M$ is
\begin{equation}
M_{\rm crit}^2 = 4 m \omega^2 \, .
\end{equation}
This corresponds in the discrete case to
\begin{equation}
\mu_{\rm crit} = \left( 1 - \sqrt{\delta t \Omega^2} \right) + O(\delta t^2)\, ,
\end{equation}
where we ignore corrections of second order in $\delta t$.

The gradient-descent equations can be obtained by simply setting $m$ to 0. The
system then is always overdamped and decays only comparatively slowly to
$x=0$.

The discretization of a differential equation cannot be reliably numerically 
integrated for arbitrary $\delta t$, as too large values can lead to 
unstable integration schemes. This is the direct analogue for differential
equations of condition Eq.~\eqref{eq:dthbcondition}; if $\Omega^2$ is large, 
then Eq.~\eqref{eq:damposc} is a so-called stiff differential equation
\cite{NumRecipes}. For such cases, the numerical integration 
requires either very small time-steps or more advanced discretization 
schemes, which is another way of interpreting the upper 
bound on the stepsizes $\alpha$ and $dt$ discussed in the main text.

%%%%%%%%%%%%%%%%%%%%%%%%%%%%%%%%%%%%%%%%%%%%%%%%%%%%%%%%%%%%%%%%%%%%%%%%%%%%%%%%

%%%%% CLEAR DOUBLE PAGE!
%\newpage{\pagestyle{empty}\cleardoublepage}

\end{document}